\def\fileversion{v2.25} \def\filedate{2005/03/25}
\documentclass[fullpage]{uiucthesis}
\usepackage{amsmath}
\usepackage{graphicx}
\usepackage{epstopdf, epsfig}

\usepackage{newtxtext}
\usepackage{newtxmath}
\usepackage{natbib}
\usepackage{hyperref}
\usepackage{xcolor}
\usepackage{subcaption}
\usepackage[symbol]{footmisc}

\hypersetup{
   colorlinks = true,
   linkcolor  = black,
    urlcolor   = blue,
    citecolor  = black,
}
\usepackage[utf8]{inputenc} 
\usepackage[english]{babel}
\usepackage{physics}
\usepackage{float}
\usepackage{enumitem}  
\usepackage{lipsum}
\usepackage{dcolumn}
\usepackage{bm}
\newcommand{\RNum}[1]{\uppercase\expandafter{\romannumeral #1\relax}}

\newcommand{\pkd}[1]{ {\color{black} #1 }}

\begin{document}
\title{NON-INERTIAL HYDRODYNAMICS OF MANIPULATING PARTICLE TRANSPORT}

\author{PARTHA KUMAR DAS}

\department{Mechanical Science and Engineering}
\phdthesis
\degreeyear{2026}
\college{Grainger College of Engineering}
\committee{
     Professor Sascha Hilgenfeldt, Chair\\
     Professor Mattia Gazzola\\
     Professor Jie Feng\\
     Professor Andres Goza}

\maketitle

\frontmatter

\begin{abstract}

Inspired by numerous lab-on-a-chip, biomedical and bioengineering applications such as cell sorting, focusing, trapping, and filtering of particles, manipulation of micron-sized particle trajectories has been of significant interest in the context of microfluidics. Systematic deflection of microparticles away from their initial streamlines is a central objective in microfluidic particle manipulation. In many widely used microfluidic platforms—including deterministic lateral displacement (DLD) devices— density-matched, force-free particles suspended in low–Reynolds-number flows encounter arrays of obstacles that potentially breaks the flow symmetry and alter their trajectories. Despite the prevalence of these devices, the physical mechanism responsible for particle deflection from encountering obstacle wall in strictly non-inertial flows (Stokes flows) remains incompletely understood and is often attributed to short-range contact interactions rather than hydrodynamic effects.

In the first part of this dissertation, we demonstrate that a net deflection of force-free particles passing an obstacle in Stokes flow can arise purely from hydrodynamic interactions, provided that the combined flow–obstacle configuration breaks fore–aft symmetry. By rigorously modeling the wall interaction from an inclined elliptic obstacle through the velocity corrections of particles being transported in Stokes flow, we show that wall-induced hydrodynamic interactions generate a systematic spatial displacement of particles, even though the governing equations of motion remain time-reversible. This establishes symmetry breaking as the essential ingredient enabling deterministic particle deflection in inertial-less transport flows around obstacles. 

We further demonstrate the obstacle's capacity in deflecting particles maximally through a rigorous analytical approach where we derive the analytical scaling laws that quantify the magnitude of this maximum deflection as a function of particle size, obstacle geometry, and flow configuration. These predictions are validated through direct numerical simulations of particle trajectories, demonstrating excellent agreement across a broad parameter range. For practically relevant conditions, the magnitude of hydrodynamic deflection is found to be comparable to that predicted by models invoking particle–surface contact or roughness-mediated interactions, indicating that hydrodynamics alone can account for a substantial fraction of the observed displacement in practical devices. 

Having systematically developed the hydrodynamics of particle motion around an obstacle in Stokes flow, this dissertation extends beyond particle deflection to examine how the developed hydrodynamic transport maneuvers particles sufficiently close to obstacle surfaces for short-range attractive forces to dominate, leading to particle capture or sticking and accumulate at specific locations, a qualitatively new type of particle manipulation central for particle filtration in porous media microfluidics. By quantifying the individual wall-normal force components on a force-free particle, we show that our asymmetric elliptic obstacle can effectively amplify the hydrodynamic force required for overcoming the short-range electrostatic repulsion making attachment possible.

Finally, addressing multiple obstacle induced asymmetry in flow structure in DLD device (analogous to porous media as well), we systematically employ our symmetry breaking strategy in Stokes flow around two-cylindrical obstacle. This work provides an elementary basis for interpreting particle transport, enriching the toolbox of passive microfluidic manipulation techniques.

\end{abstract}

\begin{dedication}
To my parents, who have long known my fondness for esoteric challenges.
\end{dedication}

\chapter*{Acknowledgments}

I would like to begin with expressing my deepest gratitude to my advisor, Professor Sascha Hilgenfeldt, for his extraordinary mentorship, intellectual guidance, and unwavering support throughout my doctoral studies without which this work would not be possible. The rich experience I have accrued through my interactions with him fundamentally shaped my research ideology and temperament, and inevitably helped me to become a more mature researcher in the vast field of fluid mechanics. I am particularly grateful for the opportunity to conduct experiments and to develop a deep theoretical understanding of the underlying physics-- long-desired pursuits that have defined my PhD journey. He has afforded me a remarkable degree of intellectual and personal freedom throughout my time here, for which I am sincerely thankful. 

I would like to acknowledge the valuable comments and insightful discussions about my work provided by the other members of my doctoral committee: Professor Mattia Gazzola, Professor Jie Feng, and Professor Andres Goza. I also thank Professor Pietro de Anna and his postdoc Filipo Miele for the opportunity of collaboration and fruitful discussion with porous media filtration. This work would scarcely have been possible without the rigorous foundation in fluid mechanics, experimental microfluidics, applied mechanics and mathematics that I was provided with through graduate classes at MechSE. I would like to especially thank Professor Randy Ewoldt for those wonderful fluid mechanics lectures which have played a crucial role during my early years as a graduate student.

I wish to thank all the people with whom I have had the pleasure of working. Specifically, I would like to thank Sidhhansh Agarwal and Xuchen Liu. The useful interactions with them have contributed significantly to the success of this work. I would be remiss if I did not acknowledge the support of the administrative staﬀ of the MechSE department, MNMS cleanroom, and Machine shop. I would especially like to thank Kathy Smith for her patience and willingness to guide countless graduate students, like me, through various stages of graduate school. I would also like to thank Professor Taher Saif and his graduate student Saeedur Rahman for the important discussion on research and career at different stages of my study here. I want to thank Professor Blake Johnson, Professor Wayne Chang, and Dr David Farrow with whom I worked as a teaching assistant. They have provided me with not only valuable teaching and mentoring experience but also an opportunity to contribute to instrumentation, technical management, and pedagogical course development.

I express my sincere gratitude to all my friends in Urbana-Champaign and beyond, who have been the source of innumerable discussions on life, politics, and everything in between, and made life outside of research enjoyable and fulfilling. 
Their friendship and support made it easier to persevere through the vicissitudes of graduate school, especially during the time of pandemic.

My warmest thanks to my wife, Medha Kundu, for her boundless aﬀection, patience, faith in me, and for being an unwavering source of emotional support through this journey. Finally, I would like to thank my parents who endured this long process with me, always offering support and love.

\tableofcontents
\listoffigures



\mainmatter

\chapter{Introduction} \label{chap intro}

\section{Microfluidics for manipulating particles}\label{section PM microfluidics}

Through a rich timeline of development in microfabrication technology \cite{fan2023celebrating, whitesides2006origins}, microfluidics has emerged as a powerful platform for the controlled manipulation of micron- and submicron-sized particles ranging from $100\ nm$ to $1\ mm$. Driven by its ability to precisely regulate fluid flow at low Reynolds numbers, characterization of flow physics on the microscale through the use of proper tracer microparticles as well as quantification of particle movement have become central topics in both academic and industrial microfluidic research. 

A wide spectrum of applications, ranging from critical activities such as the sorting and filtration of airborne contaminants to safeguard public health \cite{humes2005particle}, to the intricate mechanical testing of macromolecules such as DNA \cite{tabeling2005introduction}, to delicate procedures of the assisted fertilization of human ova with immotile sperm \cite{fleming2003micromanipulation}, as well as the precise stirring of chemical processes \cite{ault1998techniques} necessitate precise manipulation of particles suspended in fluids. Within the scope of microfluidic research and its industrial application such as disease control and diagnosis \cite{pamme2007continuous, ateya2008good, nilsson2009review, xuan2010particle, gossett2010label, puri2014particle}, drug discovery and delivery systems \cite{dittrich2006lab, kang2008microfluidics, nguyen2013design} or self-cleaning technologies \cite{callow2011trends, kirschner2012bio, nir2016bio}, controlled manipulation, encompassing both passive (synthetic) or active (biological, e.g., cells) particles, holds immense significance. Fundamentally, these manipulation strategies are quantified by forcing the particles to cross streamlines so that they do not passively follow the flow. This essence has been directly implemented in lab-on-a-chip processing as well as in the diagnostic of biological samples and biomanufacturing processes.  

Microfluidic technologies have revolutionized particle manipulation, offering a diverse array of capabilities including transportation, separation, deposition, trapping, and enrichment \cite{pratt2011rare,sajeesh2014particle,lu2017particle,pradel2024role,dersoir2015clogging,bordoloi2022structure}   through both passive and active manipulation strategies \cite{sourani2026manipulation}. Passive methods rely solely on the hydrodynamics of the flow induced by device geometry (i.e. obstacles, interface, ridges, or constrictions) \cite{karimi2013hydrodynamic,di2007continuous,di2009inertial,salafi2019review}, while active methods introduce external fields such as acoustic \cite{laurell2007chip, ozcelik2018acoustic, schmid2014sorting}, electric \cite{xuan2019recent}, optical \cite{lenshof2010continuous}, or magnetic \cite{van2014integrated} forces to influence particle motion. One of the notable microfluidic platforms for spatial separation of particles and cells that follow different trajectories based on their sizes is Deterministic Lateral Displacement (DLD) devices \cite{zhang2020concise, nasiri2020microfluidic,lin2020progress, loutherback2010improved,kruger2014deformability,liu2016particle,kabacaouglu2019sorting,kabacaouglu2018optimal}. Despite this wide range of applications particularly in the context of leveraging hydrodynamics, a notable gap persists in the quantitative understanding of these devices' operational mechanisms. 

Addressing this gap, our work pioneers a comprehensive theoretical framework elucidating the dynamics of rigid spherical particles flowing over a favorably positioned obstacle in the Stokes flow background. Through a rigorous quantification of particle displacement over such obstacles, making use of recently developed theory of the motion of particles in Stokes flow near walls, our model enhances our understanding of particle motion from the academic perspective. Based on this foundation, our work inspires novel practical design strategies for precise particle manipulation, not only within continuous flow settings such as in a DLD device (see figure ~\ref{fig DLDPMF}(a-b)) but also in confining particle mobility in porous media filtration microfluidics (see figure ~\ref{fig DLDPMF}(c)) \cite{miele2025flow, mays2005hydrodynamic}, fouling and cleaning \citep{rajendran2025recent,gul2021fouling,kumar2015fouling}, or elimination of pathogens \citep{nuritdinov2025experimental,uttam2023hypothetical,sande2020new}

\section{Particle manipulation leveraging Stokes flow hydrodynamics}\label{section PM Hydro}

There are numerous ways to design microfluidic environments for positional control of small suspended particles. Depending on the subject of such control, i.e. cell and microorganism trapping and biological sample sorting, static and dynamic trapping of active and passive particles, and controlled micro-swimming, different strategies such as, optical tweezers \cite{ashkin1986observation,grier2003revolution,chiou2005massively}, magnetic tweezers \cite{crick1950physical,wang1993mechanotransduction, gosse2002magnetic}, surface acoustic wave (SAW) trapping \cite{ghayour2018development,sesen2014microfluidic,o2012acousto,yeo2014surface}, have been proven to be effective in precise modulation and alteration of particle paths. While the basics of these techniques rely on external force fields, practices making use of diverse flows, which are generally controlled by geometrical means, such as channels or pillars, to induce hydrodynamic forces on particles have caught attention \cite{wang2013frequency,lutz2005microscopic,petit2012selective,tanyeri2011microfluidic,shenoy2016stokes,kumar2019orientation}. 

Pertaining to the low Reynolds number flows in microfluidics, particularly in a lab-on-a-chip system, the long-range hydrodynamic interaction in the Stokes limit $(Re\rightarrow0)$ can lead to enhanced modulation in particle trajectories. Carefully designed flow patterns by deploying ingenious but simple channel and interface geometries \cite{or2009geometric}, within the Stokes limit, have recently been a focal point in creating hydrodynamic traps (Stokes trap) and barriers \cite{petit2012selective,tanyeri2011microfluidic,shenoy2016stokes,kumar2019orientation,chamolly2020irreversible, shenoy2019flow}, allowing for the manipulation of particles in Stokes flow without relying on external forces. The absence of inertia in such fluid environment also provides instantaneity and reversibility and can assist in developing a fine-tuning method in maneuvering particles.  

\begin{figure}
    \centering
\includegraphics[width=\textwidth]{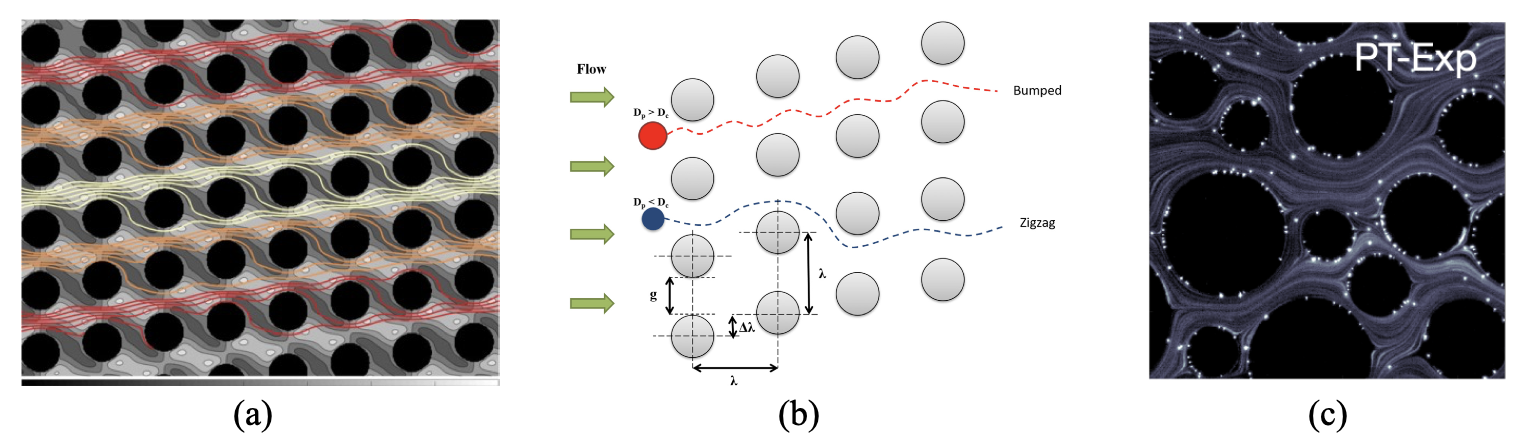}
    \caption{(a) Streamlines through an array of pillars. Flow is from left to right \cite{kabacaouglu2019sorting}. (b) Sketch of particles following different paths \cite{aghilinejad2019transport}. Particles, depending on their size $D_p$, interact with pillars and then either follow the streamlines and “swap lanes” (blue) or stay in the same lane (red). (c) Fluorescent image of particle transport (blurry white strips) through porous media while showing the evidence of attachment of some of them around the obstacles (bright white spots) \cite{miele2025flow}.} 
    \label{fig DLDPMF}
\end{figure}

Deterministic Lateral Displacement devices, commonly known as DLD devices, have been successfully demonstrated as a continuous-flow microfluidic particle separation method. Pioneered by Huang et al. in 2004 \cite{huang2004continuous}, this technique has been later optimized to separate deformable biological cells, yeasts, bacteria, viruses, DNA, and more \cite{zhang2020concise, nasiri2020microfluidic,lin2020progress, loutherback2010improved,kruger2014deformability,khodaee2016numerical,kabacaouglu2019sorting,kabacaouglu2018optimal,hochstetter2020deterministic,lu2023label,aghilinejad2019transport,davis2006deterministic,dincau2018vortex}. A DLD device consists of arrays of pillars forming a lattice. The lattice of the pillar grids is not orthogonal meaning that a unit cell of the pillar grid doesn’t align with the direction of flow. The spatial inclination of the lattice as shown in figure ~\ref{fig DLDPMF}(a) causes changes of the curvature of the streamlines due to the steric requirement of the no-penetration boundary conditions, resulting in zig-zag motion for small particles and straight motion for larger ones \cite{inglis2006critical}. Figure ~\ref{fig DLDPMF}(b) sketches the spacing of the pillars in a DLD arrangement determining this critical particle size for this behavior. No quantification of the particle dynamics, i.e. force analysis on the particles has been made though in explaining streamline crossing by the particles and the size-dependent net non-zero or zero lateral displacements. It is clear that such particle manipulation can be achieved, but it is unclear by what principles the time-reversibility of Stokes flow is broken in the presence of particles and obstacles. Is it sufficient to manipulate obstacle shape or flow geometry? Or is it necessary to take into account many separate obstacles, given the long-range character of Stokes-flow interactions? Several design strategies have been introduced to tackle deformability-based separation by incorporating variations in the shape and size of pillars \cite{kabacaouglu2019sorting,lu2023label,aghilinejad2019transport,davis2006deterministic,dincau2018vortex,loutherback2010improved,kruger2014deformability,khodaee2016numerical,kabacaouglu2019sorting,kabacaouglu2018optimal}, as well as altering the periodic orientation of pillars within the labyrinth. Although such trial-based manipulation and separation techniques are reassuring, there is often a lack to establish the complementary hydrodynamic formalism of the particle-obstacle interaction to properly understand the prescribed motion. A complete theoretical analysis to quantify the particle motion being influenced by the change in curvature of the flow lines because of, for instance, such DLD obstacles will assess the feasibility of the control parameters in such microfluidic devices and provide guidance for practical application. DLD devices do operate in the regime of Stokes flow, but a hydrodynamic analysis of their function has been lacking – modeling descriptions have relied on particles “hopping” streamlines by geometric constraints. A main focus of the present work is to go beyond these approaches.

\section{Review of previous related research}\label{section rev}


Controlled manipulation of micron-sized particle trajectories, motivated by applications in cell sorting, trapping, attachment, and filtration, is a foundational challenge in microfluidics. Among other flow actuation methods, oscillatory flow, employing inertial forces, stands out as a recently well-established strategy for controlled particle manipulation, backed by numerous theoretical investigations into inertial force dynamics \cite{thameem2017fast,agarwal2018inertial,agarwal2024density}. For the vast majority of practical flow geometries, oscillation sets up steady streaming flows simultaneously with the inertial force actuation (see appendix~\ref{appen Rich Variety of Streaming Flow} for details). In such a system, the steady streaming flow is essentially a Stokes flow generated as the second order effect driven by the oscillatory shear layer and cannot be a priori be distangled from the primary oscillatory flow. This type of oscillatory flows are often superimposed by steady transport flows (e.g./ an imposed pressure driven transport flow) for many particle manipulation applications \cite{wang2011size}. Therefore, whether the effect of manipulating particles observed in microfluidics fundamentally requires oscillation is difficult to understand without investigating the effects from oscillatory flow and steady Stokes flow individually.

\subsection{Stokes flow driven by oscillatory boundary layers: Steady streaming flow} \label{subsec streaming}

Time rectification of a periodic flow oscillating with respect to an object gives rise to steady streaming flow extracting the inertial effect to the 2nd order in the oscillation amplitude. Simple objects executing simple motion, such as the prototypical case of a cylinder oscillating translationally, can give rise to complex streaming patterns that have long been studied in the context of transport phenomena \cite{davidson1971cavitation, elder1959cavitation,riley1966sphere,bertelsen1973nonlinear,raney1954acoustical,wang1968high}. It also has great practical relevance in a variety of scientific and engineering applications \cite{wiklund2012acoustofluidics,riley2001steady,lighthill1978acoustic,marmottant2003controlled}. The geometric complexity of streaming from this conventional method is associated with patterns of pairs of vortices \cite{bertelsen1973nonlinear,riley1966sphere,lutz2005microscopic, lutz2006hydrodynamic}. The sensitivity of this complex streaming geometry is directly controlled by oscillation amplitude and frequency. Larger oscillation amplitude leads to stronger streaming flows which has led to recent interest in streaming induced by bubbles \cite{marmottant2003controlled,wang2011size, ahmed2009fast}. In the context of microfluidics, microbubble’s deformation that can be decomposed into distinct oscillation modes, induces mixed mode streaming. A systematic theoretical analysis of sessile bubble streaming flows such as those used in microfluidic devices, considering boundary conditions and modes of oscillation, was given by Rallabandi et al. \cite{rallabandi2014two,rallabandi2015three}. While the presence of the volume mode causes much stronger streaming making it an efficient tool for transport \cite{wang2012efficient}, shear force actuation \cite{marmottant2003controlled}, particle trapping and size sorting \cite{wang2011size,patel2012lateral}, or micro-mixing applications \cite{wang2013frequency,ahmed2009fast}, the robustness and sensitivity of this multi-mode micro-steaming geometry against the interfacial condition needs a thorough demonstration in comparison to the more traditional single mode cylinder streaming. We performed experiments varying the modality, channel geometry, and the dynamic boundary condition using a diffusively stable droplet oscillation. We found extremely robust fountain vortex-pair microstreaming patterns \cite{das2020robustness} (see figure ~\ref{fig streaming}(a)) which is described in more detail in appendix \ref{appen robustness and sensitivity of microstreaming} while our analytical demonstration of the rich variety of multi-modal streaming flows around interface of different boundary conditions is included in appendix \ref{appen Rich Variety of Streaming Flow}

Streaming geometries induced by the local changes in shape curvature of an obstacle, i.e. triangle, square, bullet shaped obstacles placed in a simple back and forth oscillatory flow show features of multiple complex vortex pairs and recirculation zones \cite{tatsuno1974circulatory,tatsuno1975circulatory,bhosale2020shape,bhosale2022multicurvature,parthasarathy2019streaming}. Some qualitative streaming patterns in such lateral fluid oscillation field are shown in figure ~\ref{fig streaming}(b-d). Oscillation frequency and oscillation amplitude have direct control on the qualitative variety in the streaming vortices.

Acoustic streaming induced by gently oscillating micron-sized bubbles has been noted to generate shear forces capable of rupturing lipid membranes \cite{marmottant2003controlled}. The steady deformation of vesicles in simple shear flows has been well investigated. Numerical studies employing boundary integral methods \cite{kraus1996fluid} or particle dynamics methods \cite{noguchi2004fluid}, experimental examinations using a Couette apparatus \cite{de1997deformation} or proximity to a wall \cite{kantsler2006transition,mader2006dynamics}, as well as analytical investigations \cite{seifert1999fluid}, contribute to a comprehensive understanding of this phenomenon.

\begin{figure}
    \centering
\includegraphics[width=\textwidth]{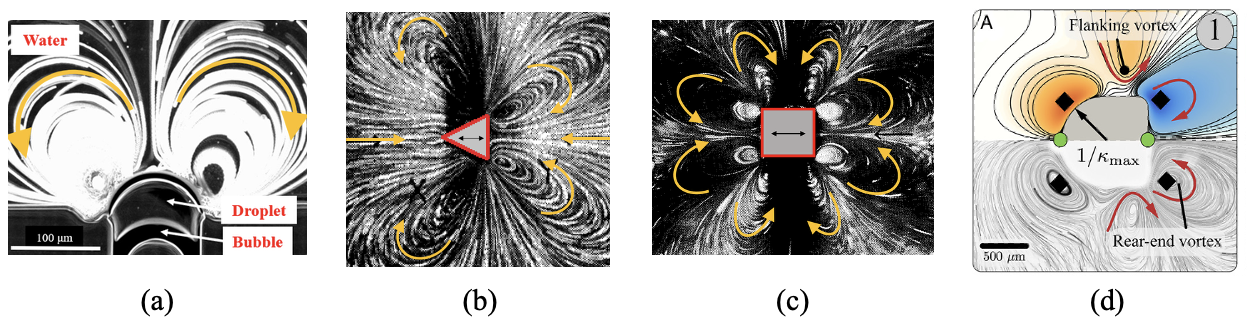}
    \caption{Stroboscopic images of complex streaming patterns. (a) droplet microstreaming, droplet is pinned in the microchannel and driven by the trapped bubble underneath \cite{das2020robustness}. (b-d) in simple back and forth oscillation around (b) an equilateral triangle with sharp edges, vortex pairs form around the sharp corners (abrupt change in local curvature) \cite{tatsuno1975circulatory}, (c) a square inducing multiple fountain and anti-fountain vortex pairs \cite{tatsuno1974circulatory}, (d) a bullet shaped multi curvature obstacle, showing agreement in simulation (upper half) and experiment (lower half) \cite{bhosale2022multicurvature}.}
    \label{fig streaming}
\end{figure}

\subsection{Inertial force in steady streaming flow}\label{subsection inertia streaming}

The steady part of the Reynolds stress in the vorticity transport equation as prescribed by (on high frequency oscillatory viscous flow) serves as the forcing function of steady streaming motion \cite{bertelsen1973nonlinear,wang1968high}. The presence of the Reynolds stress represents a small but finite inertia in microfluidic streaming motion and imparts an inertial force on a particle suspended in such steady streaming flow. The motion of such particle influenced by the streaming inertia is a fundamental fluid dynamical problem and its comprehensive solution remains as a core challenge \cite{di2009inertial,einarsson2015effect,ho1974inertial,hood2015inertial,lovalenti1993hydrodynamic,schonberg1989inertial}.  The theoretical understanding of particle force dynamics has largely relied on the seminal contributions of Maxey and Riley \cite{maxey1983equation}, first introduced four decades ago and comprising various specialized methodologies. An alternative framework has emerged, centered around acoustic secondary radiation forces (SRF), aimed at elucidating observed attractive forces towards specific features in oscillatory flows. Nevertheless, recent experimental and theoretical works have exposed that both the classical MR theory and the SRF hypothesis fall short in explaining the magnitude of attraction. Agarwal et al. \cite{agarwal2021unrecognized} have unveiled previously unrecognized significant forces that act towards oscillating boundaries, even on neutrally buoyant particles, arising from the intricate interplay of particle inertia, flow gradients, and flow curvature. 

The recent approach \cite{thameem2017fast,agarwal2018inertial,agarwal2021unrecognized} for the motion of a particle near an oscillating interface is modeled by a modified version of the Maxey-Riley equation \cite{maxey1983equation} which is an ordinary differential equation for the motion of a rigid sphere (radius $a_p$, density $\rho_p$) of mass $m_p=(4/3)\pi\rho a_p^3$ placed in a general incompressible, know background flow field $\boldsymbol{U}(\boldsymbol{r},t)$ unperturbed by the presence of the particle and with gradient $L_{\Gamma}$ and curvature $L_{\kappa}$ length scales. Density matched $(\rho_p=\rho)$ spherical particles experience an attractive force towards the object. The component of this force along the object to particle connector $\boldsymbol{e}$ takes the explicit form:
\begin{equation}
F_{\Gamma\kappa}
=
m_f \left\langle a_p^2 \, \nabla \mathbf{U} : \nabla \nabla \mathbf{U} \right\rangle
F(\lambda)\,.\mathbf{e}
\end{equation}

$m_f$ is the displaced fluid mass (which is equal to $m_p$ for a density matched particle) and the inner product represents the interaction of flow gradients and curvatures.  

The effect of oscillation frequency is quantified by the universal, analytically derived function $F(\lambda)$ of the Stokes number $\lambda$. For harmonic oscillatory flows with frequency $\omega$, is $\lambda\equiv(a_p^2 \omega)/3\nu$ and to excellent approximation $F(\lambda)$ becomes:
\begin{equation}
F(\lambda)
=
\frac{1}{3}
+
\frac{9}{16}\sqrt{\frac{3}{2\lambda}}
\end{equation}

\begin{figure}
    \centering
\includegraphics[width=0.65\textwidth]{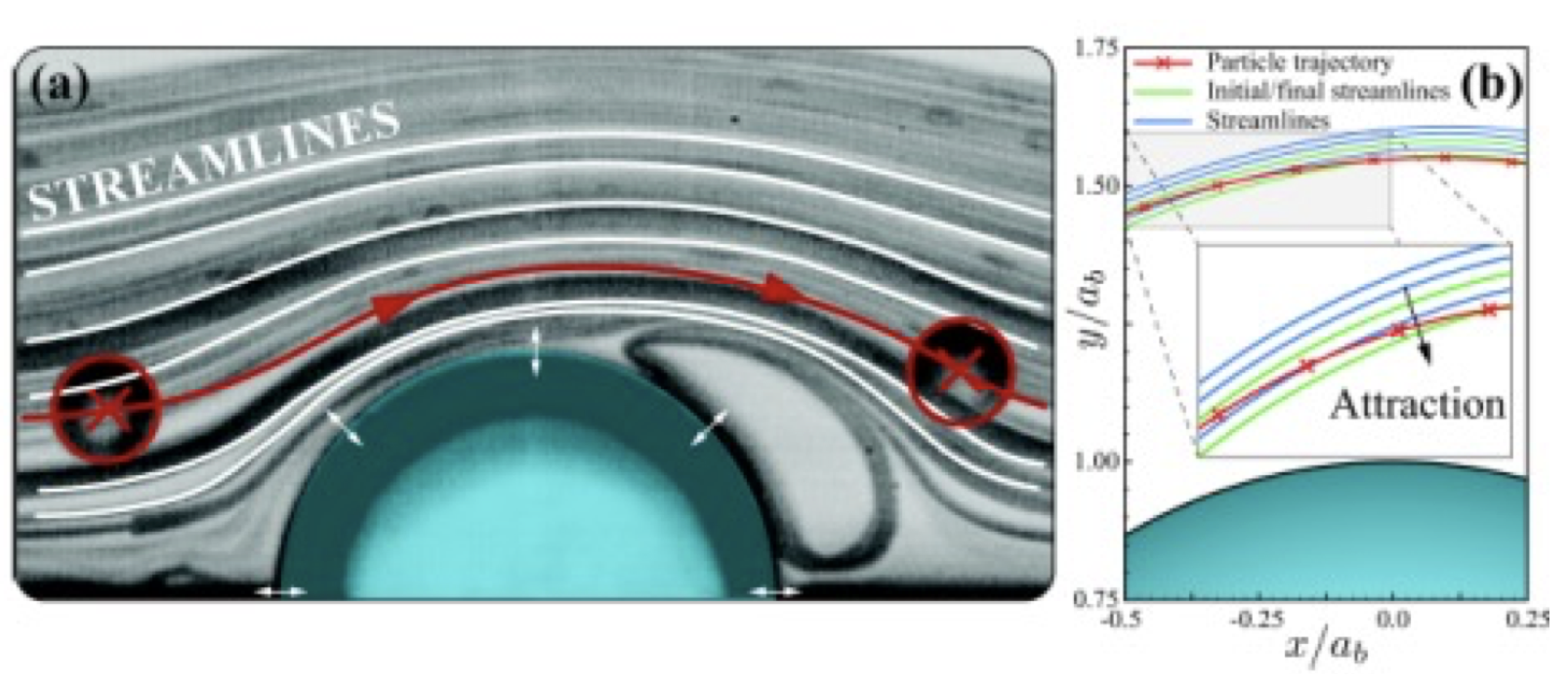}
    \caption{(a) Motion of a neutrally buoyant particle (red) in the vicinity of an oscillating bubble.
(b) Close-up shows that particle trajectory (red) crosses streamlines (blue) while being transported over the bubble, indicating a net attraction towards bubble over fast time scales of a few ms \cite{agarwal2021unrecognized}.}
    \label{fig prev streaming}
\end{figure}

Valid over the entire range from the viscous $\lambda\ll1$ to the inviscid $\lambda\gg1$ limits. In practical cases, the strongest force component of interest is exerted from localized objects, for which equation (1) mainly moves a particle along a radial coordinate measuring distance $r_p$ from the object so that the steady equation of motion becomes:
\begin{equation}\label{eq inertia force}
\frac{\mathrm{d} r_p}{\mathrm{d} t}
=
\frac{F_{\Gamma\kappa}}{6\pi a_p \mu}
\end{equation}

with  $\mu$ the dynamic viscosity of the fluid. As illustrated in \cite{agarwal2021unrecognized} and in figure ~\ref{fig prev streaming}(b), time-averaged particle trajectory described by equation \eqref{eq inertia force} in the background of a steady streaming flow is bent towards the object (a bubble in the figure ~\ref{fig prev streaming}) under the action of a continuous inertial attraction. 
For a more general streaming situation where a particle is recirculating around an obstacle under the action of streaming flow, a similar inertial attraction force can lead a particle to be sub-atomically close to the obstacle where the intermolecular force (Van der Waals attraction force) plays a dominant role and make the particle to stick. This prototypical case of sticking in a streaming flow unveils a new area of particle manipulation, i.e. localized particle sticking, accumulation, and particle cluster formation whose shape and size can be controlled by controlling the streaming orientation and local curvature of the obstacle.

\subsection{Manipulating particle motion for attachment and deposition}\label{subsec sticking}

An important type of particle manipulation aims at capturing particles by making them to stick to obstacles-- more commonly known as porous media filtration \cite{miele2025flow,mays2005hydrodynamic}. Experimental and theoretical studies have shown that filtration in porous media is fundamentally controlled by hydrodynamics at low Reynolds numbers, which maneuver particles to dive towards an obstacle where short-range attractive forces take over and lead to attachment \cite{miele2025flow}. As can be seen from the fluorescent imaging technique in figure ~\ref{fig DLDPMF}(c), these observations in porous media filtration underscore that even in purely Stokes flows, particle motion is strongly influenced by hydrodynamic interactions with nearby boundaries rather than passively following streamlines. The underlying hydrodynamic mechanism in such filtration devices closely parallels particle deflection arising from particle-obstacle interactions in DLD devices operating in Stokes flow regime.

We have also observed particle attachment for neutrally buoyant particles in an oscillatory field around an interface. A neutrally-buoyant particle near an oscillating bubble, as investigated by \citep{maxey1983equation,agarwal2018inertial,agarwal2024density,thameem2017fast}, experiences a continuous attraction towards the bubble. We aim to observe similar effects in the vicinity of an obstacle with sharp change in curvature, i.e. triangular prism, while the particle is recirculating in a streaming flow created by a simple back and forth fluid oscillation around the obstacle. We observed particles sticking to the corners orthogonal to the oscillation direction by being trapped in corner streaming as depicted in figure ~\ref{fig motivation}(a). The details of our experiment is provided in appendix~\ref{appen sticking from streaming}.

Our results serve as evidence of particles sticking selectively to regions of strong curvature in a streaming flow. Sticking is affected by close-range intermolecular forces after hydrodynamics has transported the particles very close to obstacle wall (nm scale). While our experimental demonstration offers the concept of forming specific configurations from particle accumulation leveraging the geometry of the streaming flow, the mechanism of transport close to the wall is still not well understood. The induced streaming flow by itself provides particles with a larger number of wall encounters because of circulation in the vortical structures. But the wall-normal motion of the particles that forces the particle to dive towards the wall could either be due to inertial forces in the oscillatory flow as rigorously described in \citep{maxey1983equation,agarwal2018inertial,agarwal2024density,thameem2017fast}, or due to the steady Stokes flow component of the oscillation-induced streaming flow (a 2nd order effect in oscillation amplitude) \cite{davidson1971cavitation, elder1959cavitation,riley1966sphere,bertelsen1973nonlinear,raney1954acoustical}. Thus, one still needs to answer the following fundamental questions: Can such sticking be achieved in steady Stokes flows that are not driven by oscillations? Can such Stokes flows be designed for desired particle manipulation? 
An elementary trajectory simulation from lubrication wall effect in simplified bubble microstreaming flow is performed by Xuchen Liu \cite{liu2025particlethesis} where particles are shown (see figure~\ref{fig motivation}(b)) to exhibit different spiraling behavior in the steady vortices of the streaming. While this initial test along-with our experimental evidence of streaming induced sticking indicate that particle behavior can be altered across streamlines by the wall-induced hydrodynamic effect in oscillatory boundary layer driven Stokes flow (streaming), this is a second order effect where the leading order effect comes from the oscillatory inertia as mentioned in section \ref{subsection inertia streaming}. The understanding of the accurate inertial and non-inertial contribution in displacing particles necessitates a systematic investigation of pure Stokes flow hydrodynamics as the primary effect induced by the presence of obstacle boundary.

Along-with the objective of formulating the hydrodynamic description of spatial deterministic lateral displacement of particles in DLD devices and identifying the hydrodynamic force that dives particles towards obstacle-- thereby facilitating particle sticking and deposition in Stokes flow either in pure transport through porous media or within steady Streaming vortices-- motivate my PhD work to model hydrodynamic wall effects in particle motion in a steady Stokes flow over an obstacle.

\begin{figure}
    \centering
\includegraphics[width=0.9\textwidth]{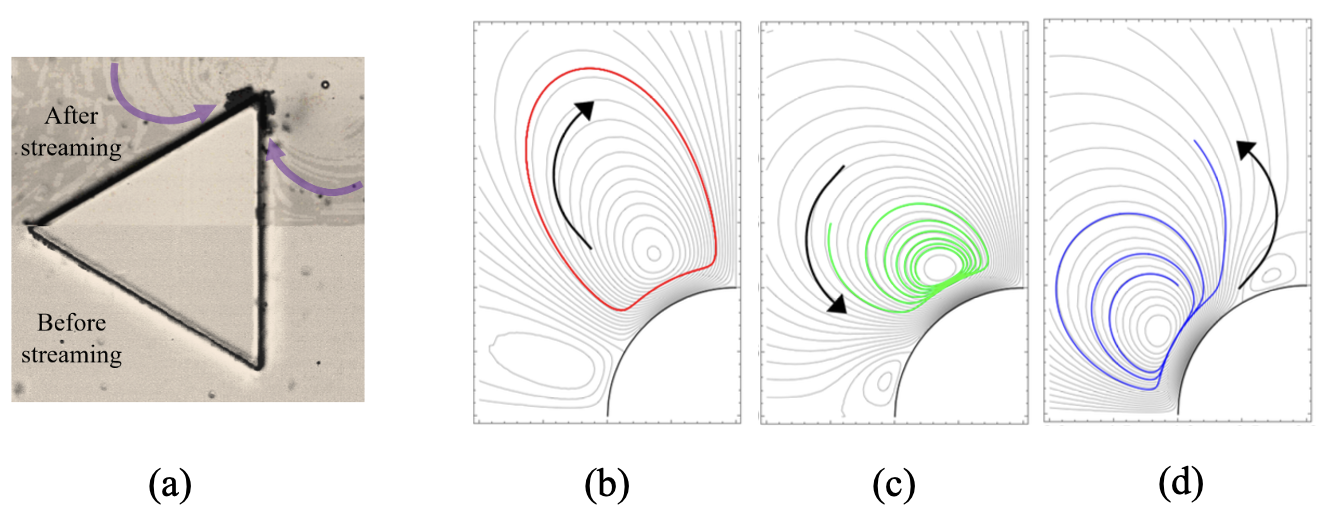}
    \caption{(a) Particle sticking and deposition induced by streaming vortices around the corner of a triangular obstacle (see appendix~\ref{appen sticking from streaming} for more details).
(b-d) particles trajectory exhibiting streamline crossing behavior (close loop (a), spiral in (b), and spiral out (c)) in simplified bubble microstreaming flows. \cite{liu2025particlethesis}.}
    \label{fig motivation}
\end{figure} 

\section{Organization of the dissertation}
The dissertation can be broadly divided as follows. In Chapter.~\ref{chap 2}, we establish the equations of motion for spherical neutrally-buoyant force-free particles in both wall-parallel and wall-normal directions in arbitrary Stokes flows. Built upon previous literature, we extend the theory to refine the  expression of particle's wall-normal and wall-parallel motion for a generalized obstacle geometry.

In Chapter~\ref{chap 3}, we explain the importance of breaking the fore-aft flow symmetry in a Stokes transport to induce a net wall effect on particle trajectories that encounter the obstacle. Using an analytically known Stokes transport flow over an inclined elliptic obstacle, we demonstrate that permanent displacement of a single spherical particle by hydrodynamic interactions in Stokes flow is possible, provided the flow symmetry is sufficiently broken. We rigorously compute the magnitude of displacement of particles across streamlines depending on their initial position and unveil the capacity of the asymmetric obstacle to deflect particles hydrodynamically, comparing with existing theories of non-hydrodynamic displacement of a particle by a (symmetric) obstacle.  

In Chapter~\ref{chap 4}, through a thorough derivation of the analytical framework of the asymmetric obstacle's capability to displace particles, we establish modeling derivations of the scaling behavior of the effect as a function of the control parameters.

In Chapter~\ref{chap 5}, addressing the contribution of hydrodynamic force in particle sticking applications, we quantify particles dive and minimum approach towards the elliptic obstacle in this inclined Stokes flow that breaks the fore-aft symmetry of the flow geometry. We quantify and compare the effect of this symmetry breaking in magnifying the hydrodynamic force competing to short-range repulsive forces to lead particle very close to obstacle and stick.

In Chapter~\ref{chap 6},  understanding the importance of symmetry breaking hydrodynamics in particle manipulation applications, we employ this symmetry breaking principle by modeling an inclined Stokes flow over two circular cylinders-- a geometry more analogous to the unit cell of a DLD device or a porous media. We show that non-intuitive flow structures form between the cylinders depending on the gap between them and on the inclination angle of the flow that effectively breaks the symmetry of the flow geometry.

Chapter~\ref{chap: conclusions} provides a concise discussion of the main results and conclusions, along with suggestions for future work that expand on both theoretical developments and practical experimental design strategies for precise particle manipulation in Stokes flow.

\section{Key Accomplishments}

The key accomplishments of the research all elucidate the so-far overlooked particle dynamics in Stokes transport flows as a new strategy for particle manipulation applications. Furthering our technique of breaking the fore-aft geometry to achieve this goal opens up qualitatively non-intuitive flow geometry with different ideas being exploited.

\begin{itemize}
    \item \textbf{Hydrodynamic model of spherical-particle-wall interaction in Stokes flow}: We present the equation of motion for a rigid spherical particle in a general Stokes background flow. We improve the wall-normal correction from previous literature by developing a variable expansion method for small distances between the particle and the wall, and the wall-parallel correction by properly enforcing limits. The main accomplishment here is the systematic construction of uniformly valid expressions for wall interaction effects on particle trajectories for all distances from the wall. We present the equation of motion for a rigid spherical particle in a general Stokes background flow around an inclined elliptic obstacle. This specific orientation breaks the fore-aft geometry of the flow which makes a net particle displacement possible.

    \item \textbf{Quantify spherical particle motion in Stokes transport flow over an inclined elliptic obstacle}: Here we gain a fundamental understanding of manipulating spherical particle transport over an obstacle by obstacle wall interaction effects in a Stokes flow. In analytically known Stokes flow around an elliptic obstacles,  we first show that in a symmetric flow, the upstream wall effect is completely neutralized by downstream wall effect resulting zero net displacement of particles. We then demonstrate a fore-aft symmetry-breaking condition by inclining the elliptic obstacle with a non-trivial angle where a net particle displacement by wall effect becomes possible. We numerically quantify the obstacle's capacity in displacing particles across streamlines leveraging this hydrodynamic fore-aft symmetry breaking technique. This rigorous analysis provides clear evidence that hydrodynamic wall effect can be leveraged to achieve precise displacement of passive particles in many continuous particle transport situations at low Reynolds numbers.
    \item \textbf{Theory of particle deflection by interacting an asymmetric elliptic obstacle in Stokes transport}: We formulate the theoretical foundation of the displacement effect resulting from particle's close interaction with the elliptic obstacle wall. By using the appropriate lubrication model of particle motion at small distances, we first develop a semi-analytical method to compute the net displacement of particles undergoing near-wall interactions and to quantify the maximum displacement attainable through this mechanism. Building on this foundation, we then develop a fully analytical scaling law parameterizing this maximum displacement effect. This theoretical work establishes the design criteria for controlled particle displacement around an asymmetric obstacle that can be systematically tuned by hydrodynamic control parameters, potentially opening up new avenues for particle manipulation in a continuous transport microfluidic system.

    \item \textbf{Hydrodynamic description of localized particle sticking to obstacle}: By quantifying the closest approach of a particle diving towards the obstacle, we obtain the physical gap between particle and obstacle at which the short-range intermolecular force takes over and induces localized sticking. We further compute the hydrodynamic forces coming from the presence of the obstacle wall and show that the forces get magnified if the symmetry of the flow is broken. This insight makes a valuable prediction in porous media filtration applications.

    \item \textbf{Symmetry breaking Stokes hydrodynamics with multiple obstacles}: Motivated by the geometry in DLD devices and in heterogeneous porous media, we utilize an inclined Stokes flow around two cylindrical obstacles that effectively brakes the fore-aft symmetry of the flow structures-- the technique we employed for obtaining the net wall effect on  by a single asymmetric elliptic obstacle. We compute particle trajectory simultaneously encountering both obstacles. This finding sets up the foundation to understand the asymmetric hydrodynamics of particle manipulation in DLD arrays and heterogeneous porous media.

\end{itemize}

\chapter{Hydrodynamic formalism of particle-wall interaction in arbitrary Stokes flow}\label{chap 2}

In this chapter\footnote[1]{This chapter is adapted from Das et al. \cite{das2025controlled}} by making use of the results from \cite{liu2025principles}, we present a model that quantifies the hydrodynamic interaction between a rigid spherical particle with an obstacle having a radius curvature much larger than that of the particle. Specifically, we focus on the motion of force-free, density-matched spherical particles, which move solely in response to the local background flow and the hydrodynamic disturbances caused by nearby boundaries. We aim to describe how these particles deviate from passive advection due to the presence of one or more walls. Building upon wall-parallel velocity corrections inspired by previous literature, we refine them by incorporating properly enforced limits. We also develop a variable expansion method for the wall-normal correction smoothly connecting the wall expansion model with particle expansion model. Eventually, we construct uniformly valid expressions for particle velocities at all wall-particle distances.

\section{Introduction}

The motion of particles suspended in viscous flows near boundaries is a fundamental problem in low Reynolds number hydrodynamics with direct relevance to microfluidic particle-laden flows, colloidal suspensions, and biomedical systems. In most particle-laden microfluidic flow setups, at least a subset of particles placed into the flow will either be located or be transported near boundaries, which could be solid (no-slip) or fluid/fluid interfaces (e.g.\ immiscible liquids, droplets, bubbles). The presence of a nearby boundary generally induces particle displacements that depend on parameters such as the Reynolds number, density contrast, and flow geometry. In steady inertial microfluidics, boundaries and the associated shear flow gradients lead to slow inertial migration effects that persist even at relatively large distances from the wall \citep{segre1962behaviour, di2009inertial,di2007continuous}. In oscillatory inertial microfluidics, including systems driven by acoustically excited microbubbles, boundary effects become dominant only at very small separations and may often be treated using lubrication theory \citep{thameem2017fast,agarwal2018inertial}. 

In contrast, in flows with negligible inertia (Stokes flow), although boundary effects are expected to extend over longer ranges and appear more prominent, whether a practically useful net displacement can occur after a particle’s interaction with a wall or obstacle remains an unresolved question. While the motion of a force-free particle in an unbounded fluid is well understood, the effects of proximity to rigid walls are still lacking. The ability to quantify these wall-induced effects is critical for predicting transport, mixing, and sorting behaviors in confined systems.

Microparticles placed in a background Stokes flow around an interface experience forces at varying distances from the boundary. Even at large distances (in bulk flow), particle trajectories deviate from the background flow because of Fax\'en's correction \citep{happel1965low}, but close to the interfaces, the hydrodynamic interaction between boundaries is dominant.

Different situations can be addressed in the modeling of particles in Stokes flow \citep{brenner1961slow,goldman1967slow,goldman1967slow2}, particularly (i) the forces on a particle moving at a given speed, (ii) the forces on a particle held fixed in a certain location, or (iii) the motion of a force-free particle. The present work focuses on the third scenario, which is particularly relevant for density-matched particles (absent the effects of gravity) in microfluidic devices. 
Setting thus the total force on the particle to zero, we quantify the modification of the particle velocity due to the effects of nearby boundaries altering particle motion both in the wall-parallel and the wall-normal direction. 

In general, studies of force-free particle dynamics predict deviations of the particle velocity from the background fluid velocity in which the particle is embedded, representing the non-passive component of its motion. These velocity corrections can be naturally decomposed into components parallel and perpendicular to the boundary. Analyses of the equation of motion for force-free particles incorporating wall-normal velocity corrections have been presented by \citep{rallabandi2017hydrodynamic}, while wall-parallel velocity corrections, both far from and in close proximity to a boundary, have been examined in other studies \citep{ekiel2006accuracy,pasol2011motion}.

Building on this prior work, we decompose the wall-induced velocity corrections into distinct wall-parallel and wall-normal contributions. For the wall-parallel correction, we adopt an approach inspired by lubrication theory and asymptotic matching to derive an improved uniformly valid expression that accurately captures both near-wall and far-field behavior.

A detailed analysis of the motion of force-free particles with wall-normal velocity corrections has been provided by \citep{rallabandi2017hydrodynamic}. Other work has described wall-parallel velocity corrections far from and close to the wall \citep{ekiel2006accuracy, pasol2011motion}. In our very recent study on particle motion in internal Stokes flow \citep{liu2025principles}, we have constructed uniformly valid expressions for all particle-wall distances from this precursor work. Here, we extend the formalism to determine particle trajectories transported over an obstacle, representing a boundary whose orientation and curvature near the particle both change as the particle moves. 

Our approach offers a novel, unified framework for determining the hydrodynamic displacement of rigid spherical particles near arbitrarily curved boundaries in Stokes flow. It expands on existing work to arrive at an equation of motion applicable to all particle-wall distances.

\begin{figure}
    \centering
\includegraphics[width=0.8\textwidth]{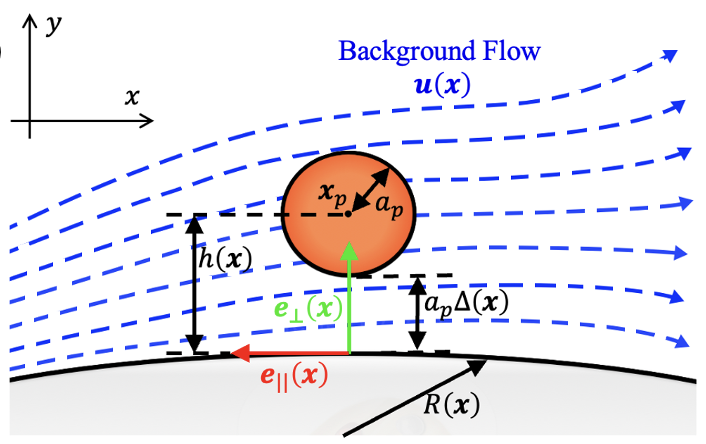}
    \caption{Schematic of a small spherical particle of radius $a_p$ located at $\boldsymbol{x}_p=(x_p,y_p)$ in the vicinity of an obstacle boundary of radius of curvature $R(\boldsymbol{x}$). The particle is immersed in a density-matched Stokes background flow $\boldsymbol{u}(\boldsymbol{x})$.}
    \label{Fig flow setup}
\end{figure}

\section{Particle velocity in the presence of a wall} \label{section wall correction}
Reflecting common situations in microfluidic setups with structures that span the entire height of a channel, we here describe 2-D Stokes flows $\boldsymbol{u}(\boldsymbol{x})$ in a Cartesian coordinate system $\boldsymbol{x}=(x,y)$. Although any 2-D Stokes flow around an obstacle is subject to the Oseen paradox \citep{proudman1957expansions} and does not exist as a consistent solution to arbitrary distance from the obstacle wall, a unique background Stokes flow exists in the vicinity of the obstacle, and the range and accuracy of that solution can be arbitrarily increased by lowering the Reynolds number. We place a spherical, force-free inertia-less particle in such a flow $\boldsymbol{u}(\boldsymbol{x})$ as sketched in figure~\ref{Fig flow setup}. 

We consider particle radii $a_p$ much smaller than the radius of curvature of the obstacle wall $R(\boldsymbol{x})$. The unit vector $\boldsymbol{e}_\perp(\boldsymbol{x_p})$ normal to the wall pointing towards the particle center $\boldsymbol{x}_p$ defines the closest distance $h(\boldsymbol{x_p})$ between particle center and boundary.  The tangential unit vector $\boldsymbol{e}_{||}(\boldsymbol{x})$ represents the wall-parallel direction. 

Tracking the particle trajectory under the influence of the obstacle needs careful modeling of the corrections to both the wall-parallel and the wall-normal velocity components of the particle particularly close to the obstacle, where the effects are most prominent. In such proximity, we have verified that effects of finite obstacle curvature (quantified in \cite{rallabandi2017hydrodynamic}) are small and do not alter the outcomes presented here (see section 4 for details), so that we restrict ourselves to the flat-wall approximation $(a_p/R\rightarrow 0)$ here.

Far from any walls, the motion of a spherical, neutrally buoyant particle is described by
\begin{equation}\label{vpfar}
    \boldsymbol{v}_{p,{far}} =\boldsymbol{u}(\boldsymbol{x}_p(t))+\boldsymbol{u}_{Faxen} (\boldsymbol{x}_p(t))\,,
\end{equation}
with the effect of streamline curvature
on the length scale of the particle size addressed by the Fax\'en correction, 
\begin{equation}\label{faxen}
    \boldsymbol{u}_{Faxen} (\boldsymbol{x})=\frac{a_p^2}{6}\nabla^2\boldsymbol{u}(\boldsymbol{x})\,.
\end{equation}
The presence of an obstacle imposes wall effects, and thus an additional velocity correction, $\boldsymbol{W}$. With no inertia, the particle equation of motion remains a first-order  dynamical system,
\begin{equation}\label{vp}
    \frac{d\boldsymbol{x}_{p}(t)}{dt}=\boldsymbol{v}_{p} (\boldsymbol{x}_p(t))=\boldsymbol{u}(\boldsymbol{x}_p(t))+\boldsymbol{u}_{Faxen} (\boldsymbol{x}_p(t))+\boldsymbol{W}(\boldsymbol{x}_p(t))\,.
\end{equation}

In the far-field limit, $\boldsymbol{W}$ must decay to zero, leaving only the Fax\'en correction in effect. 
Decomposing the ambient velocity field 
$
\boldsymbol{u}=
u_{||} \boldsymbol{e}_{||} + u_\perp \boldsymbol{e}_{\perp}
$ 
in the wall-parallel and wall-normal directions, we write
\begin{equation}\label{vpparallel}
    \boldsymbol{v}_{p||} = \left(\boldsymbol{u} + \frac{a_p^2}{6}\nabla^2\boldsymbol{u}\right)\cdot\boldsymbol{e}_{||} + W_{||}\,,
\end{equation}
\begin{equation}\label{vpperp}
    \boldsymbol{v}_{p\perp} = \left(\boldsymbol{u} + \frac{a_p^2}{6}\nabla^2\boldsymbol{u}\right)\cdot\boldsymbol{e}_{\perp} + W_{\perp}\,,
\end{equation}
and quantify in the following the particle velocity corrections parallel to $(W_{||})$ and normal to $(W_{\perp})$ the wall.
While we largely quote results from previous work \citep{liu2025principles}, we point out where the present problem requires particular care and greater modeling effort.

\section{Wall-parallel corrections to the particle velocity}\label{subsection: wall parallel particle}
The components of $\boldsymbol{W}(\boldsymbol{x}_p)$ depend on the wall distance, conveniently described by the dimensionless parameter
\begin{equation}\label{eq Delta}
    \Delta=\frac{h(\boldsymbol{x}_p)-a_p}{a_p}\,,
\end{equation}
which is the surface-to-surface distance relative to the particle radius, cf.\ \citep{rallabandi2017hydrodynamic, thameem2017fast, agarwal2018inertial}. 
The wall-parallel velocity correction $W_{||}$ slows down the wall-parallel velocity of a force-free particle by a fraction $f(\Delta)$, so that
equation~\eqref{vpparallel} takes the form
\begin{equation}\label{eq vpparallelFull}
    v_{p||}(x_p,y_p) = \left[\left(1-f(\Delta)\right)\left(\boldsymbol{u} + \frac{a_p^2}{6}\nabla^2 \boldsymbol{u}\right)\cdot\boldsymbol{e}_{||} \right]_{\boldsymbol{x}_p}\,.
\end{equation}

The wall-parallel velocity correction coefficient $f(\Delta)$ has been worked out in detail for specific cases such as linear shear flow \citep{o1968sphere, jeffrey1984calculation,  stephen1992characterization, williams1994particle, chaoui2003creeping}, quadratic flow \citep{goren1971hydrodynamic, ekiel2006accuracy, pasol2006sphere}, or modulated shear flow \citep{pasol2006sphere}, with asymptotic expressions available for $\Delta\to 0$ and $\Delta\to\infty$. We note that (i) the wall interactions are most prominent for small $\Delta$ and (ii) the linear shear part of any flow dominates as $\Delta\to 0$. In particular, when $\Delta=0$ (particle touching the wall), the sphere has to come to rest.

In order to obtain a uniformly valid expression for $f(\Delta)$, we follow the expansion approach of \citep{pasol2011motion} but modify it to enforce exact matching with known asymptotic results. For $\Delta\ll 1$, linear shear flow is dominant, and Williams's near-wall expression \citep{williams1994particle} must be recovered. Far from the wall, $f(\Delta)\to c \Delta^{-3}$, where the positive constant $c={\cal O}(1)$ depends on the type of flow \citep{goldman1967slow2,  ghalia2016sphere}. The exact value of $c$ makes no qualitative difference to the effects explored here, and we enforce $c=5/16$ to agree with the far-field asymptote provided by \citep{goldman1967slow2} for linear shear flow. 

\begin{figure}
    \centering
\includegraphics[width=0.7\textwidth]{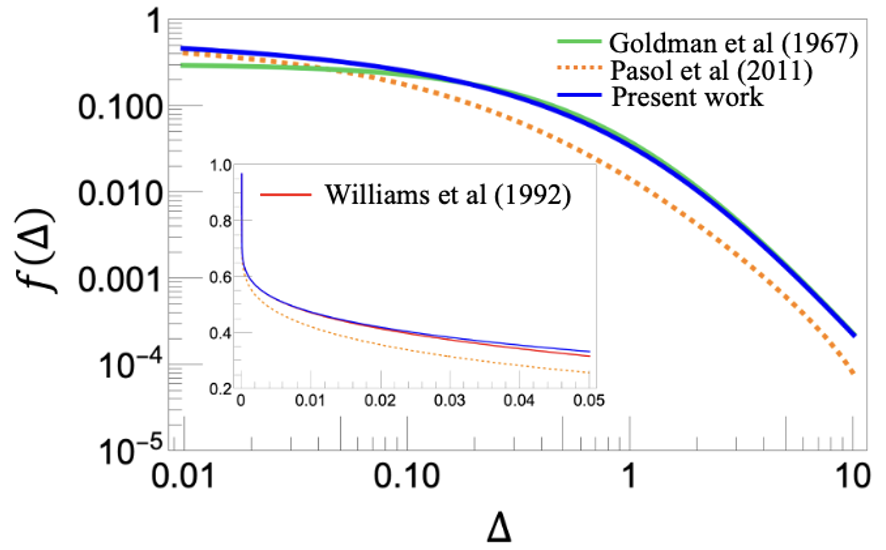}    \caption{Wall parallel velocity correction factor $f(\Delta)$ as a function of non-dimensional gap $\Delta$ showing good agreement with \cite{goldman1967slow} in the far-field limit ($\Delta\gg1$) and with \cite{stephen1992characterization} (inset) near the wall ($\Delta\ll1$), including the regime of lubrication theory}
    \label{fig f}
\end{figure}

Appendix~\ref{appen wall-parallel} details the derivation leading to the following expression:

\begin{equation} \label{f}
f(\Delta)=1-\frac{(1+\Delta)^{4}}{0.66+3.15\Delta+5.06\Delta^2+3.73\Delta^3+\Delta^4-0.27(1+\Delta)^{4} \log\left(\frac{\Delta}{1 + \Delta}\right)}\,.
\end{equation}
employed for all $\Delta$. Fig.~\ref{fig f} illustrate the agreement with the asymptote at $\Delta\gg 1$ \citep{goldman1967slow2} as well as the logarithmic lubrication theory approach to $f=1$ at $\Delta\to 0$ \citep{stephen1992characterization, williams1994particle}.
 
Note that this logarithmic behavior means that $1-f$ only drops to $\approx 0.32$ at $\Delta= 10^{-4}$, which for a typical particle of $a_p=5 \mu$m translates into a sub-nanometer gap, where continuum theory breaks down. Thus, in practical situations, $f$ will slow the wall-parallel motion significantly, but never dramatically. Furthermore, by its nature, this wall-parallel velocity modification is much less important than the wall-normal effect in pushing particles across streamlines. Equation~\eqref{f} is derived for a linear wall-parallel background velocity profile, which is asymptotically accurate for small $\Delta$, the case of primary interest in this study.

\section{Wall-normal corrections to the particle velocity}\label{subsection: wall normal particle}
Employing a quadratic expansion of the background flow about the spherical particle's center, a general expression for the normal component of the hydrodynamic force on a spherical, neutrally buoyant particle was obtained by Rallabandi et al. \cite{rallabandi2017hydrodynamic}:
\begin{equation}\label{Flambda}
    F_{\perp}=6\pi\mu a_p[\{-\mathcal{A}(\boldsymbol{v}_p-\boldsymbol{u})-a_p\mathcal{B}(\boldsymbol{e_{\perp}}\cdot \nabla\boldsymbol{u})+\frac{a_p^2}{2}\mathcal{C}(\boldsymbol{e_{\perp}}\boldsymbol{e_{\perp}}:\nabla\nabla\boldsymbol{u})+\frac{a_p^2}{2}\mathcal{D}\nabla^2\boldsymbol{u}\}\cdot\boldsymbol{e_{\perp}}]_{\boldsymbol{x}_p}\,
\end{equation}
Here $\mu$ is the viscosity of the fluid, $\boldsymbol{v}_p$ and $\boldsymbol{u}$ are the particle and ambient flow velocity, respectively.
The first correction term, proportional to $\mathcal{A}$, is due to the translation of the particle relative to the mean surrounding background flow and is identical to the general expression from \citep{brenner1961slow}. The term proportional to $\mathcal{B}$ is due to extensional gradients of the flow field. This contribution is zero for a sphere in an infinite flow field \citep{happel2012low, batchelor1970stress} but is generally non-zero for a finite distance to the wall. The second moments of the background flow result in two separate contributions to the force: one (proportional to $\mathcal{C}$) dependent on the curvature of the background flow velocity normal to the wall ($\boldsymbol{e_{\perp}}\boldsymbol{e_{\perp}}:\nabla\nabla\boldsymbol{u}$); another (proportional to $\mathcal{D}$) proportional to $\nabla^2\boldsymbol{u}$. This latter term asymptotes to the Faxen correction as $\Delta\to\infty$.

The wall-normal velocity correction $W_{\perp}$ for a force-free particle follows from setting $\boldsymbol{F}_\perp=0$ and separating the Faxen part (cf.\ \eqref{faxen}) from the Laplacian:
\begin{equation} \label{eq normal correction}
    W_{\perp}(x_p,y_p) = \left[\left\{-a_p\frac{\mathcal{B}}{\mathcal{A}}(\boldsymbol{e_{\perp}}\cdot \nabla\boldsymbol{u}) + \frac{a_p^2}{2}\frac{\mathcal{C}}{\mathcal{A}}(\boldsymbol{e_{\perp}}\boldsymbol{e_{\perp}}:\nabla\nabla\boldsymbol{u}) + \frac{a_p^2}{2}\left(\frac{\mathcal{D}}{\mathcal{A}}-\frac{1}{3}\right)\nabla^2\boldsymbol{u} \right\} \cdot\boldsymbol{e_{\perp}}\right]_{\boldsymbol{x}_p}
\end{equation}

Note that the background flow field $\boldsymbol{u}(\boldsymbol{x})$ at particle position $\boldsymbol{x}=\boldsymbol{x}_p$ can be treated as the Taylor expansion of the flow field at the particle position: $\boldsymbol{u}(\boldsymbol{x}_p)=\boldsymbol{u}(\boldsymbol{x})|_{\boldsymbol{x}=\boldsymbol{x}_p}$. Thus, with the normal component of the relative velocity from \eqref{eq normal correction}, the particle wall-normal velocity expanded from particle position as per \eqref{vp} results in the following form:

\begin{equation} \label{eq vpPE}
    \begin{split}
    v_{p\perp}^{PE}(x_p,y_p)= &\left[\left\{\boldsymbol{u} + \frac{a_p^2}{6}\nabla^2 \boldsymbol{u} - a_p\frac{\mathcal{B}}{\mathcal{A}}(\boldsymbol{e_{\perp}}\cdot \nabla \boldsymbol{u}) + \frac{a_p^2}{2}\frac{\mathcal{C}}{\mathcal{A}}(\boldsymbol{e_{\perp}}\boldsymbol{e_{\perp}}:\nabla\nabla \boldsymbol{u}) + \right.\right. \\ & \left.\left. \frac{a_p^2}{2}\left(\frac{\mathcal{D}}{\mathcal{A}}-\frac{1}{3}\right)\nabla^2 \boldsymbol{u}\right\}\cdot\boldsymbol{e}_{\perp} \right]_{\boldsymbol{x}_p}
    \end{split}
\end{equation}

The full analytical expressions for the $\Delta$-dependent functions $\mathcal{A}$, $\mathcal{B}$, $\mathcal{C}$ and $\mathcal{D}$ are provided in \cite{rallabandi2017hydrodynamic} whose asymptotic behaviors for large separations $\Delta \gg 1$ ($\mathcal{A}_{large}$ etc.)
and for small separations $\Delta \ll 1$ ($\mathcal{A}_{small}$ etc.) are included in \ref{appen res coeff asymp}. For arbitrarily large separations $(\Delta\rightarrow\infty)$, $\frac{\mathcal{B}}{\mathcal{A}}\to 0$, $\frac{\mathcal{C}}{\mathcal{A}}\to 0$, and $\frac{\mathcal{D}}{\mathcal{A}}\to 1/3$, so that $W_{\perp}\to 0$. We use the superscript $PE$ (``particle expansion") to emphasize that all background flow velocities and derivatives are evaluated at ${\bf x}_p$.

As the particle approaches the wall $(\Delta \ll 1)$, this particle expansion formalism \eqref{eq vpPE} relying on Taylor expansion of the flow around the particle center becomes inaccurate -- note that \eqref{eq vpPE} is not guaranteed to vanish when the particle touches the wall $(\Delta=0)$. If instead we employ Taylor expansion of the background flow field $\boldsymbol{u}(\boldsymbol{x})$ around the point on the wall closest to the particle, 
\begin{equation}\label{eq xw}
\boldsymbol{x}_w
=\boldsymbol{x}_p
-a_p\boldsymbol{e}_{\perp}(\boldsymbol{x}_p)\,,
\end{equation}
as suggested by Rallabandi et al. \citep{rallabandi2017hydrodynamic}, the no-penetration condition is enforced (``wall expansion"). The resulting particle velocity is linear in $\Delta$ to leading order,
\begin{equation}\label{eq wall xpnsn}
    v_{p\perp}^{WE} = 1.6147 
 a_p^2\kappa(\boldsymbol{x}_p)\Delta+O(\Delta^2)\,.
\end{equation}
Here $\kappa(\boldsymbol{x}_p)=\partial_{\perp}^2u_{\perp}(\boldsymbol{x}_w)$ is the background flow curvature at the wall point $\boldsymbol{x}_w$ from \eqref{eq xw}.
corresponding to the closest distance from the particle position $\boldsymbol{x}_p$. The prefactor can be derived from the result of \cite{goren1971hydrodynamic} and was confirmed in \cite{rallabandi2017hydrodynamic}. 
Although \eqref{eq wall xpnsn} accurately determines the particle wall-normal velocity for $\Delta\ll1$, we find that it does not smoothly transition to \eqref{eq vpPE} when $\Delta\sim1$. We therefore generalize and refine the wall-expansion procedure in this intermediate region by constructing second-order expansions of $\boldsymbol{u}(\boldsymbol{x})$ around variable expansion points $\boldsymbol{x}_E=\boldsymbol{x}_E(\boldsymbol{x}_p)$, i.e., 
\begin{equation} \label{upVE}
    \begin{split}
    \boldsymbol{u}^{VE}(\boldsymbol{x}) = \boldsymbol{u}|_{\boldsymbol{x}=\boldsymbol{x}_E} + (\boldsymbol{x}-\boldsymbol{x}_E)\cdot(\nabla \boldsymbol{u})_{\boldsymbol{x}=\boldsymbol{x}_E} + \frac{1}{2}(\boldsymbol{x}-\boldsymbol{x}_E)(\boldsymbol{x}-\boldsymbol{x}_E):(\nabla \nabla \boldsymbol{u})_{\boldsymbol{x}=\boldsymbol{x}_E}\,.
    \end{split}
\end{equation}
Replacing $\boldsymbol{u}$ by $\boldsymbol{u}^{VE}$ in \eqref{eq vpPE} gives the generalized expression
\begin{equation} \label{eq vpVE}
    \begin{split}
        v_{p\perp}^{VE}(x_p,y_p) = &\left[\left\{\boldsymbol{u}^{VE} + \frac{a_p^2}{6}\nabla^2 \boldsymbol{u}^{VE} - a_p\frac{\mathcal{B}}{\mathcal{A}}(\boldsymbol{e_{\perp}}\cdot \nabla \boldsymbol{u}^{VE}) + \frac{a_p^2}{2}\frac{\mathcal{C}}{\mathcal{A}}(\boldsymbol{e_{\perp}}\boldsymbol{e_{\perp}}:\nabla\nabla \boldsymbol{u}^{VE}) + \right.\right. \\ & \left.\left. \frac{a_p^2}{2}\left(\frac{\mathcal{D}}{\mathcal{A}}-\frac{1}{3}\right)\nabla^2 \boldsymbol{u}^{VE}\right\}\cdot\boldsymbol{e}_{\perp} \right]_{\boldsymbol{x}_p}\,,
    \end{split}
\end{equation}
where the derivatives and resistance coefficients are still evaluated at the particle center. 

The functional form of the expansion point 
$\boldsymbol{x}_E(\boldsymbol{x}_p)$
is constructed for \eqref{eq vpVE} to obey the WE and PE limits, i.e 
$\boldsymbol{x}_E\to \boldsymbol{x}_w$ for touching particles $(\Delta \to 0)$ and 
$\boldsymbol{x}_E = \boldsymbol{x}_p$ for all $\Delta\geq \Delta_E$.
For simplicity, we choose the linear relation 
\begin{equation}\label{eq xE}
    \boldsymbol{x}_E=\boldsymbol{x}_p-a_p\left(1-\frac{\Delta}{\Delta_E}\right)\boldsymbol{e}_{\perp}\,,
\end{equation}
for $\Delta\leq \Delta_E$ and $\boldsymbol{x}_E=\boldsymbol{x}_p$ otherwise. Here, $\Delta_E$ denotes a specific choice of $\Delta$ that ensures a smooth interpolation between $v_{p_\perp}^{WE}$ and $v_{p\perp}^{PE}$ as shown in figure~\ref{fig DeltaE}(a) in section~\ref{appen modeling}. In particular the normal derivative of $v_{p\perp}$, and thus $W_\perp$,  needs to be smoothly varying.  While the internal vortical flows in \cite{liu2025principles} (where we worked with the generic $\Delta_E=1$) are bounded by infinite straight walls and vary on gradient scales given by the wall distance (channel width) only, leading to shallow gradients, the variation of slope and curvature along the surface of the elliptical obstacle in the present study of external transport flow leads to larger tangential gradients and, by continuity, larger normal gradients for certain ranges of angular positions. As a consequence, a smooth transition of $W_{\perp}$ requires a longer blending zone which we achieve for $\Delta_E=3$. We adopt this choice for subsequent sections where we illustrate the smooth transition from particle expansion model to wall expansion model (see figures~\ref{fig DeltaE}(b), and~\ref{fig DeltaE}(c)) , while carefully checking that the results are robust against other ${\cal O}(1)$ choices of $\Delta_E$ (section~\ref{appen modeling}).

Using equations \eqref{eq vpparallelFull} and \eqref{eq vpVE} in the dynamical system \eqref{vp} constitutes our formalism for computing particle motion in the presence of wall effects for arbitrary Stokes background flow.

\section{Conclusion}

In this chapter, we quantify the hydrodynamic interactions between rigid spherical particles and a general curved boundaries in an arbitrary Stokes flow. The presence of a nearby wall significantly alters the velocity of a particle from just passively following the background flow, leading to pronounced effects on its translational dynamics even in the absence of inertia. 

Through asymptotic approximations, we characterized the corrections to particle velocity in both wall-normal and wall-parallel directions uniformly valid for all distances to the wall. Particular care is taken in evaluating the wall-normal component of the velocity from the correction terms to ensure proper physical limits. These corrections will then be used to compute particle trajectories near boundaries in Stokes flow over a weakly curved obstacle relative to the particle to reveal the fate of the particle while being transported over the obstacle.

This formalism establish the framework to compute the obstacle wall affected particle velocity as well as to the study of interactions among multiple particles near boundaries. 
The results presented herein are critical for accurately modeling particle motion in confined geometries, commonly encountered in microfluidic devices, sedimentation processes, and biological environments. When encountering an obstacle, steric requirements forbidding penetration enforce crossing streamlines, leading to well-defined changes in normal and tangential velocity components of passive particles. By elucidating how such presence of boundaries alters the aforementioned hydrodynamic interactions not only provides fundamental insights into low Reynolds number fluid mechanics but also informs the design of engineered systems where wall interactions play a dominant role.

\chapter{Asymmetric hydrodynamics for inertialess deterministic particle deflection}\label{chap 3}

In the previous chapter, we have developed a dynamical system to describe the motion of a rigid, neutrally-buoyant, force-free, spherical particle immersed in an arbitrary Stokes flow near a curved wall. The framework includes systematic derivations of wall-induced velocity corrections in both the wall-normal and wall-parallel directions. In this chapter\footnote[1]{This chapter is adapted from Das et al. \cite{das2025controlled}}, we solve the dynamical system to obtain particle trajectory being transported over an obstacle in a Stokes flow. By breaking the fore-aft symmetry of the transport flow geometry, we rigorously model the displacement of spherical particles across streamlines originating from the hydrodynamic particle-wall interaction satisfying the steric requirement of no-penetration of particle to obstacle boundary. Particles moving mostly far from the obstacle are negligibly affected by the wall effect and follows primarily the Faxe\'n flow, whereas the wall effect grows as particles travel near the obstacle and results in a net spatial displacement that nonetheless preserves the time-reversibility of Stokes flow. However, owing to the physical constraint that stagnation streamlines cannot be crossed, there exists a critical trajectory starting from a certain initial streamline that exhibits the maximum displacement. This characteristic displacement quantifies the capacity of an obstacle to laterally deflect particles, highlighting its promise in inertial-less transport flow devices such as deterministic lateral displacement systems, porous media filtration.

\section{Introduction}

Inspired by numerous lab-on-a-chip, biomedical and bioengineering applications such as cell sorting, focusing, trapping, and filtering of particles, manipulation of micron-sized particle trajectories has been of significant interest in the context of microfluidics. Fundamentally, manipulation strategies force particles to cross streamlines so that they do not passively follow the flow. This essence of controlled manipulation, encompassing both passive (synthetic) or active (e.g.\ biological cells) particles, is implemented today in lab-on-a-chip processing as well as in the diagnosis of biological samples and biomanufacturing processes 
\citep{pamme2007continuous,ateya2008good,nilsson2009review,xuan2010particle,gossett2010label,puri2014particle}, drug discovery and delivery systems \citep{dittrich2006lab,kang2008microfluidics,nguyen2013design} or self-cleaning technologies \citep{callow2011trends,kirschner2012bio,nir2016bio}.  
Many techniques of precise control of suspended microparticles rely on the particle response to external forces including 
electrical \citep{xuan2019recent}, optical \citep{lenshof2010continuous,ashkin1986observation,grier2003revolution,chiou2005massively}, and magnetic techniques \citep{van2014integrated,crick1950physical,wang1993mechanotransduction,gosse2002magnetic}. However, not all particles are susceptible to these, prompting a continuous interest in manipulation strategies solely based on hydrodynamic forces. In practical applications, these approaches use customized flow geometries, such as channels or pillars 
\citep{lutz2006hydrodynamic,wiklund2012acoustofluidics,petit2012selective,tanyeri2011microfluidic,shenoy2016stokes,kumar2019orientation,chamolly2020irreversible}. In recent years, many such techniques leverage particle inertia \citep{di2007continuous,di2009inertial} particularly from oscillatory flow \citep{wang2011size,wang2012efficient,thameem2017fast,agarwal2018inertial} and have prompted an advance of theoretical frameworks beyond long-standing approaches \citep{maxey1983equation,gatignol1983faxen} to include new important effects \citep{rallabandi2021inertial,agarwal2021unrecognized,agarwal2021rectified,agarwal2024density}. 

Inertial effects are not present in the Stokes limit $($Reynolds number $Re\rightarrow0)$, where hydrodynamic effects on particles are notoriously long-range and
have been described in fundamental detail \citep{happel1965low, brady1988stokesian, kim2013microhydrodynamics, pozrikidis1992boundary, pozrikidis2011introduction,rallabandi2017hydrodynamic}, while systematic modulation of particle trajectories in Stokes flow has not been the subject of a detailed study. 
Fundamentally, this is because of the instantaneity and time reversibility of Stokes flow that at first glance seems to preclude lasting particle displacements, despite the success of Deterministic Lateral Displacement (DLD) in sorting particles by size, forcing them onto different trajectories through the interaction with a forest of pillar obstacles at very low $Re$
\citep{huang2004continuous,kruger2014deformability,kabacaouglu2019sorting,hochstetter2020deterministic,lu2023label,aghilinejad2019transport,davis2006deterministic}. 

Size-based particle sorting in a typical DLD set-up is determined by whether a particle can cross separating streamlines (row-shift).  \citep{inglis2006critical,kulrattanarak2011analysis,pariset2017anticipating,cerbelli2012separation,DLDhydro1} heuristically developed critical length-scales for particle size above which lateral displacement by this row-shifting happens.
Modeling approaches for the interaction of particles with individual obstacles
generally appeal to contact forces or other strong short-range forces \citep{lin2002distance,dance2004collision} that break the time reversibility of the Stokes flow and the hydrodynamic forces following from it. In particular, \citep{frechette2009directional,balvin2009directional,bowman2013inertia} model the effects of a non-hydrodynamic repulsive force originating from surface roughness \citep{ekiel1999hydrodynamic} and its dependence on particle size. This approach has been informed by similar ideas
in the modeling of suspension rheology
\citep{da1996shear,metzger2010irreversibility,blanc2011experimental,pham2015particle,lemaire2023rheology}. 

\begin{figure}  
    \centering
\includegraphics[width=\textwidth]{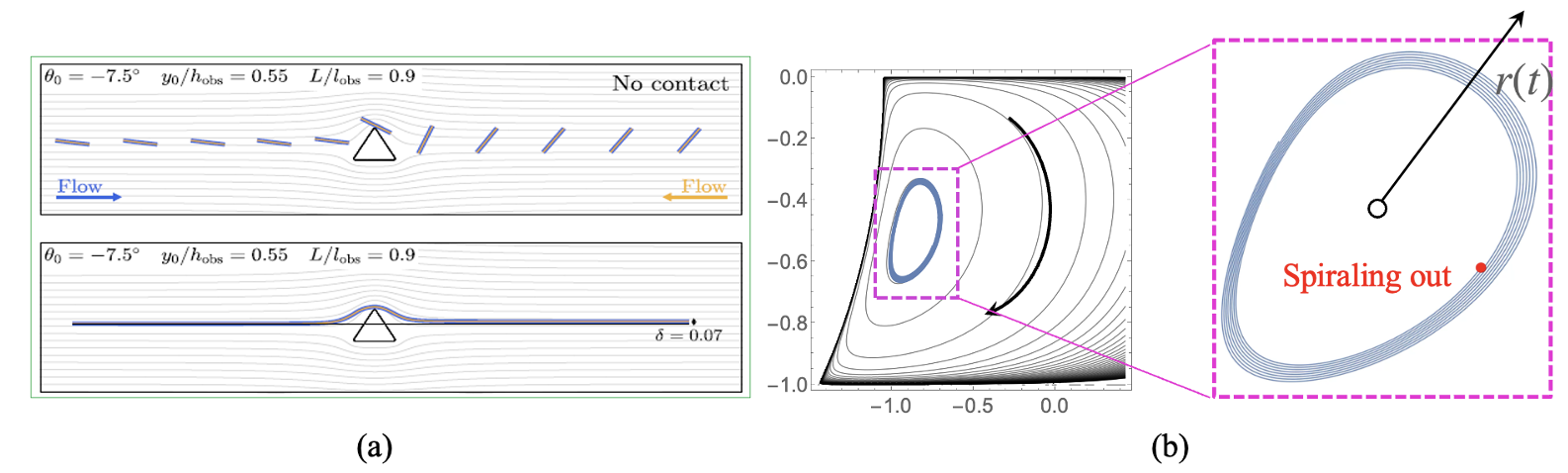}    \caption{(a) Asymmetry in fiber orientation while being transported over an triangular obstacle in Stokes flow \cite{li2024dynamics}. The fiber is spatially displaced by a net amount even without having any direct contact to the obstacle and without violating the time reversibility of Stokes flow. 
(b) Spherical particle motion in a symmetry broken Stokes vortical flow demonstrating that hydrodynamic particle-wall interaction alone can manipulate particle in Internal Stokes flow if the flow symmetry is broken. \cite{liu2025principles}}  \label{fig fiber_vortex}
\end{figure}

Contact forces almost invariably contain an ad-hoc element \citep{ekiel1999hydrodynamic}, fundamentally because in strict Stokes flow, there is no surface contact in finite time \citep{brady1988stokesian,claeys1989lubrication,claeys1993suspensions}.
On top of that, in a strict unidirectional Stokes flow, force-free particles cannot cross streamlines \citep{bretherton1962motion}. However with the presence of a wall-normal velocity component, streamline crossing should generally be expected. A simple example is a particle transported very close to a wall in a contracting channel flow where the particle cannot remain on its initial streamline without penetrating the wall and must therefore deviate from it. A recent study by Li et al.  \citep{li2024dynamics} shows that fibers passing over a triangular obstacle can show net displacements in Stokes transport flow even when not experiencing direct contact (figure~\ref{fig fiber_vortex}(a)). The fibers are extended, non-spherical objects and their interactions with obstacles are modeled by effective forces  \citep{dance2004collision}, leaving open the question whether rigorously described hydrodynamic interactions can cause net displacement. Very recent work by Liu et  al. \citep{liu2025principles} describes particle motion in {\em internal} (vortical) Stokes flow and finds that purely hydrodynamic interactions with confining channel walls  can indeed result in lasting displacements through repeated wall encounters (figure~\ref{fig fiber_vortex}(b)). 
Such vortex flows have, however, not been practically implemented yet for this purpose. The present work, by contrast, aims at a rigorous hydrodynamic description of particle motion in Stokes flow {\em external} to an obstacle (the elementary process of DLD), establishing the required geometry of flow and obstacle to effect a lasting displacement, as well as bounds on its magnitude.

\section{Symmetry-breaking non-inertial transport around an obstacle}\label{sec flow field}

\begin{figure}  
    \centering
\includegraphics[width=\textwidth]{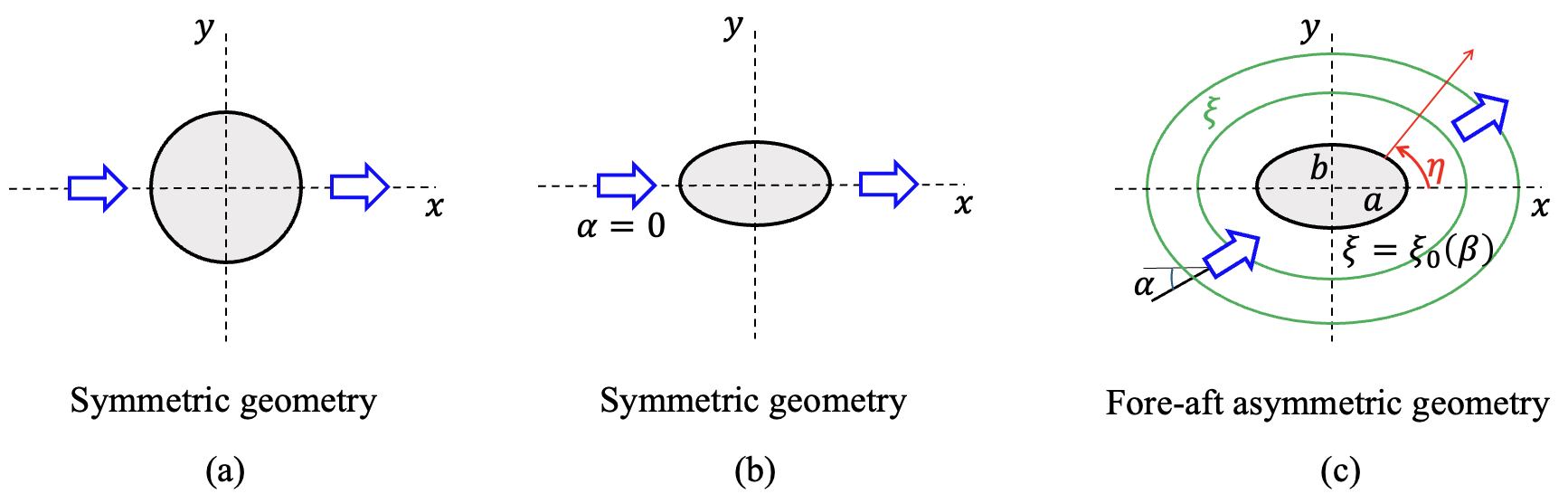}    \caption{Sketches of the principle of fore-aft symmetry breaking in a Stokes flow over an obstacle showing fore-aft symmetric obstacle-flow geometries with a circular cylinder (a) or a symmetrically placed elliptic cylinder (b).  This fore-aft symmetry breaks in (c), where the flow has non-trivial angle of attack $\alpha$. Also sketched is the elliptic coordinate system $(\xi,\eta)$ used to determine the flow field as in equations \eqref{eq flow}-\eqref{eq faxen flow}.
}  \label{fig fore-aft asymmetry}
\end{figure}

We apply the modeling detailed in chapter~\ref{chap 2} to particles transported towards and past an obstacle. When a density-matched, force-free particle in Stokes flow encounters an obstacle or boundary, steric requirements forbidding penetration enforce deviations from the initial streamline, leading to well-defined changes in normal and tangential velocity components.
If the particle’s approach to and departure from the interface on its trajectory occur in a symmetric fashion, the gradients normal to the wall cancel out during approach and departure, so that no net displacement (relative to the streamline the particle starts on) will be observed. 
This is clearly the case for a single circular cylinder obstacle (figure~\ref{fig fore-aft asymmetry}(a)). The same principle applies, even if we deform this circle into an ellipse met head on by the flow (figure~\ref{fig fore-aft asymmetry}(b)).
However, if the combined geometry of the boundary and the flow field break this symmetry, net displacement and thus meaningful manipulation of particle transport is possible. We investigate here the case of an elliptic cylinder obstacle with aspect ratio $\beta=\frac{b}{a}$ (with $a$ and $b$ the major and minor axis, respectively) placed in a uniform Stokes flow $U$ whose direction makes an angle $\alpha$ with the major axis. This situation is shown in figure~\ref{fig fore-aft asymmetry}(c) together with associated cartesian and elliptic coordinate systems. Note that symmetry arguments again preclude net displacement if $\alpha=0$ or $\alpha=\pi/2$ (figure~\ref{fig fore-aft asymmetry}(b)).

We use the single elliptic obstacle as the simplest case study of fore-aft symmetry breaking, which in 
traditional DLD setups is achieved by inclining a forest of circular pillars relative to the uniform stream. 
The Stokes background flow around the elliptic cylinder is known analytically as a solution to the 
biharmonic equation
    $\nabla^4\tilde{\psi}_B=0$ 
for the stream function $\tilde{\psi}_B$.
We choose the semimajor axis $a$ and the uniform flow speed $U$ as our length and velocity scales, so that $\tilde{\psi}_B$ is non-dimensionalized by $aU$. 

The Oseen paradox restricts this solution to an inner region in the vicinity of the obstacle surface.
It is implicit in the work of
\cite{berry1923steady} and was explicitly derived
by \cite{shintani1983low} in elliptic coordinates $(\xi,\eta)$ (since the flow close to the cylinder is strongly dependent on the body shape curvature) as the infinite sum $\tilde{\psi}_B=\sum_{n=1}^\infty\tilde{\psi}_n$ with

\begin{equation}\label{eq flow}
    \begin{split}
        \tilde{\psi}_n= \frac{1}{(\ln Re)^n}[&\Lambda_n \{(\xi-\xi_0)\cosh {\xi} + \sinh \xi_0 \cosh\xi_0\cosh\xi - \cosh^2\xi_0 \sinh\xi \}\cos\eta - \\& \Omega_n \{(\xi-\xi_0)\sinh\xi -\sinh\xi_0\cosh\xi_0 \sinh\xi + \sinh^2 \xi_0 \cosh\xi\}\sin\eta]\,.
    \end{split}
\end{equation}
Here $x=\sqrt{1-\beta^2} \cosh\xi\cos\eta$ and  $y=\sqrt{1-\beta^2} \sinh\xi\sin\eta$ relate cartesian and elliptic coordinates, and $\xi=\xi_0=\frac{1}{2} \ln\frac{1+\beta}{1-\beta}$ defines the elliptic cylinder surface. The prefactors $\Lambda_n$ and $\Omega_n$ are functions of $\alpha$ and $\beta$ obtained from asymptotic matching \citep{proudman1957expansions,kaplun1957low}. The analytical form of  \eqref{eq flow}, by construction, does not contain fluid inertia except an implicit dependence on Reynolds number $Re$ which defines the Oseen distance -- the range of validity of the solution around the obstacle which is $1/Re$ 
\citep{proudman1957expansions, batchelor2000introduction},
and can be made arbitrarily large by choosing a very small $Re$. Since our goal is to investigate the wall effect on particle trajectory and effective particle-wall interaction always occurs in the proximity of the obstacle, this flow description reflects reality to a well-defined degree of accuracy. In the limit $Re\rightarrow0$, the $n=1$ term of \eqref{eq flow} is dominant, and we scale out the $Re$ dependence by defining the rescaled background flow as
\begin{equation}\label{eq psiB}
\begin{split}
    \psi_B=(-\ln Re)\tilde{\psi}_{n=1}= &\Lambda_1 \{(\xi-\xi_0)\cosh {\xi} + \sinh \xi_0 \cosh\xi_0\cosh\xi - \cosh^2\xi_0 \sinh\xi \}\cos\eta - \\& \Omega_1 \{(\xi-\xi_0)\sinh\xi -\sinh\xi_0\cosh\xi_0 \sinh\xi + \sinh^2 \xi_0 \cosh\xi\}\sin\eta  \,,
\end{split}
\end{equation}
with $(\Lambda_1,\Omega_1)
    =\sqrt{1-\beta^2}
    (\sin\alpha, \cos\alpha)$. 

\begin{figure}[t]  
    \centering
\includegraphics[width=\textwidth]{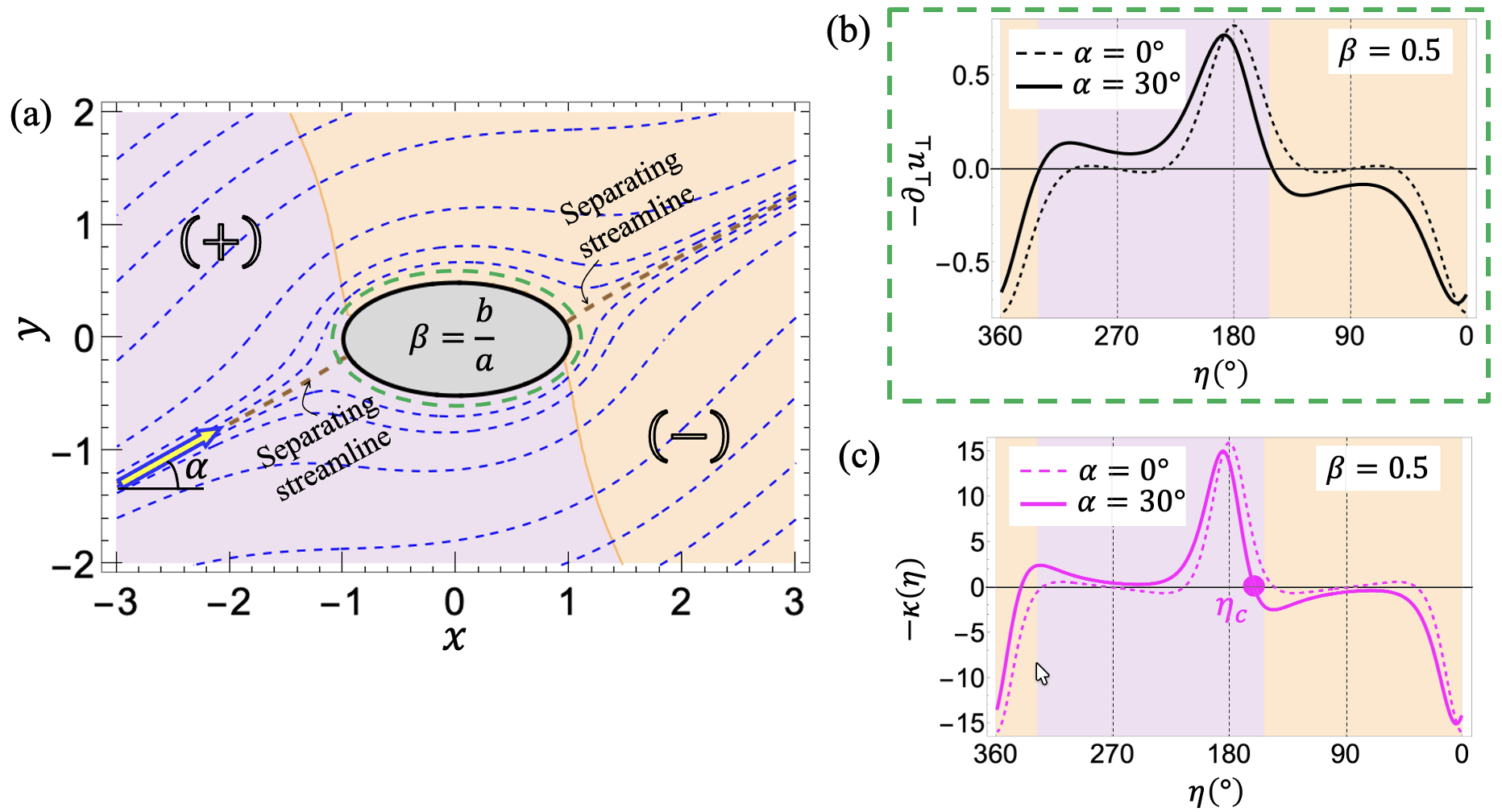}    \caption{(a) Fore-aft asymmetric Stokes flow (dashed blue streamline contours from \eqref{eq psiB}) impacting on an elliptic obstacle ($\beta=1/2$) under an inclination $\alpha=30^{\circ}$. 
$W_\perp$ is positive in the purple shaded zone and is negative in the orange shaded zone indicating particle repulsion from the obstacle and attraction towards the obstacle, respectively. 
The separating streamlines $\psi=0$ are indicated in dashed brown. 
(b) The sign changes of the quantity $\partial_{\perp} u_{\perp}$ reflect, to  leading order, those of $W_{\perp}$ (shading). This quantity is measured along the green dashed line in (a), locations at a distance $\Delta = 1$, here for $a_p=0.1$, where $\partial_{\perp} u_{\perp}$ dominates the corrections of particle motion. (c) The background flow curvature $-\kappa=-\partial^2_{\perp}u_{\perp}$ evaluated at the obstacle wall determines normal particle motion in the wall expansion model \eqref{eq wall xpnsn} for $\Delta\ll 1$. The upstream zero of this quantity, $\eta_c$, indicates a point of closest approach 
to the obstacle as discussed in section 3.
}  \label{fig flow}
\end{figure}

This analytical expression agrees with the description of Stokes flow over an inclined elliptic fiber by  \cite{raynor2002flow}. It could be improved in accuracy by taking into account terms of higher $n$ successively smaller by powers of $1/|\ln{Re}|$, but we take \eqref{eq psiB} to allow for clearer analytical scaling results. Figure~\ref{fig flow}(a) shows streamline contours $(\psi_B)$ for an angle of attack  $\alpha = 30^{\circ}$ and obstacle aspect ratio $\beta=0.5$.  Lastly, in order to isolate and explicitly quantify the wall effect on particle motion only, we use as our reference streamfunction the sum of $\psi_B$ and the Fax\'en streamfunction $\psi_{Faxen}=\frac{a_p^2}{6}\nabla^2\psi_B$, i.e.,
\begin{equation}\label{eq faxen flow}
    \begin{split}
        \psi = \psi_B+\psi_{Faxen}\,.
    \end{split}
\end{equation}

We have confirmed (see appendix \ref{appen bulk Faxen}) that particle displacement effects from the Fax\'en correction are negligible compared to those from particle-wall interaction at small $\Delta$, while they are absolutely small when $\Delta\gg 1$ and thus do not affect our findings on net displacement resulting from particle-obstacle encounters.

\section{Motion of a spherical particle in a symmetry breaking Stokes transport over an elliptic obstacle} \label{sec: particle motion}
\subsection{Fore-aft asymmetry analysis}\label{sub sec fore-aft symmetry analogy}

As implied in section~\ref{sec flow field}, it is fundamental that in order to have net particle displacement you need to break the symmetry of a Stokes flow between upstream and downstream of the obstacle that particle passes by. Li et  al. \cite{li2024dynamics} observed a net displacement of fiber passing around a triangular obstacle in Stokes flow even without having a direct contact between the fiber-obstacle surface (cf figure~\ref{fig fiber_vortex}(a)). In this scenario , it is the orientation of the fiber with respect to obstacle that breaks the fore-aft symmetry of fiber transport. In the recent work of Liu et  al. \cite{liu2025principles}, the flow geometry itself breaks that symmetry where spherical particle is placed inside the Stokes vortex flow confined by walls (cf figure~\ref{fig fiber_vortex}(b)).

Therefore, we aim to obtain the motion and net displacement of a spherical particle in transport by only hydrodynamic particle-wall interaction in a symmetry-breaking Stokes flow over an obstacle. The hydrodynamic formalism developed in chapter~\ref{chap 2} is now applied to compute the particle trajectory of transport around an inclined elliptic obstacle in the inertialess flow $\psi$ from \eqref{eq faxen flow}. The equations of motion \eqref{eq vpparallelFull}, \eqref{eq vpVE} use the background flow velocity $\boldsymbol{u}$ derived from $\psi$ in \eqref{eq psiB} via
\begin{equation}\label{eq ueta}
    u_\eta(\xi,\eta)=-g\partial_\xi\psi_B\,,
\end{equation}
\begin{equation}\label{eq uxi}
    u_\xi(\xi,\eta)=g\partial_\eta\psi_B\,
\end{equation}
where $g=g(\xi,\eta)=[(1-\beta^2)(\cosh^2\xi-\cos^2\eta)]^{-1/2}$  is the scale factor of the elliptic coordinate system \citep{shintani1983low,raynor2002flow}. For numerical computations, we transform all equations into cartesian reference coordinates, while for some analytical arguments, we will use elliptic coordinates directly. 


Qualitatively, the particle is transported towards the obstacle surface on the upstream side and away from it on the downstream side. If the initial position of the particle leads to a very close approach to the obstacle wall (or even a hypothetical overlap with passive transport), $W_{\perp}$ will act to repel the particle from the wall on the upstream side. Conversely, $W_{\perp}$ represents an attraction downstream. Figure~\ref{fig flow}(a) confirms these observations and shows that for non-trivial $\alpha,\beta$ the zones of repulsion and attraction are strongly asymmetric.

We will be most interested in particle trajectories that follow the obstacle outline closely, i.e., where $\Delta$ is not large. It can be verified that at $\Delta\sim 1$, the magnitude of $W_{\perp}$ is dominated by the first normal derivative term $-a_p\frac{\mathcal{B}}{\mathcal{A}}\partial_{\perp}u_{\perp}$ in \eqref{eq vpPE}. Indeed, figure~\ref{fig flow}(b) confirms that the sign of $-\partial_{\perp}u_{\perp}$ determines total attraction $(-)$ or repulsion $(+)$ to good accuracy for points at a distance of $h=2a_p$ from the wall ($\Delta=1$).

In the limit $\Delta \to 0$, the net effect of $W_{\perp}$ on the particle motion is determined to leading order of $\Delta$ by the wall flow curvature $\kappa=\partial_{\perp}^2 u_{\perp}$,
cf.\ \eqref{eq wall xpnsn}. 
Figure~\ref{fig flow}(c) demonstrates that $\kappa$ changes sign in very similar angular positions as the full $W_{\perp}$. We focus on information about $W_\perp$ here, as its effect dominates particle streamline crossing, while $W_\parallel$ mainly serves to slow particles down on their path.

We will illustrate particle trajectories and displacements in the following with particles that travel from left to right ``above'' the obstacle, where the repulsion region $(+)$, encountered first, is shorter than the attraction region $(-)$. All behavior of particles traveling ``below" the obstacle can be inferred by symmetry as discussed later.

The cases ``above" and ``below" have two separating streamlines ($\psi=0$) as boundaries,  representing a remaining flow symmetry: because of time reversibility of the Stokes flow (and the resulting hydrodynamic effects, which are linear in the Stokes flow), no particle can cross over these separating streamlines, which intersect the obstacle at the angular coordinate  $\eta^{sep}=\arctan(\beta\tan\alpha)$ downstream and $\pi+\eta^{sep}$ upstream, respectively.

\subsection{Particle trajectories around the obstacle}\label{subsec trajectory}

The equation of motion of the particle is solved numerically to obtain the trajectory $\boldsymbol{x_p}(t)$. We define a complete journey of a particle from left to right as  starting from an initial $x=x_i$ position and ending at $x_f=-x_i$ position. 

It is  important to note that: \\
(i) The inherent linearity in Stokes flow causes the wall effects $\boldsymbol{W}$ to be fully time-reversible, meaning that particles will come back exactly to their starting points in reversing the flow. \\
(ii) Stokes flow does not possess any memory from inertia. Thus, the effect on the particle trajectory by $\boldsymbol{W}$ is instantaneous, and an initial particle position determines the full trajectory.\\
(iii) Because of the anti-symmetry of the flow about the separating streamlines, a trajectory transported above the obstacle from an initial stream function value $\psi=\psi_i$ is equivalent to a trajectory starting at $\psi=-\psi_i$ that is transported below in the opposite direction.\\

\begin{figure}
  \centering
  \begin{subfigure}[t]{\textwidth}
    \centering
    \includegraphics[width=\linewidth]{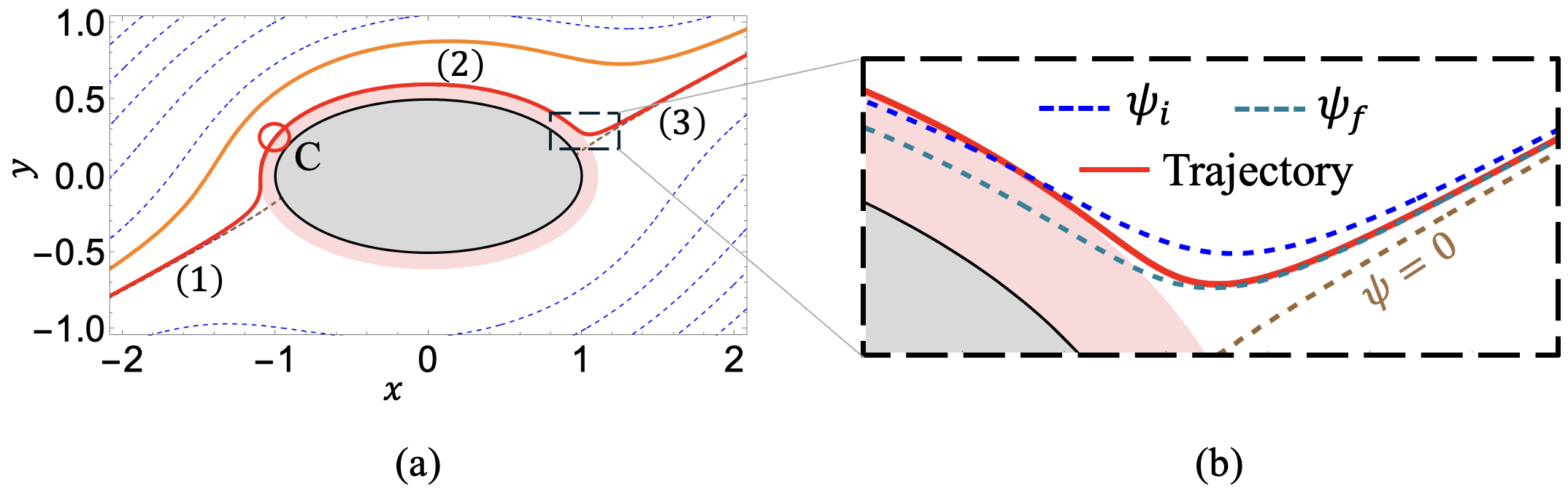} 
  \end{subfigure}

  \vspace{2em}

 \begin{subfigure}[t]{\textwidth}
    \centering
    \includegraphics[width=\linewidth]{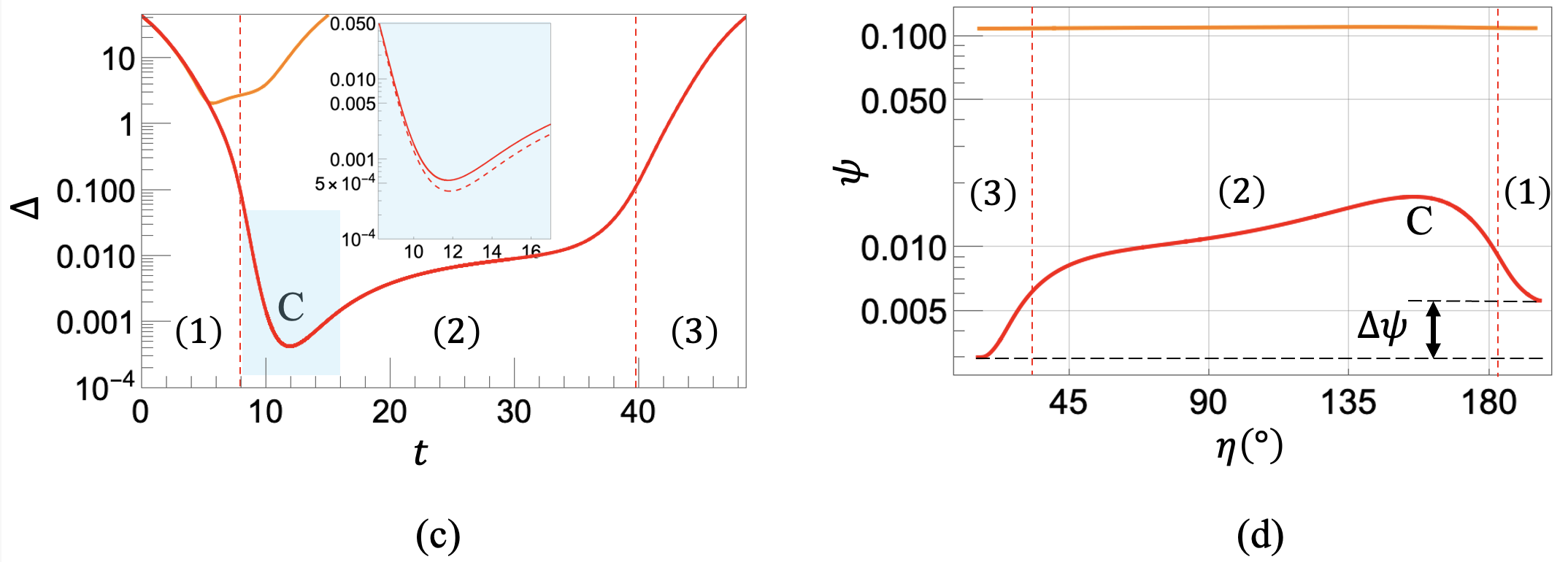}
  \end{subfigure}
  \caption{(a) Two computed trajectories (orange and red) of an $a_p=0.1$ particle above an obstacle ($\beta=0.5$) and $\alpha=30^{\circ}$ inclined Stokes flow ($\psi$ isolines in dashed blue). Red shading indicates the exclusion zone for the hard-sphere particle center. (b) Magnified downstream view shows that the final streamline the particle asymptotes to is lower than its initial streamline ($\psi_f<\psi_i$) manifesting the effect of obstacle wall interaction in deflecting particles in such a fore-aft asymmetric travel.
(c) Particle-wall gap $\Delta$ as a function of travel time. The orange trajectory stays far from the wall ($\Delta>1$), while the red trajectory remains very close ($\Delta\ll1$). The inset shows good agreement around the minimum with computations using the wall expansion model \eqref{eq wall xpnsn} (red dashed line).
(d) Local value of stream function on the trajectories as a function of angular position $\eta$. The orange trajectory remains essentially undeflected, while the red trajectory shows the effects of strong deflection away (1-2) and towards (2-3) the obstacle. 
Asymmetry of the flow ensures a net change in streamfunction $\Delta\psi=|\psi_i-\psi_f|$. 
}
  \label{fig trajectory}
\end{figure}

For empirical computations, unless otherwise stated, we choose $a_p=0.1$ and track the particle center positions to get  pathlines.  Two representative trajectories are shown in figure~\ref{fig trajectory}(a): one (in orange) never gets close to the obstacle (as $\Delta>1$ throughout), while the other particle (in red) spends the majority of time creeping around the obstacle at very small gap values ($\Delta\ll 1$), a trajectory called a ``dive" in porous media literature \citep{miele2025flow}. 
Figure~\ref{fig trajectory}(c) shows the dynamics of the gap $\Delta$. The inset magnifies the closest approach of the dive trajectory towards the wall (position C), and verifies that the dynamics there is closely approximated by the leading order wall expansion model \eqref{eq wall xpnsn}.

With an initial condition 
\pkd{far}  from the obstacle (a situation common in microfluidics devices; we take $x_i=-5$, \pkd{although results are insensitive to this choice}), at first the particle moves along its initial streamline $\psi=\psi_i$ (stage (1) in figure~\ref{fig trajectory}). When closer to the obstacle, the wall-normal correction counteracts the background flow still pointing towards the obstacle. At position C, the normal flow velocity $u_{\perp}$ and the wall effect $W_{\perp}$ cancel, so that
 the particle normal velocity $v_{p\perp}=0$ and the particle-obstacle gap $\Delta$ reaches its minimum $\Delta_{min}$ (figure~\ref{fig trajectory}c).

Proceeding further in the dive stage (2), the wall effect $W_\perp$ becomes negative (cf.\ figures~\ref{fig flow}(b) and ~\ref{fig flow}(c)) and thus pushes the particle towards the obstacle while the background flow causes transport away from the wall. This continues until the particle ends its travel downstream on a well-defined final streamline $\psi=\psi_f$ (stage (3) in figure~\ref{fig trajectory}). 
For the situation depicted here ($\alpha<\pi/2$), we see that the 
phase of inward pull is of greater extent than that of outward repulsion, and that indeed there is a net downward displacement of the particle  ($\psi_f<\psi_i$, see figure~\ref{fig trajectory}b).

It is convenient to quantify displacements by changes of the instantaneous stream function value of the particle position.
Figure~\ref{fig trajectory}(d) shows $\psi$ as a function of angular elliptic coordinate $\eta$. While the trajectory staying far from the obstacle (orange) shows negligible changes, the dive trajectory (red, very small $\psi_i$) registers a rapid increase of $\psi$ in the dive phase up to C, and then a gradual decrease to a lasting displacement with well-defined
$\Delta \psi = |\psi_f-\psi_i|$, which quantifies the strength of net deflection due to the symmetry-broken flow geometry. 

At first glance, this implies larger $\Delta\psi$ as $\psi_i$ decreases. However, when the initial position approaches the separating streamline $\psi_i\rightarrow 0$, this tendency cannot continue: particles cannot cross the separating streamline for symmetry reasons (see above), so $\psi_f\to 0$ is required and also the (downward) deflection must vanish, $\Delta\psi\to 0$. 
This reasoning guarantees a characteristic  maximum value of net displacement $\Delta\psi$ for a particular $\psi_i$.
In the following subsections we systematically describe this surprising effect both numerically and analytically. 

\subsection{Particle net displacement across streamlines}\label{subsec computational net displacment}

Numerical integration of the particle trajectory was performed as described above for varying initial conditions. We fix $x_i=-5$ and vary $y_i$ in $\psi_i=\psi(x_i,y_i)$. The final stream function value is then assessed as $\psi_f=\psi(-x_i,y_f)$ when the particle's $x$-position reaches $x_p=-x_i$. We choose $|x_i|$ large enough that it does not influence the result, i.e., changes in $\psi$ for particle positions $x<x_i$ or $x>x_f$ are negligible.

\begin{figure}[t]
    \centering
\includegraphics[width=\textwidth]{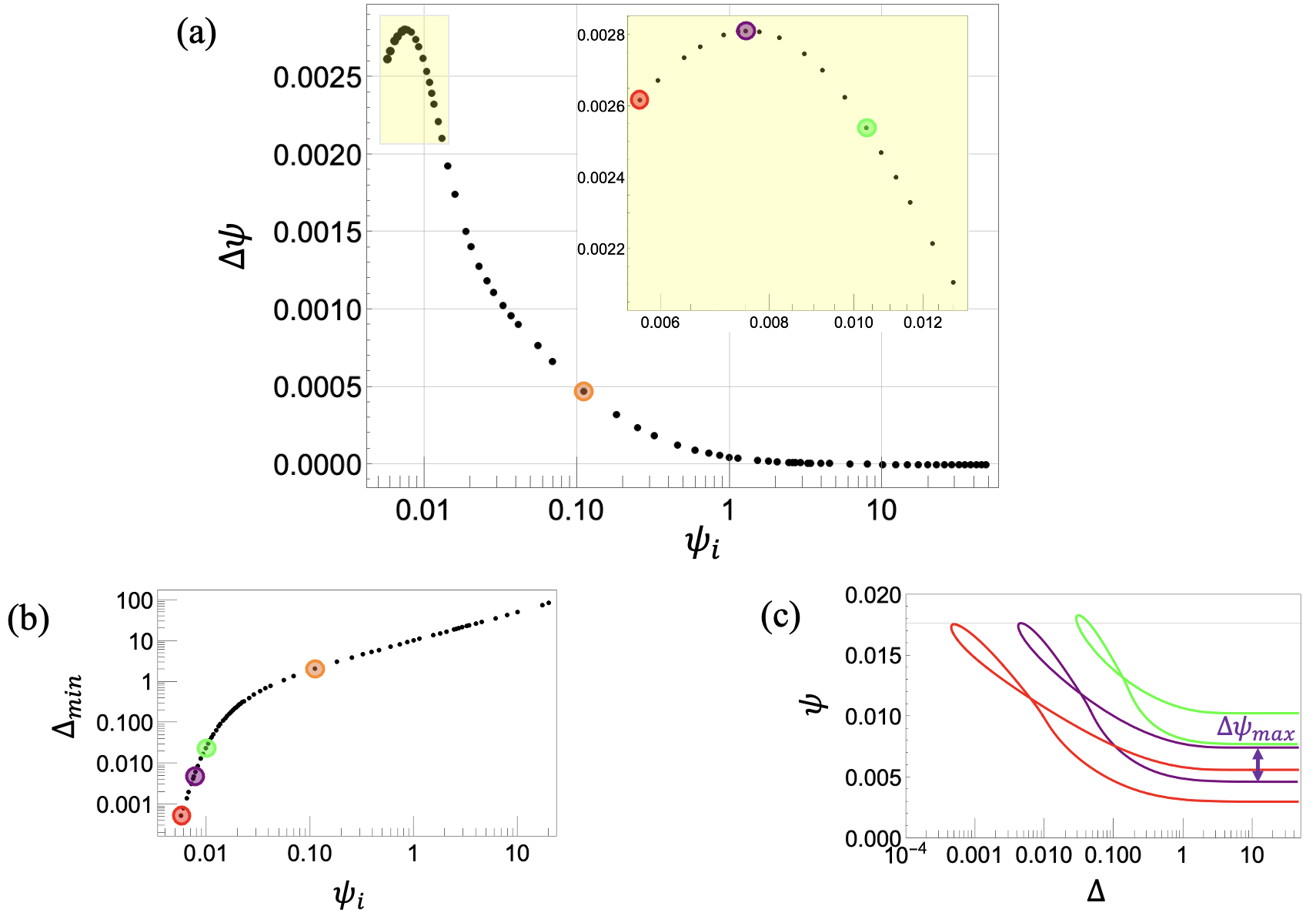}    \caption{Particle net displacement results for $(\alpha,\beta,a_p) = (30^{\circ},0.5,0.1)$. (a) Plot of displacement $\Delta\psi$ for particle trajectories released from different initial streamlines $\psi_i$. The inset magnifies the region around the maximum value $\Delta\psi_{max}$. (b) Variation of $\psi$ with $\Delta$ along the trajectories corresponding to the colored points in (a), including the $\Delta\psi_{max}$ trajectory in purple. (c) Minimum approach $\Delta_{min}$ for different particle trajectories of initial streamline $\psi_i$. The orange and red data points in the plot (a) and (c) represents the orange and red trajectories, respectively in figure~\ref{fig trajectory} 
}\label{fig deflection}
\end{figure}

As discussed in section~\ref{sub sec fore-aft symmetry analogy} and referred from figure~\ref{fig flow}(a), particle accumulates the wall-effect until it reaches an angular position when the corresponding first normal derivative of the normal background flow (cf \ref{fig flow}(b)) changes sign. The particle then experiences an opposite force which tries to compensate for the deflection already accumulated by the particle. We identify a nonzero net push on the particle by the wall meaning that the compensation cannot fully balance the displacement already accrued by the time particle creeps to the min gap point. The imbalance in such competition between the two opposite wall effects originating from breaking the fore-aft symmetry determines the net fate of the particle as shown in figure~\ref{fig deflection}(a). 

Figure~\ref{fig deflection}(a) shows the variation of the deflection measure $\Delta\psi$ with $\psi_i$ for $\beta=1/2, \alpha=30^{\circ}$.
Figure~\ref{fig deflection}(b) shows the relationship between $\psi_i$ and $\Delta_{min}$ determined from trajectory computation. Particles that start far away from the separating streamline ($\psi_i \gtrsim 1$) experience minimal wall interaction throughout $(\Delta_{min}\gg1)$ and $\Delta\psi$ remains nearly zero. Particles accumulate meaningful wall effect, and thus net displacement, as they encounter the obstacle in close proximity in $\Delta\lesssim1$ when starting at smaller $\psi_i$. The computations confirm the general argument stated in section~\ref{subsec trajectory} that crossing a stagnation streamline is not possible and particle starting from the stagnation streamline must come back to the same stagnation streamline in the downstream, exhibiting a prominent maximum in $\Delta\psi$ at a small value of $\psi_i$. The inset highlights three initial condition before (green), at (purple), and beyond (red) the maximum.
In figure~\ref{fig deflection}(c) we show the  changes in $\psi$ and $\Delta$ on these three particle trajectories. 

The dynamical system is solved in Mathematica with controlled accuracy, which for initial positions  $\psi_i \lesssim 0.005$
becomes computationally too demanding. The maximum in $\Delta \psi$ is, however, well resolved and we will show in chapter 5 that we can understand the behavior for $\psi_i\to 0$ analytically. The value of the maximum, $\Delta\psi_{max}$, and the corresponding initial position $\psi_{i,max}$ are main results of the current work that characterize the maximum net deflection that can be expected from a given symmetry-broken obstacle.

Figure~\ref{fig deflection}(a)  shows a net positive displacement $(\psi_i>\psi_f)$ meaning that they set to a downward position relative to their initial streamlines, $\psi_i$. We present particle deflection above the obstacle from left to right while other scenarios can be well-understood. Since the Stokes flow does not have any memory and the wall corrections are entirely determined by it, the nature of the forces is also local. 

We stress that the stream function changes quantified here are with respect to the 
reference flow $\psi$ from \eqref{eq faxen flow} and thus explicitly quantify wall effects only. The Fax\'en term in the reference flow describes the bulk flow curvature effects on particle trajectories (relative to passive transport in the background Stokes flow). We confirm in section~\ref{appen bulk Faxen} that deflection by the Fax\'en term in the absence of the wall is at least one order of magnitude smaller than the  wall effect, so that we isolate the main physics here.

\section{Sensitivity of the computation}

\subsection{Effect of the bulk Fax\'en correction}\label{appen bulk Faxen}

We have quantified the displacement along particle trajectories by stream function changes $\Delta\psi$ with respect to the reference flow $\psi$ from \eqref{eq faxen flow} in order to explicitly quantify the wall effects only. Even without wall effects, the presence of the Fax\'en term (the effects of bulk flow curvature) can lead to changes along the particle trajectory between the initial and final background stream function values, $\Delta\psi_B=\Delta\psi_{B,i}-\Delta\psi_{B,f}$.  Figure~\ref{fig faxen_DeltaE}(a) shows that this Fax\'en contribution has a noticeable effect only for initial conditions that keep the particle far from the obstacle \pkd{$\Delta\gtrsim 1$ throughout)}; yet in these cases, the displacement remains extremely small. For initial conditions of smaller $\psi_i$, which yield the important displacement effects discussed in this work, the Fax\'en effect is at least one order of magnitude smaller than the wall effects.

\subsection{Robustness of results against choice of the expansion gap parameter $\Delta_E$}\label{appen modeling}

\begin{figure}[t]
    \centering
\includegraphics[width=\textwidth]{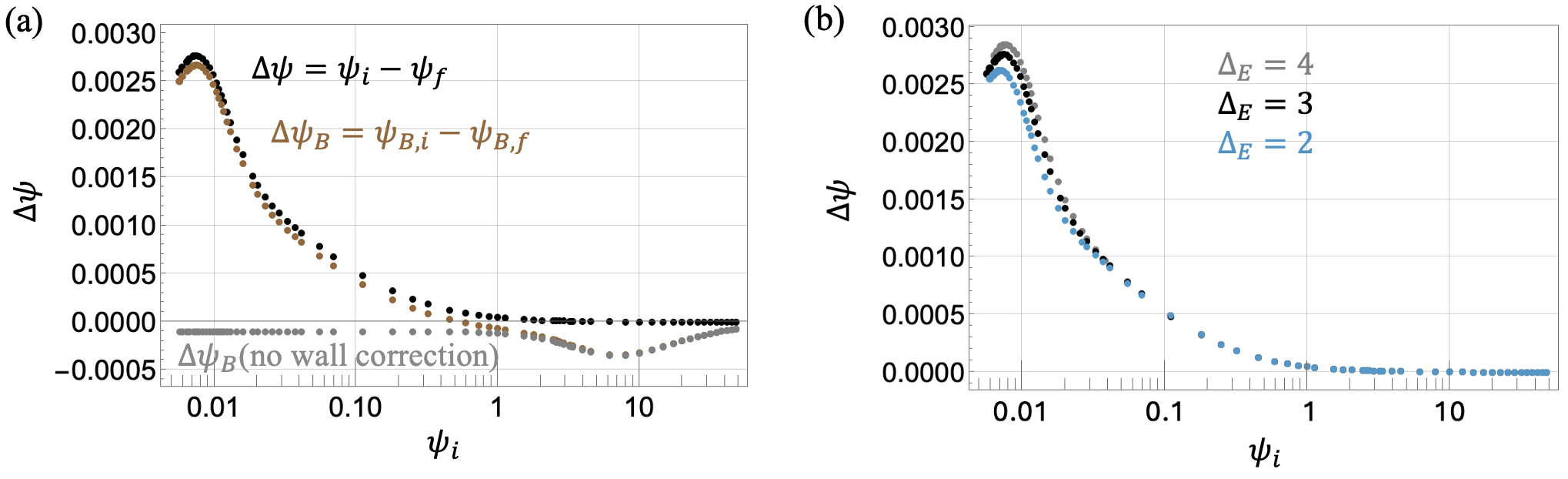}    \caption{
(a) Comparison of displacement effects (difference between final and initial stream function values along a trajectory) with the Fax\'en contribution 
subtracted out ($\Delta\psi$) or taken into account ($\Delta\psi_B$). Fax\'en effects are
only noticeable for trajectories \pkd{traveling at larger distances from the obstacle} (having larger $\psi_i\gtrsim 1$), where displacements are very small. For smaller initial conditions $\psi_i$ the displacements caused by the Fax\'en term alone (with no wall correction, gray) are insignificant compared to those caused by the wall effect. (b) Changing the value of the modeling parameter $\Delta_E$ in the variable expansion approach by ${\cal O}(1)$ factors only weakly affects the outcome of particle displacement. All computations are for $(a_p,\alpha,\beta) = (0.1,30degree,0.5)$. }
    \label{fig faxen_DeltaE}
\end{figure}

As mentioned in section~\ref{subsection: wall normal particle} , the general approach to determining the wall-normal correction, $\boldsymbol{W_\perp}$ of the particle velocity at moderate to large $\Delta$ involves expanding the background velocity field around the particle center position, cf.\ \eqref{eq vpPE}, while when the particle gets very close to touching the obstacle ($\Delta\to 0$) the description asymptotes to \eqref{eq wall xpnsn}, where the velocity
is expanded around a point at the obstacle wall closest to the particle center \cite{rallabandi2017hydrodynamic,adamczyk1983resistance,maude1961end,brenner1961slow}. However, there is no analytically established model in the intermediate $\Delta$ region. In order to smoothly transition from one description to the other, we introduce the variable expansion formalism \eqref{eq vpVE}, moving the expansion point continuously with the value of $\Delta$, an approach already utilized in our recent study on particle dynamics in internal Stokes flows \citep{liu2025principles}. 
We tested the robustness of this approach in the present case against changing the functional form of the expansion point dependence \eqref{eq xE} \pkd{to various nonlinear forms} and found that variations are small as long as the overall smoothness of the transition is preserved. In figure~\ref{fig faxen_DeltaE}(b), we show small quantitative differences upon changing the transition parameter $\Delta_E={\cal O}(1)$ 
in \eqref{eq xE}, demonstrating that the displacement effect as a whole, and its maximum magnitude, are insensitive to such modeling changes. 

Our quantitative choice of $\Delta_E$ is motivated by evaluating the smoothness of the dependence of $W_\perp$ on $\Delta$. This quantity is informed by flow gradients, which are locally stronger in the present case of external Stokes flow than in the work (Liu et  al. \cite{liu2025principles} used a generic $\Delta_E=1$) as discussed in section~\ref{subsection: wall normal particle}, so that a slightly longer transition region works best here.

\begin{figure}[t]
    \centering
\includegraphics[width=\textwidth]{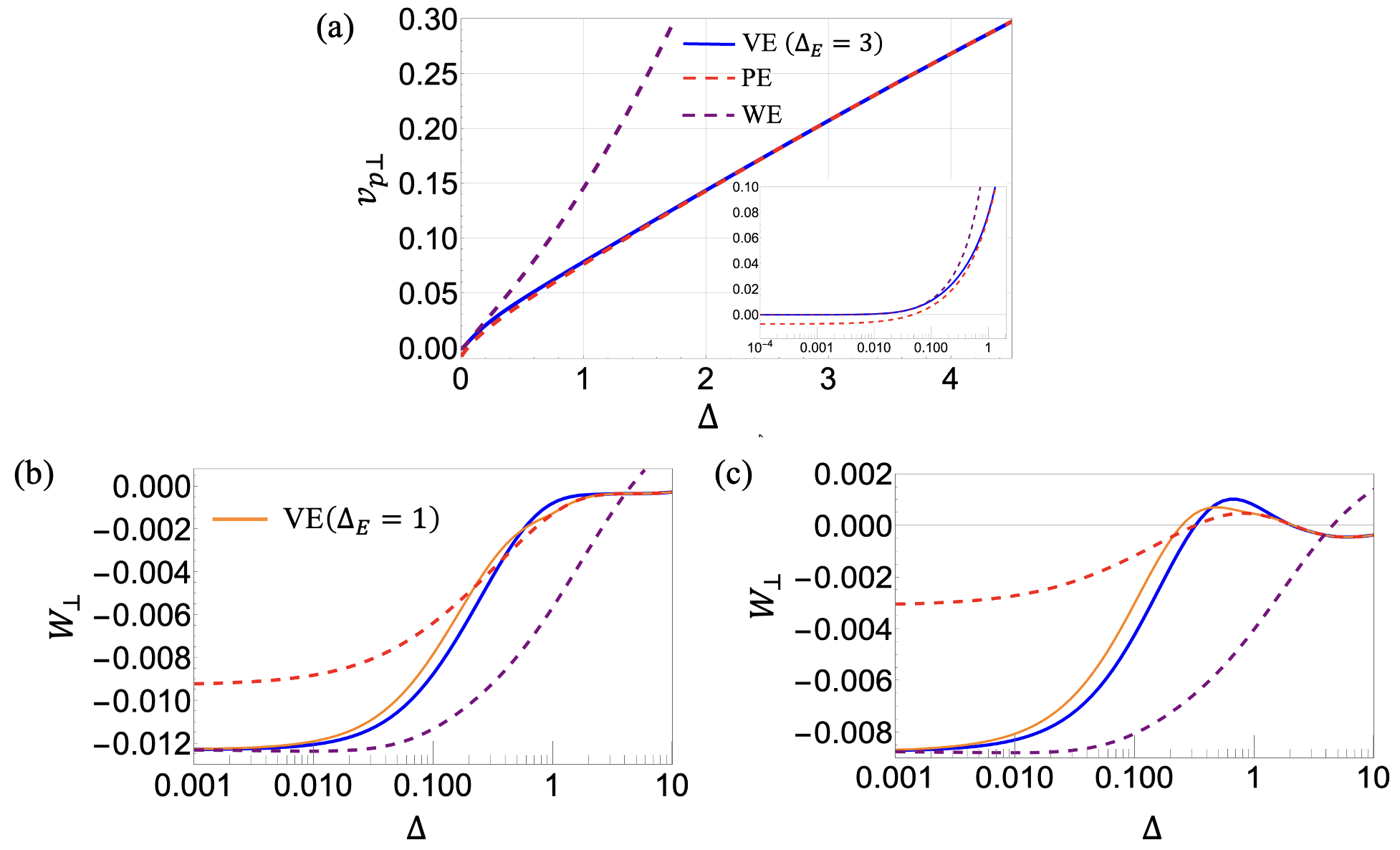}    \caption{
(a) An example of wall-normal particle velocity as a function of the gap coordinate $\Delta$ at an angular position $\eta=20^\circ$ 
using variable expansion modeling (cf. \eqref{eq vpVE}) with $\Delta_E=3$ showing a smooth transition \pkd{to the particle expansion for large $\Delta\gtrsim 1$ and to the wall expansion (inset) for $\Delta\ll 1$}. (b) and (c) Wall normal velocity correction with $\Delta_E=1$ and $3$ showing smoother transition from particle expansion model to wall expansion model for $\Delta_E=3$, in a range of $\eta$ values $(\eta=145^\circ$ and $155^\circ$ in (b) and (c) respectively$)$ where flow gradients are strong.}
    \label{fig DeltaE}
\end{figure}

Figures~\ref{fig DeltaE}(b) and (c) illustrates the greater smoothness of the variable-expansion normal velocity corrections $W_\perp$ as a function of normal coordinate for $\Delta_E=3$ compared to $\Delta_E=1$. We emphasize here a  range of $\eta$ that is particularly important for the particle dynamics and where the aforementioned gradients are strongest. While at other $\eta$ the value of $\Delta_E$ matters much less, we apply $\Delta_E=3$ uniformly. Note that the location of smooth transition to the particle expansion along the $\Delta$ axis is well separated from  that to the wall expansion at much smaller $\Delta$.

\subsection{Effect of wall curvature}\label{appen flat wall}

The results presented in the main text are obtained from wall corrections assuming a flat wall \citep{rallabandi2017hydrodynamic}, which is self-consistent in the limit of small gaps, $\Delta<1$. While it is intuitive that the majority of the wall interaction effects happen under this condition, one has to verify whether interactions at larger $\Delta$ accumulate to appreciable deflections and, if so, whether obstacle curvature must then be taken into account.

\begin{figure}
    \centering
\includegraphics[width=\textwidth]{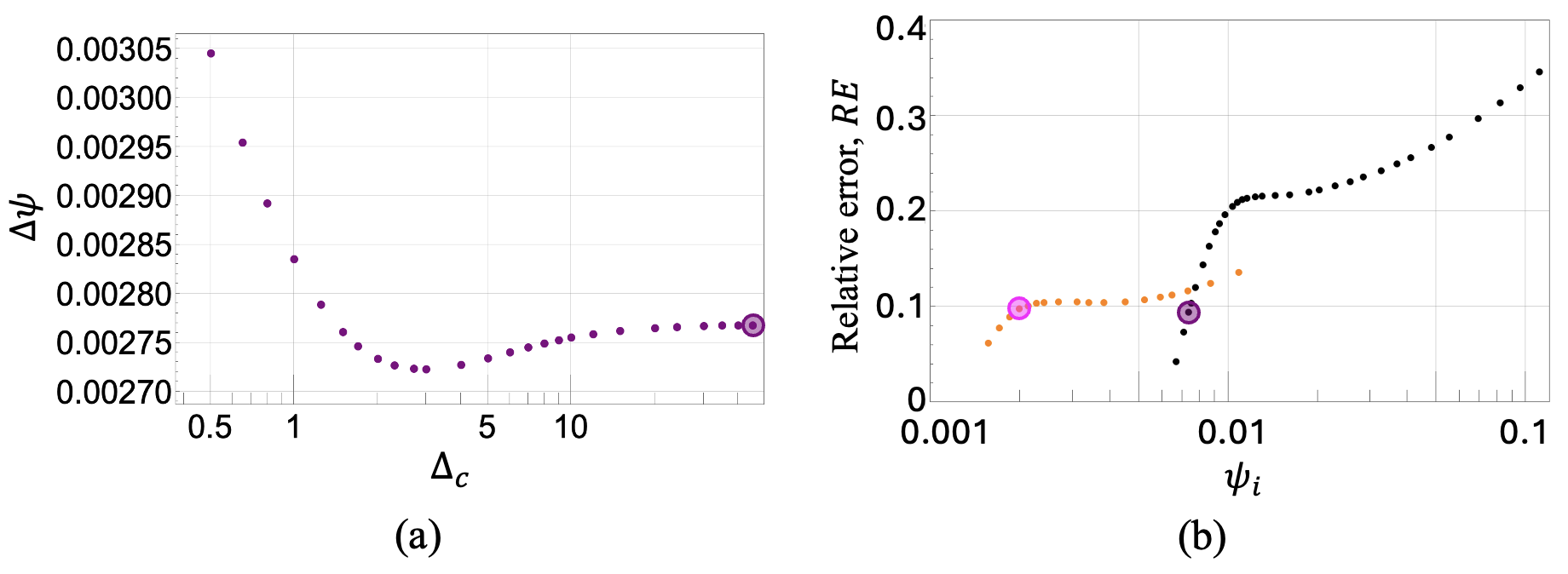}    \caption{(a) Computation of $\Delta\psi$ for $\psi_i$ corresponding to the ``purple" trajectory $(\Delta\psi_{max})$ in figure~\ref{fig deflection} with wall effect $W_{\perp}$ turned on for $\Delta<\Delta_c$; for results in the main text $W_{\perp}$ was turned on everywhere (purple circle on the right). The wall effects accumulated \pkd{at $\Delta\gtrsim 1$} have negligible effect on the displacement, as choosing $\Delta_c\gtrsim1$ does not affect the final outcome appreciably while a very late activation of $W_{\perp}$  misses some important effects. Therefore, we recomputed $\Delta\psi$ for trajectories with $\Delta_{min}\lesssim 1$ taking wall curvature into account (b). 
Plotted is the relative error $RE=(\Delta\psi_{curved\, wall}-\Delta\psi_{flat\, wall})/\Delta\psi_{curved\,wall}$ for two particle sizes, $a_p=0.1$ in black and $a_p=0.05$ in orange (magenta data corresponds to $\Delta\psi_{max}$ for $a_p=0.05$). Ignoring wall curvature when the wall effect is important $(\Delta_{min}\leq\Delta_c)$  underestimates the deflection, but RE remains small $(<10\%)$ in computing the maximum deflection. The control parameters are $(\alpha,\beta) = (30^{\circ},0.5)$. 
}
    \label{fig flat wall}
\end{figure}

To address the first point, we present in figure~\ref{fig flat wall}(a) computations of $\Delta\psi$ along the ``purple" type trajectory in figure~\ref{fig deflection} (exhibiting maximum displacement) where we only activate the wall effects for $\Delta<\Delta_c$, setting $W_\perp$ to zero for larger distances. The data presented in the main text are for $\Delta_c\sim 50$ (wall effects are activated everywhere, larger purple circle). The figure shows that wall effects are negligible for $\Delta_c\gtrsim 1$, confirming that (i) the vast majority of wall effect displacement happens at small $\Delta$, and (ii) the modeling of a distant elliptical obstacle at large distance as a flat wall does not introduce significant errors.

For the parts of the trajectory where $\Delta\leq\Delta_c$, is it important to model the finite curvature of the obstacle wall? We recomputed $\Delta\psi$ for the trajectories with $\Delta_{min}\leq 1$ using the curved-wall formalism developed for spherical obstacles in \cite{rallabandi2017hydrodynamic} by evaluating the radius of curvature $R$ of the wall point closest to the particle by 
\begin{equation}
R = \frac{1}{\beta}\left(\beta ^2 \cos^2{\eta}+\sin^2{\eta}\right)^{3/2} \,.
    \label{eq rcurv}
\end{equation}
Figure ~\ref{fig flat wall}(b) shows that the relative error in $\Delta\psi$ using the flat-wall formalism sharply drops for particles passing the obstacle at very close distance (in particular for trajectories near maximum deflection), while it becomes small throughout for smaller particles. Note that this computation likely overestimates  the influence of curvature, as \cite{rallabandi2017hydrodynamic} describe the effect of a nearby spherical obstacle (two radii of curvature); qualitatively, our computation indicates that the displacements are slightly enhanced by curvature. However, because the curvature formalism is computationally expensive and does not introduce large errors, we use the flat-wall formalism for all computations.

\section{Size based particle manipulation: Comparison with contact theory}\label{sec roughness}

We also compare the impact on particle separation by size due to the hydrodynamic effects modeled here to that inferred from the detailed modeling of short-range roughness effects \citep{frechette2009directional}. The present work treats interaction of a single particle and obstacle, and thus cannot provide critical particle sizes for crossing separating streamlines between multiple obstacles in a DLD array, but we can ask by how much the particle-obstacle interaction makes the trajectories of two particles of different sizes deviate from each other, setting up further downstream separation. We choose typical dimensional microparticle radii $a_{p1}=4\mu$m and $a_{p2}=8\mu$m, interacting with obstacles of circular cross section with $r=32\mu$m. In \cite{frechette2009directional}, the obstacles are themselves spherical and the interaction depends weakly on the  symmetry-breaking surface roughness -- we assume an experimentally relevant scale of $\sim 100$\,nm \citep{smart1989measurement,yang2007correction,hulagu2024towards}. The results in \cite{frechette2009directional} then allow for an evaluation of the displacements of the two particles when they are initially on the same streamline. Among all initial conditions, the maximum expected difference for the example parameters above is $\Delta n_{12}\approx 1.4\mu$m perpendicular to the uniform flow direction. Evaluating by comparison the hydrodynamic effects of encountering an elliptic cylinder, we choose scales of $a=40\mu$m and $b=20\mu$m (resulting in nearly the same cross-sectional area as the circular obstacle), and a flow angle of attack $\alpha=30^\circ$. At a distance of $\sim a$ behind the obstacle, we find a maximal displacement difference of the same particle sizes normal to the far-field flow of $\Delta n_{12}\approx 0.9\mu$m. Thus, the short-range roughness interactions for symmetric obstacles and the hydrodynamic effects modeled here for symmetry-breaking obstacles have a comparable effect on particle separation by size.   We emphasize here that our work does not invalidate the existing DLD explanation through non-hydrodynamic short-range interactions. Rather, we argue that both effects can naturally coexist and should both be taken into account when modeling DLD displacement and when designing DLD devices, 
as we expect significant quantitative changes when obstacle cross sections are made asymmetric.

\begin{figure}
    \centering
\includegraphics[width=\textwidth]{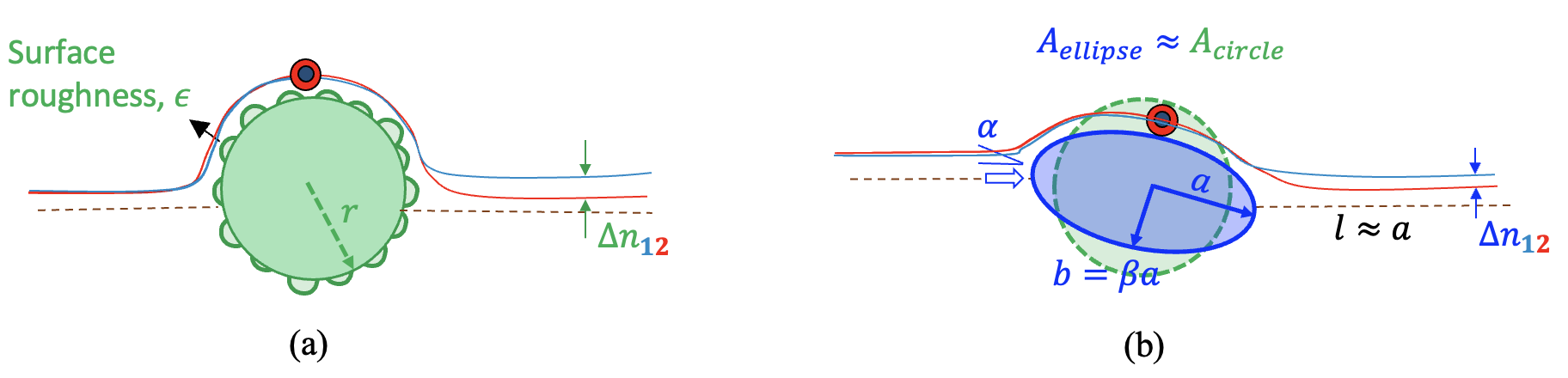}    \caption{Schematic of size-based particle displacement comparison between the non-hydrodynamic effect of roughness oriented particle contact with symmetric obstacle (a) vs the hydrodynamic effect of asymmetric obstacle that we have formulated (b). We compare the difference of the net displacements $\Delta n_{12}$ normal to the uniform stream of two particles of different sizes $a_{p1}$ (blue trajectory) and $a_{p2}$ (red trajectory) induced by the symmetric circular (radius $r$) and asymmetric elliptic obstacles (aspect ratio $\beta$) of nearly the same cross--sectional area. We measure the hydrodynamic $\Delta n _{12}$ in (b) at around obstacle size distance away $(l\approx a)$ which is typical with respect a DLD obstacle structure.}

\end{figure}

\section{Conclusion}

We have quantitatively demonstrated that force-free spherical particles transported in a Stokes flow over an obstacle can be systematically displaced through purely hydrodynamic interactions with the obstacle wall. Since a particle cannot contact a wall in finite time, a typical trajectory has a portion approaching the wall and a portion receding, governed by the background flow and the wall corrections derived from it. If the approaching and receding parts of the flow are symmetric, displacement effects off a streamline will cancel out. Thus, a Stokes-flow-inducing net particle displacement must break fore-aft symmetry of travel. 

To realize that symmetry breaking in a simple tractable way, we employ an inclined uniform Stokes flow past a 2D elliptic obstacle with a nontrivial angle of attack, which produces a fore-aft asymmetric distribution of the hydrodynamic wall effects. As a result, we observed that after excursing the obstacle, particles end to a final streamline different than the streamline they were released from. And we quantified $\Delta\psi$-– net displacement across streamlines.  Particle trajectories which are at a larger initial $\psi$, the net displacement is negligible but grows for trajectories with smaller $\psi_i$ as we approach to trajectories that have the dive phase where $\Delta$ is very small. We obtain trajectories traveling around the obstacle in close proximity when being released from an initial streamline close to the separating streamline. These trajectories experienced major wall effect asymmetry and as a result were displaced by a net amount downstream. However, if we were to start at precisely the separating streamline, for symmetry reasons we would have to end up on it again because it cannot be crossed so $\Delta\psi $ must logically decrease again and we find that from our simulated trajectories and a specific trajectory shows the maximum displacement. This maximum characterizes the capability of this obstacle to spatially displace and separate particles in an inertialess framework inheriting time-reversibility.

We quantify the sensitivity and uncertainty associated with choosing modeling parameters. Throughout our computation we have modeled particle motion around the obstacles as it sees the obstacle wall as flat. This is indeed true for the region $\Delta\ll1$ where particle goes around the wall in very close proximity. Figure~\ref{fig flat wall}(a) does show that, for a trajectory with overall significant displacement (trajectories with smaller $\psi_i$), the relative contribution of $\Delta\gtrsim 1$, is negligible. For much larger $\psi_i$ (and much smaller overall displacements), relative contributions of this far-field range are larger (but practically less relevant, in particular in comparison to the displacements $\Delta\psi_B$ stemming from the Fax\'en term, see figure~\ref{fig faxen_DeltaE}(a)). Trajectories with $\Delta_{min}\gtrsim 1$ show very small overall displacements and relative errors in their quantification have multiple sources, from the flow field modeling to the curvature effects, while the results are not very sensitive to the choice of $\Delta_E$. Our uncertainty quantification is thus meant to focus on the cases of strongest displacement, which characterize a particular obstacle's ability to displace particles.

We carefully modeled the computational trajectory by computing the wall-normal and wall-tangential components of background flow velocity and its gradients. Particular emphasis was given on accurately modeling the wall-normal velocity, which is critical for displacing particles obeying the steric requirement that particle can not penetrate the obstacle. We developed variable expansion model that involves expanding the background flow field at points varying between the center point of the particle to the closest point on the obstacle wall depending on the particle position. This approach showed a smooth and consistent transition between the existing particle expansion model and wall expansion model.  It is to note that, besides this variable expansion approach, one can apply a direct approach by first determining the offset of particle expansion velocity \eqref{eq vpPE} at $\Delta=0$, and then subtracting the offset from the original expression \eqref{eq vpPE} with a suitable exponential factor for decaying the offset effect as $\Delta\gg1$. We conducted a thorough investigation of the qualitative and quantitative robustness of $\Delta\psi$ comparing exponentially decaying offsets to asymptote to particle expansion velocity at different $\Delta$. We found that while the transition is not as smooth as that obtained from the variable expansion models, all the models are qualitatively same and quantitatively similar confirming that, regardless of the modeling approach in the transition region, a fore-aft asymmetric Stokes flow will displace particles while transporting them around an obstacle while a particular position will displace particles maximally.

This work places hydrodynamic wall interactions in the newest context of particle manipulation in microfluidic devices such as DLD devices. We perform a fair comparison between size dependent displacement effect using the existing theory of particle-wall contact by roughness around a circular obstacle and the hydrodynamic formalism that we have formulated around an elliptical obstacle. By comparison, we find that purely hydrodynamic interactions arising from symmetry-broken Stokes flows near solid boundaries can produce particle displacements of comparable magnitude, even in the absence of inertia or surface contact. This study therefore establishes a hydrodynamic framework for interpreting obstacle induced particle displacement in micro-scale transport problems.

Microfluidics with inclined arrays of multiple cylinders in a DLD device and heterogeneous pore configuration in porous media, the effects on the flow of groups (e.g. two) of circular cylinders by those of an effective elliptic cylinder, as the geometric motivation for symmetry breaking remains the same. In the proceeding chapter (chapter~\ref{chap 6}) we perform an analogous treatment of deflection from two circular cylinders should be possible, as flow solutions and trajectory geometry can be described in bipolar coordinates similar to the present-case elliptic coordinate system. Separating streamlines then connecting two obstacles become crucial boundaries that particles of different sizes do or do not cross, leading to DLD-like separation of particle paths by row-shifting.

In a microfluidic application with many particles, the present single-particle formalism is implicitly applicable to dilute  concentrations. An important extension of this approach would be to incorporate particle-particle interactions, allowing for the assessment of non-inertial effects in particle-laden flows, which are crucial for many practical applications \citep{guha2008transport}. Another valuable generalization of the current approach is to treat non-spherical particles, whose additional degrees of freedom allow for a wider range of qualitative trajectory behaviors \citep{yerasi2022spirographic,li2024dynamics,liu2025particlethesis}. In all cases, taking symmetry-breaking hydrodynamic interactions with boundaries into account will add a previously overlooked component to particle manipulation in any viscous flow situation.

The 2D flow geometry in the present work was chosen both for simplicity and for its fundamental importance in a large class of microfluidic set-ups. Many other applications of force-free, inertialess particle manipulation use 3D geometries, particularly in porous media filtration \citep{davis2006deterministic, bordoloi2022structure,miele2025flow}). We remark  that the modeling of boundary interactions employed here would be unchanged in this 3D case as long as 
the particle sizes remain much smaller than both  radii of curvature of the 3-D obstacle geometry. In order to quantitatively model a particular 3D flow situation, though, it will be advantageous to connect the fundamental hydrodyamic interaction formalism developed here with a numerical Stokes flow simulation, which could deal with arbitrary geometries and also address 
challenges like noise and fluctuations.

\chapter{Particle net displacement theory}\label{chap 4}

In previous chapter, we have developed the computational model that have systematically quantified the net displacement of a single, spherical, and neutrally buoyant particle in a Stokes flow. Computation of the time integral of the trajectory that numerically ensures that a single obstacle will deflect particles to varying degrees depending on the closeness of the streamline to the separating streamline. Our empirical computation was successful in resolving the maximum net displacement effect with controlled numerical accuracy and precision which becomes too demanding for computing all possible trajectories passing the obstacles in close proximity and thus for capturing the asymptotics behavior of  the net displacement. This chapter\footnote[1]{This chapter is adapted from Das et al. \cite{das2025controlled}} presents a rigorous theoretical framework in understanding the displacement effect better through the lens of analytics that eventually offers method to obtain its asymptotic limits and develops the scaling laws that quantify the capacity of an obstacle to displace particles.

\section{Analytical formalism for the net displacement}\label{sec analytical}

In this section, we develop a method of predicting $\Delta\psi_{max}$, particularly in the limit $\psi_i\to 0$, which proves computationally challenging. We shall see that only information directly derived from the background flow field is needed.

All trajectories for $\psi_i\lesssim \psi_{i,max}$ are of the ``red" type in figure~\ref{fig trajectory}, i.e., they (1) approach the obstacle in close proximity of the upstream separating streamline (the angular particle position is $\eta_i\approx \pi+\eta^{sep}$), (2)
accumulate meaningful deflection while moving along the obstacle surface maintaining very small gaps $\Delta\ll 1$ (``dives",  \cite{miele2025flow}), and (3) leave on a final streamline again close to the downstream separating streamline ($\eta_f\approx \eta^{sep}$). To good approximation, the important ``dive" phase is described by the wall expansion model \eqref{eq wall xpnsn}, see the inset of figure~\ref{fig trajectory}(c). Noting $v_{p\perp}=a_p d\Delta/dt$, and dividing by $d\eta/dt$ along the trajectory, we write
\begin{equation} 
        \frac{d(\log\Delta)}{d\eta}=1.6147\frac{a_p\kappa}{\frac{d\eta}{dt}}\,,
    \end{equation} 
an equation we want to integrate from a starting point at the end of phase (1) $(\Delta_i,\eta_i)$ to an end point at the start of phase (3) $(\Delta_f,\eta_f)$. As particle displacements are strongly dominated by the (2) phase, the results are insensitive to the choice of $\Delta_i$ and $\Delta_f$, as long as both are $\ll 1$. In particular, we can choose $\Delta_i=\Delta_f=\Delta^*$ and thus require 

\begin{equation} \label{eqdeltaeta}
\int_{\eta_i}^{\eta_f} 1.6147 \frac{a_p\kappa}{\frac{d\eta} {dt}} d\eta = 0 \,.
\end{equation} 
To make analytical progress, we observe that $d\eta/dt$ along the trajectory is the rate of change in the $\eta$ direction of a particle moving at a distance of $\approx a_p$ from the obstacle surface, which translates to the wall-parallel particle velocity $v_{p \parallel}$, and further to the $\eta$-component of the background velocity $u_\eta$ as defined in section~\ref{sub sec fore-aft symmetry analogy} (equation \eqref{eq ueta}), 
\begin{multline}\label{eqetadotvpar}
\frac{d\eta}{dt} = g(\xi_{a_p},\eta) v_{p \parallel} = g(\xi_{a_p},\eta) (1-f(\Delta)) u_{\parallel}\\= g(\xi_{a_p},\eta) (1-f(\Delta)) u_{\eta}(\xi_{a_p},\eta)=-g^2(\xi_{a_p},\eta) (1-f(\Delta))\partial_\xi\psi_B(\xi_{a_p},\eta)\,.
 \end{multline} 
These equations, accurate to relative order ${\cal O}(a_p^2)$, contain the scale factor of the elliptic coordinate system $g =g(\xi,\eta)$ and the $\xi$-derivative of the background streamfunction $\psi_B$ (cf. sections \S\ref{sec flow field} and \S\ref{sub sec fore-aft symmetry analogy}) both evaluated at $\xi_{a_p}(\eta) = \xi_0 + a_p g(\xi_0,\eta)$, representing points a distance $a_p$ from the obstacle wall.  
The factor $f(\Delta)$ from \eqref{eq vpparallelFull} depends very weakly on $\Delta$ over the range of interest here, so that we replace it by a constant average value $\tilde{f}={\cal O}(1)$. The choice of  this constant $\bar{f}$ is irrelevant for solving \eqref{eqdeltaeta}. 
Combining the constants into $\mathcal{S}=-\frac{1.6147}{1-\tilde{f}}$, the integrand of \eqref{eqdeltaeta} becomes
\begin{equation} \label{eq phi}
       \phi(\eta)=\mathcal{S}\frac{a_p\kappa}{g^2 \partial_\xi\psi_B}\,,
    \end{equation}
which is now entirely defined by the background flow, and is a function of $\eta$ only. Its  behavior is discussed in more detail in the succeeding section \ref{scaling}.

\begin{figure}[t]
    \centering
\includegraphics[width=0.7\textwidth]{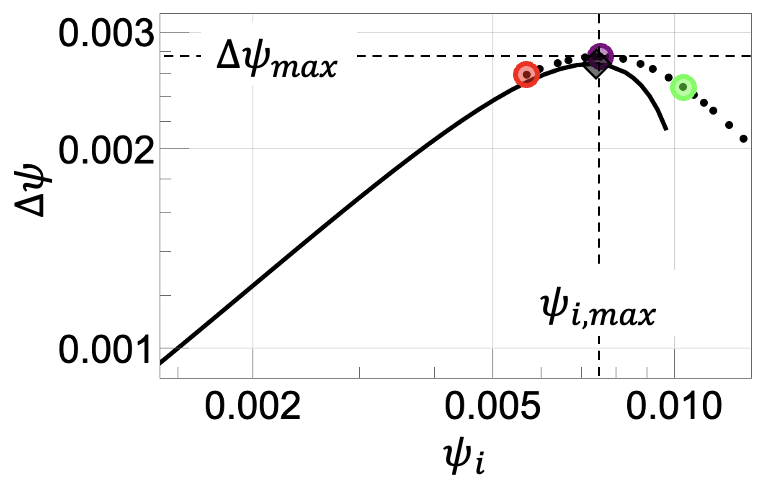}    \caption{
Particle net displacement across streamlines $\Delta\psi$. The solid line depicts  $\Delta\psi(\psi_i)$ from the analytical model equation~\eqref{eq etafinal etaini} for small $\psi_i$ while the black dots are the computational trajectory data taken from figure~\ref{fig deflection}(a). Red, purple, and green data points represent the ``red", ``purple", and ``green" trajectories respectively in figure~\ref{fig deflection}. Both the magnitude of $\Delta\psi_{max}$ and its location $\psi_{i,max}$ are in good agreement with the empirical calculations.  }
    \label{fig analytical}
\end{figure}

Introducing for convenience the indefinite integral
$I(\eta)\equiv \int^\eta\phi(\tilde{\eta})d\tilde{\eta}$\,,
the condition \eqref{eqdeltaeta} becomes
    \begin{equation} \label{eq etafinal etaini}
        I(\eta_f)=I(\eta_i)\,,
    \end{equation}
relating valid pairs of initial and final $\eta$ coordinates. These translate into initial and final stream function values 
\begin{equation}\label{eq psistar}
\psi_i^*=\psi(\Delta=\Delta^*,\eta=\eta_i)\,,
\end{equation}
\begin{equation} \label{eq psfstar}
    \psi_f^*=\psi(\Delta=\Delta^*,\eta=\eta_f)\,.
\end{equation} 
The results of this calculation are insensitive to the precise value of $\Delta^*$ as long as it is $\ll 1$ as described in the next section \ref{sec Deltastar}. Figure~\ref{fig analytical} 
uses $\Delta^*=0.05$ and shows good agreement with the small-$\psi_i$ trajectory data from figure ~\ref{fig deflection}(a). It is remarkable that this small-$\psi_i$ theory yields not only an asymptote for $\Delta\psi\to 0$, but captures the position of $\Delta\psi_{max}$.

\section{Sensitivity of the analytical result against the choice of parameter $\Delta^*$}\label{sec Deltastar}

\begin{figure}
    \centering
\includegraphics[width=0.7\textwidth]{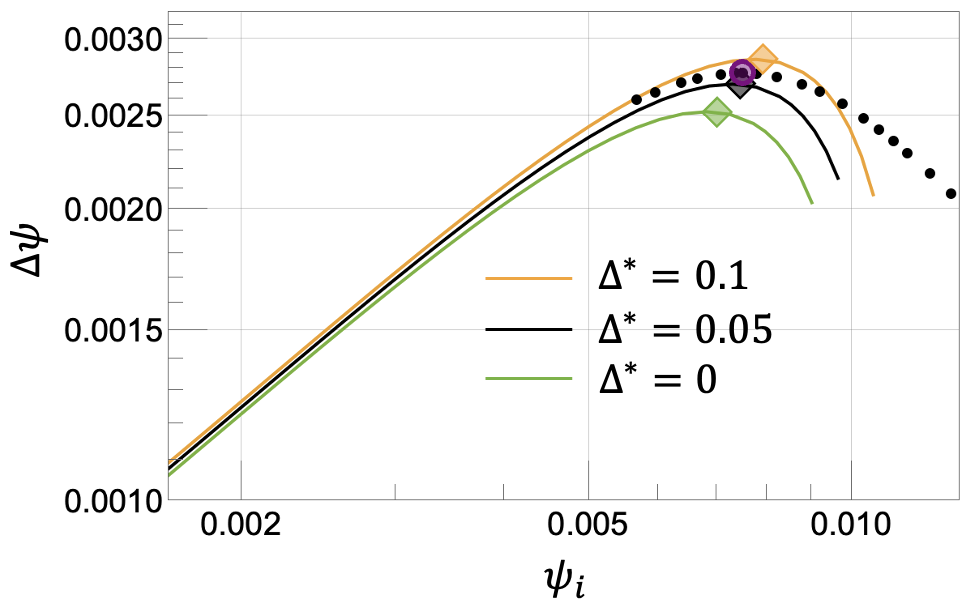}    \caption{Results for $\Delta\psi(\psi_i)$ computed from equations \eqref{eq etafinal etaini}--\eqref{eq psfstar} with different choice of $\Delta^*$ according to the methodology discussed in \S \ref{sec analytical}. Black dots are the results obtained from direct numerical analysis of trajectories, cf.\ \S\ref{subsec computational net displacment}. Colored symbols identify the positions of $\Delta\psi_{max}$. Here $(a_p,\alpha,\beta) = (0.1,30degree,0.5)$.
}
    \label{fig Deltastar}
\end{figure}

As part of establishing the analytical model of $\Delta\psi$ from \eqref{eq etafinal etaini}, we need to specify a gap value $\Delta^*$ to compute $\psi^*_i$ and $\psi_f^*$ from \eqref{eq psistar} and \eqref{eq psfstar}, respectively, as described in the preceding section   \ref{sec analytical}. We see from figure ~\ref{fig Deltastar} that the choice of $\Delta^*$ in evaluating $\psi^*_i$ and $\psi^*_f$ only weakly affects  the ultimate displacement $\Delta \psi=\psi^*_i-
\psi_f^*$, as long as $\Delta^*\ll 1$, as the wall-expansion modeling formalism needs $\Delta\ll1$ throughout on trajectories. In particular, the existence and position of a maximum in $\Delta\psi$ are robust against that choice. This motivated the simplified analytical treatment in \ref{scaling}, where most quantities are evaluated at a distance of $a_p$ from the obstacle, i.e., at $\Delta=0$.  

\section{Scaling laws for the displacement capacity of the elliptic obstacle}\label{scaling}

\subsection{Analytical simplification}\label{subsec analytical simplification}

The analytical approach in section \ref{sec analytical} is successful, but still evaluates the necessary functions and integrals numerically. In order to understand systematically how maximum deflection depends on the parameters of the particle-obstacle encounter, we employ further simplifications.  
First, we replace $\psi$ by $\psi_B$ in the definition of $\Delta\psi$ (cf.\  \eqref{eq psiB}
), again using $\xi=\xi_{a_p}$, and furthermore expand $\psi_B$ in small $\xi-\xi_0$ (these approximations are consistent with $a_p\ll 1$).
To leading order in $a_p$, the simplified background stream function reads
\begin{equation}\label{eq streamfunction expansion in xi0}
    \hat{\psi}_{B}(\eta)=(\sin\eta\cos\alpha-\beta\cos\eta\sin\alpha)a_p^2g^2(\xi_0,\eta)\,.
\end{equation}
This stream function expression is used to evaluate initial and final values $\hat{\psi}_{i,f}=\hat{\psi}_B(\eta_{i,f}$) for given angular arguments, and also to obtain simplified versions of the functions $\phi\to\hat{\phi}$ from \eqref{eq phi} and $I\to \hat{I}$ resulting in an algebraically simplified analog of \eqref{eq etafinal etaini},
    \begin{equation} \label{eq ihat}
        \hat{I}(\eta_f)=\hat{I}(\eta_i)\,.
    \end{equation}

To directly compute the maximum of $\Delta\psi(\hat{\psi}_i)$, we need to derive a second equation. We isolate the values for maximum particle deflection by starting with the definition of the extremum 
\begin{equation}\label{eq max Deltapsi psi}
    \frac{\partial\Delta\psi}{\partial\psi_i}=0
\end{equation}
and using the chain rule and the simplified streamfunction $\hat{\psi}_B$ from \eqref{eq streamfunction expansion in xi0} to obtain
\begin{equation}\label{eq max Deltapsi psi 2}
  \frac{\partial\Delta\psi}{\partial\hat{\psi}_B}=\frac{\partial\Delta\psi}{\partial \hat{I}}\frac{\partial_\eta \hat{I}}{\partial_\eta\hat{\psi}_B}\,.
\end{equation}
One can find that both $\partial_\eta \hat{I}=\hat{\phi}$ and $\partial_\eta \hat{\psi}_B$ are non-zero over the range of $\eta$ of interest for evaluating the maximum. The monotonicity of
$I(\eta)$ and $\psi(\eta)$ around $\Delta\psi_{max}$ allows to first write the maximum condition as
\begin{equation} \label{eq max delfection}
    \frac{\partial\Delta\psi}{\partial \hat{I}}=0\,,
\end{equation}
and further,  defining $\zeta(\eta)\equiv \partial_{\eta}\hat{\psi}_B/\hat{\phi}$, we conclude that \eqref{eq max delfection} implies
\begin{equation}\label{eq zeta}
\zeta(\eta_f)=\zeta(\eta_i)\,.
\end{equation}

Solving \eqref{eq ihat} and \eqref{eq zeta} simultaneously yields a pair of values $(\eta_i,\eta_f)=(\eta_{i,max},\eta_{f,max})$ that determine 

\begin{equation}\label{eq Deltapsimax analytical}
    \Delta\psi_{max}=\psi_{i,max}-\psi_{f,max}\, ,
\end{equation}
from
\begin{equation}\label{eq psiimax analytical}
    \psi_{i,max}=\hat{\psi}_B(\eta_{i,max})\, ,
\end{equation}
\begin{equation}\label{eq psifmax analytical}
    \psi_{f,max}=\hat{\psi}_B(\eta_{f,max}).
\end{equation}
To make analytical progress, it is crucial to acknowledge that the upstream and downstream angular positions $\eta_i$ and $\eta_f$ are not equidistant from the separation streamlines, but that in writing $\eta_i=\pi+\eta^{sep}-\Delta\eta_i$ and  $\eta_f=\eta^{sep}+\Delta\eta_f$, the angular deviations have a leading-order asymmetry $\delta = \Delta\eta_i-\Delta\eta_f$, which vanishes as $\alpha\to 0$. 

\begin{figure}
    \centering
\includegraphics[width=0.6\textwidth]{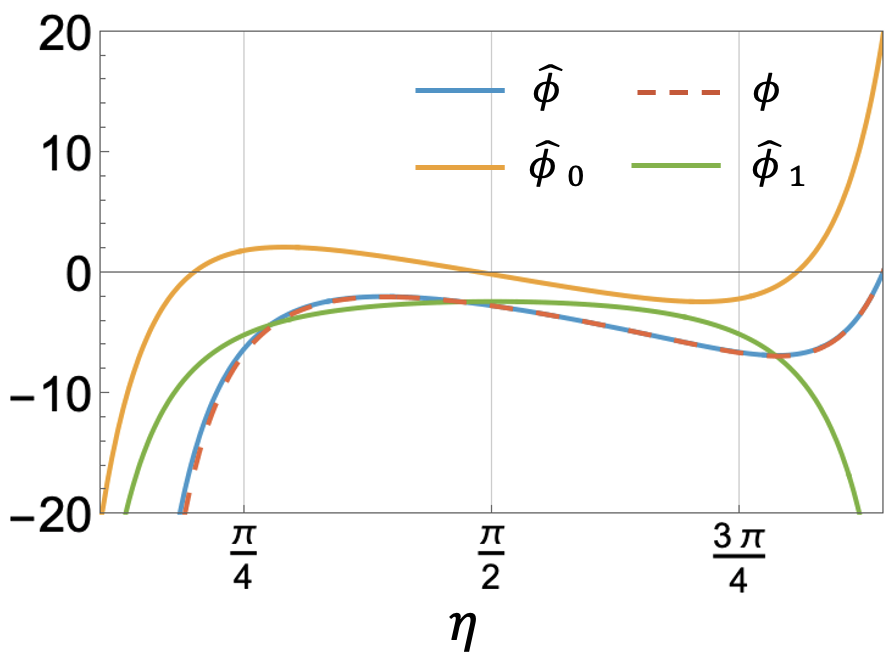}    \caption{The analytically integrable function $\hat{\phi}$ as introduced in section \S\ref{scaling} is in excellent agreement with the $\phi$ function \eqref{eq phi} developed in section 3.   We expand $\hat{\phi}$ in $\sin{2\alpha}$ as $\hat{\phi}=\hat{\phi}_0+\hat{\phi}_1\sin{2\alpha+O(\sin^2{2\alpha})}$ where the leading order term $\hat{\phi}_0$ is antisymmetric but the first order term $\hat{\phi}_1$ is symmetric around $\eta=\pi/2$. Here $(a_p,\alpha,\beta)\equiv(0.1,30degree,0.5)$.}
    \label{fig phi}
\end{figure}

The central quantity on which analytical computation of displacement hinges is the function $\phi(\eta)$ from \eqref{eq phi} depicted in figure~\ref{fig phi}. We know that $\phi$ must be symmetric with respect to $\eta=\pi/2$ when $\alpha=0$, and proceed to expand it for small $\alpha$. Consistent to leading order, we choose $\sin 2\alpha$ as our expansion parameter, which conforms with the expected behavior at larger $\alpha$. Expanding as far as $\phi=\phi_0+\phi_1\sin 2\alpha$, equation~\eqref{eq etafinal etaini} reads
\begin{equation}\label{eqinitial}
    \int_{\eta_f}^{\eta_i}(\phi_0+\phi_1\sin{2\alpha}) d\eta=0\,.
\end{equation}
Making use of the simplified background stream function \eqref{eq streamfunction expansion in xi0} and consistent evaluation of the terms in \eqref{eq phi} at $a_p$ distance to the obstacle ($\Delta\to 0$), 
we obtain analytically integrable versions of the two leading-order functions in \eqref{eqinitial}, namely 
\begin{multline}
    \hat\phi_0=\frac{(1-\beta ^2)\left(3 \beta ^2-1+\left(1-\beta ^2\right) \cos {2 \eta}\right)}{
   \left(1-\left(1-\beta ^2\right) \cos {2\eta}+\beta ^2\right)^{5/2}\tan {\eta}}\\\left(\cos {2 \eta}-\cosh 
   \left(\frac{2\sqrt{2} a_p}{\sqrt{1-\left(1-\beta^2\right) \cos {2 \eta}+\beta^2}}+2\tanh
   ^{-1}\beta\right)\right)
\end{multline}
and
\begin{multline}
    \hat\phi_1=\frac{\beta (1-\beta^2) }{(1-(1-\beta^2 )\cos {2 \eta}+\beta^2)(1-\cos{2\eta})}\\\left(\cos {2 \eta}-\cosh 
   \left(\frac{2\sqrt{2} a_p}{\sqrt{1-\left(1-\beta^2\right) \cos {2 \eta}+\beta^2}}+2\tanh
   ^{-1}\beta\right)\right)\,.
\end{multline}
As can be seen in figure ~\ref{fig phi}, $\hat\phi_0$ is odd and $\hat\phi_1$ is even with respect to $\eta=\pi/2$ (and the sum of both terms is an excellent approximation of $\phi$).
However, integration over $\hat\phi_0$ from $\eta_i=\pi+\eta^{sep}-\Delta\eta_i$ to $\eta_f=\eta^{sep}+\Delta\eta_f$ does not respect these symmetries for two reasons: $\eta^{sep}\not = 0$ and $\delta=\Delta\eta_i-\Delta\eta_f\not = 0$. As already stated, both quantities do vanish as $\alpha\to 0$; from its definition we have $\eta^{sep} =(\beta/2)\sin 2\alpha$ to leading order and we write $\delta=\delta_1 \sin 2\alpha$ with a $\delta_1$ to be determined. Substituting into \eqref{eq ihat}, \eqref{eq zeta} and further expansion in the small quantities $a_p$ and $\Delta\eta_f$ results in a system of two equations that can be solved for $\delta_1$ and $\Delta\eta_f$ to determine $\Delta\psi_{max}$ from equations~\eqref{eq Deltapsimax analytical}-\eqref{eq psifmax analytical} as detailed next.

\subsection{Determination of $\delta_1$ and $\Delta\eta_f$: Parameterization of $\Delta\psi_{max}$}

Applying the definitions of $\eta_i,\,\eta_f,\,\eta^{sep},$ and $\delta$ as described in the preceding section~\ref{subsec analytical simplification} into \eqref{eqinitial}, all non-zero terms are proportional to $\sin 2\alpha$ and we obtain an explicit equation for $\delta_1$,

\begin{equation}\label{eq delta1 from phi}
    \delta_1=\frac{\int_{\eta=\Delta\eta_f}^{\eta=\pi-\Delta\eta_f}\hat\phi_1 d\eta}{\hat\phi_0(\eta=\pi-\Delta\eta_f)}+\beta\,.
\end{equation}

Furthermore, we expand \eqref{eq delta1 from phi} in small $a_p$ and small $\Delta\eta_f$ to obtain

\begin{equation}\label{eq delta1 from phi leading}
    \begin{split}
        \delta_1=2\bigg[& a_p\frac{(2-\beta^2) E\left(1 - \frac{1}{\beta^2}\right)- 
        K\left(1 - \frac{1}{\beta^2}\right)}{\beta}\Delta\eta_f \\& -\frac{(1- \beta^2) (\beta^2 + 3 a_p)}{\beta^3}\Delta\eta_f^2+O(\Delta\eta_f^3)\bigg]+O(a_p^2)
    \end{split}
\end{equation}
with the complete elliptic integrals of first ($K$) and second ($E$) kind. Equation~\eqref{eq delta1 from phi leading} computes $\delta$, and thus $\eta_i$, from a given $\Delta\eta_f$, and thus $\eta_f$. 

We now utilize \eqref{eq zeta} to obtain a second equation that will solve $\delta_1$ and $\Delta\eta_f$ simultaneously from \eqref{eq delta1 from phi leading}. 
We first define $\zeta(\eta)\equiv \partial_{\eta}\hat{\psi}_B/\hat{\phi}$ (cf. section~\ref{subsec analytical simplification}) and rewrite \eqref{eq zeta} as
\begin{equation}\label{eq zeta Deltaeta}
    \zeta(\eta_i=\pi+\eta^{sep}-\Delta\eta_f-\delta)=\zeta(\eta_f=\eta^{sep}+\Delta\eta_f)\,.
\end{equation}
Employing the same expansion in small $\delta$ as above, we find
\begin{equation}\label{eq delta from zeta}
\delta=\delta_1\sin{2\alpha}=\frac{\zeta(\eta=\pi+\eta^{sep}-\Delta\eta_f)-\zeta(\eta=\eta^{sep}+\Delta\eta_f)}{\partial_{\eta}\zeta|_{\eta=\pi+\eta^{sep}-\Delta\eta_f}}\,,
\end{equation}
and expansion of the RHS in small $\eta^{sep}$ again yields terms proportional to $\sin 2\alpha$. Expanding in small $a_p$ and $\Delta\eta$ results in another leading-order expression for $\delta_1$,
\begin{equation}\label{eq delta zeta leading}
    \delta_1=\left[6a_p\frac{1-\beta^2}{\beta^3}\Delta\eta_f^2+O(\Delta\eta_f^3)\right]+O(a_p^2)
\end{equation}

Solving \eqref{eq delta1 from phi leading} and \eqref{eq delta zeta leading} simultaneously for $\delta_1$ and $\Delta\eta_f$, we obtain 

\begin{equation}\label{eq DeltaEtaf full}
    \Delta\eta_f=\frac{a_p\beta^2[(2-\beta^2) E(1 - \frac{1}{\beta^2})- 
        K(1 - \frac{1}{\beta^2})]}{(1-\beta^2)(6a_p+\beta^2)} \,,
\end{equation}

\begin{equation}\label{eq delta1 full}
    \delta_1=\frac{6a_p^3\beta[(2-\beta^2) E(1 - \frac{1}{\beta^2})- 
        K(1 - \frac{1}{\beta^2})]^2}{(1-\beta^2)(6a_p+\beta^2)^2} \,.
\end{equation}
The same consistent expansions in $\hat{\psi}_B$ eventually provide an analytical expression for $\Delta\psi_{max}$ in terms of $\delta_1$ and $\Delta\eta_f$, namely

\begin{equation}\label{eq Deltapsimax full in delta1 Deltaetaf}
    \Delta\psi_{max}=a_p^2 \left[\frac{\delta_1}{\beta^2}+\frac{2(1-\beta^2)}{\beta^3}\Delta\eta_f^2+\delta_1\frac{(-6+5\beta^2)}{2\beta^4}\Delta\eta_f^2+O(\Delta\eta_f^3)\right]\sin{2\alpha}\,.
\end{equation}
Plugging \eqref{eq DeltaEtaf full} and \eqref{eq delta1 full} into \eqref{eq Deltapsimax full in delta1 Deltaetaf} allows for evaluation $\Delta\psi_{max}$ at arbitrary $\beta$ values. Note that only the $\delta_1$ terms in \eqref{eq Deltapsimax full in delta1 Deltaetaf} contribute in this limit.  

\begin{figure}
  \centering
  \begin{subfigure}[t]{\textwidth}
    \centering
    \includegraphics[width=\textwidth]{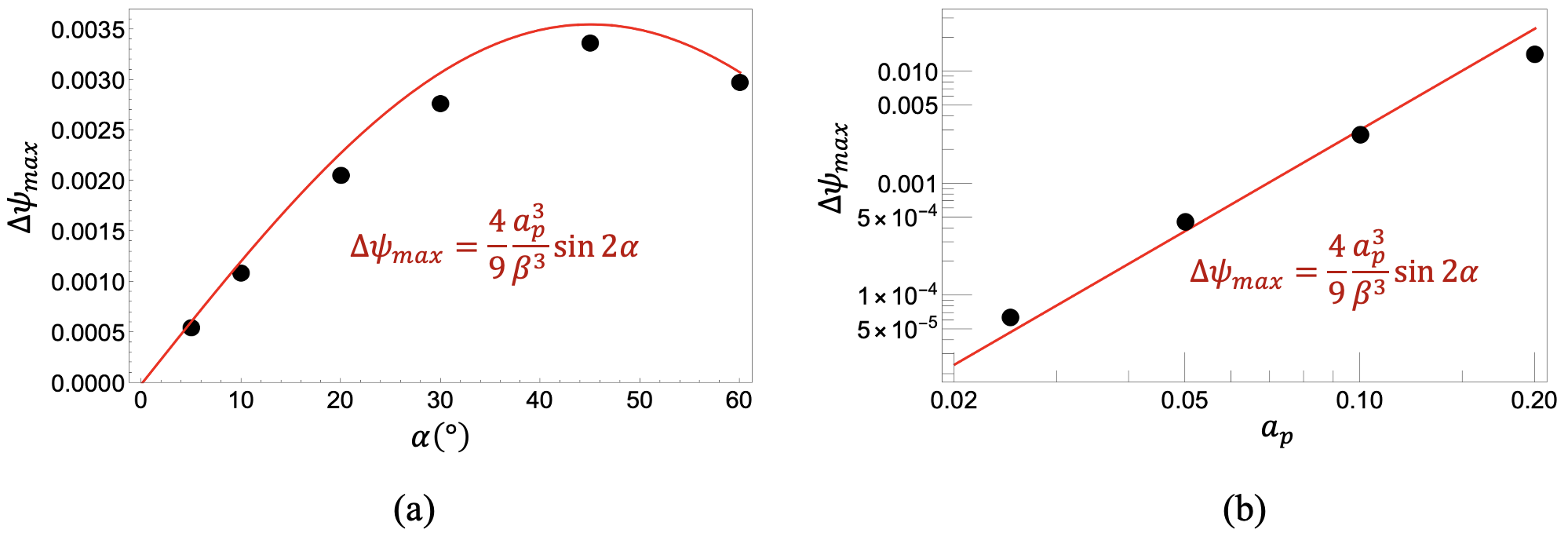}
    \end{subfigure}

  \vspace{1em}

 \begin{subfigure}[t]{0.52\textwidth}
    \centering
    \includegraphics[width=\textwidth]{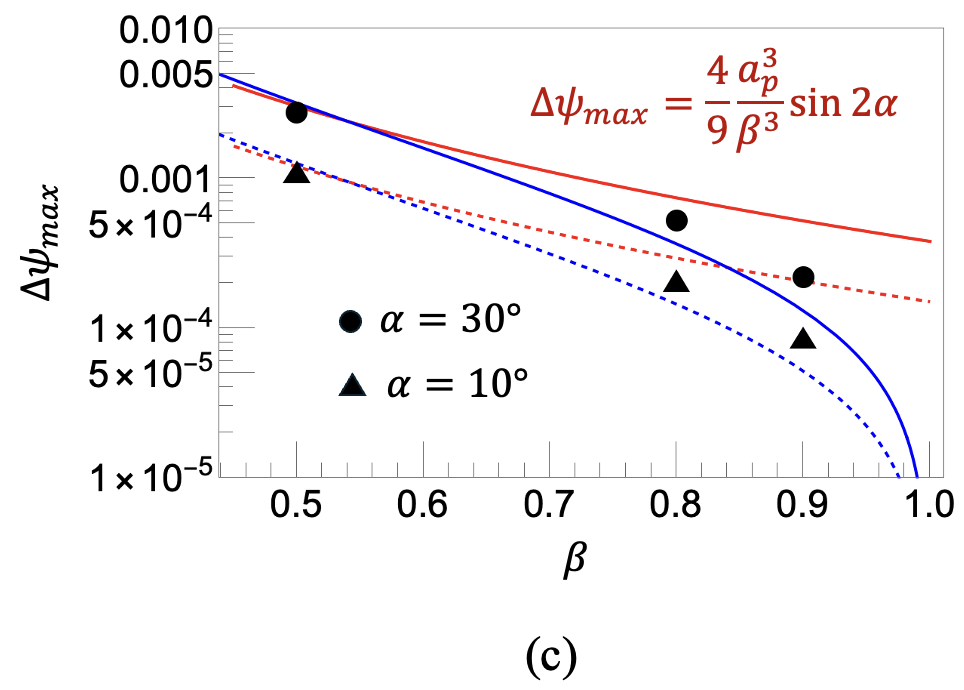}
    \end{subfigure}
  
  \caption{Scaling of $\Delta\psi_{max}$ with (a) flow angle $\alpha$ $(a_p=0.1,\beta=0.5)$, (b) particle size $a_p$ $(\alpha=30degree,\beta=0.5)$, and (c) aspect ratio $\beta$ $(a_p=0.1,\alpha=10degree$ and $30degree$). The red lines are obtained from the analytical scaling theory \eqref{eq scaling psiB} for $\beta\ll 1$. In (c), 
  the small-$\beta$ theory is successful even for $\beta=0.5$, but to capture displacements for near-spherical obstacles ($\beta\lesssim 1$), the complete theory of 
  \eqref{eq DeltaEtaf full} -- \eqref{eq Deltapsimax full in delta1 Deltaetaf} is needed (blue lines). 
  }\label{fig scaling}
\end{figure}

The expressions of $\Delta\eta_f$ and $\delta_1$ as in equations \eqref{eq DeltaEtaf full} and \eqref{eq delta1 full}, respectively, can be evaluated in the limit of  $\beta\ll 1$ to yield
\begin{equation}\label{eq DeltaEtaf}
    \Delta\eta_f=\frac{\beta}{3}\,,
\end{equation}
\begin{equation}\label{eq delta1}
    \delta_1=\frac{2}{3}\frac{a_p}{\beta}\,,
\end{equation}
which after insertion in
\eqref{eq Deltapsimax full in delta1 Deltaetaf}  obtains the analytical limit of $\Delta\psi_{max}$  for small $\beta$ and takes a much simpler form as

\begin{equation}\label{eq scaling psiB}
   \Delta\psi_{max}= \frac{4}{9}\frac{a_p^3}{\beta^3}\sin2\alpha\qquad (\beta\ll 1)\,.
\end{equation}
Remarkably, we find that for our preferred study case $\beta=1/2$ the small-$\beta$  solution \eqref{eq scaling psiB} still reproduces trajectory data very accurately, as demonstrated by the red curves in figure~\ref{fig scaling}.

Figure~\ref{fig scaling}(a), (b) shows that
\eqref{eq scaling psiB} compares very well
with direct trajectory calculations varying flow angle of attack $\alpha$ and particle size $a_p$. The expansions \pkd{in small $\sin{2\alpha}$, $a_p$, and $\beta$} behind the analytical expression maintain good accuracy beyond their strict limits: The leading-order angle dependence $\sin 2\alpha$ appears valid for all angles, and is indeed the expected dependence from the elliptic symmetry of the obstacle geometry. Particle sizes as large as $a_p=0.2$ still yield good agreement. Figure~\ref{fig scaling}(c) does find that approximating displacements around near-circular obstacles requires the full scaling theory \eqref{eq DeltaEtaf full} -- \eqref{eq Deltapsimax full in delta1 Deltaetaf}, but the displacement effect is also dramatically smaller as $\beta\to 1$ and the obstacle fore-aft symmetry vanishes. The surprisingly strong $\beta^{-3}$ dependence in \eqref{eq scaling psiB}, and its accuracy for $\beta=1/2$, suggests 
that employing even slightly more eccentric obstacles could strongly enhance particle deflection. For much smaller $\beta$, however, effects of obstacle surface curvature  on the hydrodynamic wall corrections (equation \eqref{eq rcurv} in section~\ref{appen flat wall}) will likely become important and modify the results. 

\section{Conclusion}\label{sec concl chap 4}

In this chapter, we developed a rigorous analytical framework to understand and quantify particle displacement induced by hydrodynamic wall interactions in steady Stokes flow transporting particle over an elliptic obstacle. Building upon the computational observations of the previous chapter and addressing the challenges posed by numerically demanding near-wall dynamics, the analysis presented here provides a comprehensive understanding of how net particle displacement arises from close particle–obstacle interactions. By modeling particle motion in the near-wall regime with appropriate lubrication approximation, we derived a semi-analytical description that captures the dominant contribution to cross-streamline displacement during close encounters with an obstacle. This approach allowed us to isolate the portion of the trajectory responsible for the maximum displacement and to explicitly identify how wall-induced hydrodynamics govern the net effect.

Somewhat non-intuitively, there is a particular initial condition for given parameters for which the net displacement is maximal. Extending our analysis, we derived fully analytical scaling laws for the net displacement by evaluating the asymptotic behavior of the governing expressions. These scaling results quantify the “displacement capacity” of an obstacle in terms of geometric and flow parameters, and provide explicit predictions for how particle size, obstacle curvature, and flow orientation influence the magnitude of cross-streamline transport.  For the practical case of particles much smaller than the obstacle ($a_p\ll 1$), this initial condition is very close to the separating streamline associated with the obstacle, and the maximum displacement (in terms of changes of stream function) is proportional to $a_p^3$ for eccentric obstacles. The effect is thus small, but also strongly dependent on particle size. A quantitative comparison with the effects of short range roughness modeling shows comparable influence of obstacle symmetry breaking on the ability to separate microparticles by size. Thus, obstacle shape should be taken into account in set-ups pursuing such goals. 

The wall-induced displacement capacity of our obstacle is explicitly independent of Reynolds number-- meaning that this effect is inherently present and becomes significant at small but finite Reynolds number flow in the near wall region. A common practical example of finite Reynolds number effect on particle transport is the inertial migration of microparticles across streamlines in a steady channel flow confined by straight walls which has been extensively studied in the context of steady inertial microfluidics \cite{di2007continuous, martel2014inertial,di2009inertial}. In such steady transport, near-wall particles experience no-slip wall induced inertial lift forces that drive cross-stream motion which is shown to be proportional to $a_p^6$ compared to that of non-inertial wall effect $\mathcal{O}(a_p^3)$ developed here. While this wall confined flow and the associated boundary layer structure is not fully similar to our asymmetric external flow, this scaling behavior suggests that any asymmetry in the channel wall structure (i.e. sudden constriction/expansion, deformability of the channel wall) can induce comparable non-inertial wall effects. Therefore in applications involving small particles and weak inertia, the non-inertial wall interaction described here can significantly influence near-wall particle migration dynamics and should be incorporated alongside steady inertial wall effects.

We find that the hydrodynamic displacement effects can be maximized by choosing a flow angle $\alpha = \pi/4$ and by decreasing the aspect ratio $\beta$ of the ellipse, though corrections from obstacle surface curvature must be taken into account for very eccentric ellipses. These results do not contradict time reversibility: If time is reversed, all hydrodynamic forces also change sign together with the background flow, and particle paths are traced backwards.

We here want to mention that, since our analytical result is obtained from the wall-expansion model only,  figure~\ref{fig deflection}(b) suggests that the trajectories dominated by the wall-expansion model have $\Delta_{min}$ significantly smaller than $0.1$.  Figure~\ref{fig deflection}(b) does reflect a change between two regimes: $\Delta_{min}$ scales differently with $\psi_i$ depending on its magnitude. While we observe this separation of regimes consistently with parameter changes, it results from trajectory computations and we have not found an a priori analytical criterion. 

Our analytical approach as developed in this chapter establish a predictive, analytically tractable description of hydrodynamic particle displacement near an elliptic obstacles in inertial-less flows. By establishing the scaling laws, this work translates the numerical displacement effect into a design principle. The analytical insights developed here form a critical foundation for the subsequent chapters as well, where these ideas are extended to more complex geometries including multiple obstacles, and experimentally realizable configurations relevant to porous media filtration and deterministic lateral displacement devices.

\chapter{Dynamics of particle sticking and deposition}\label{chap 5}

\begin{figure}
    \centering
\includegraphics[width=\textwidth]{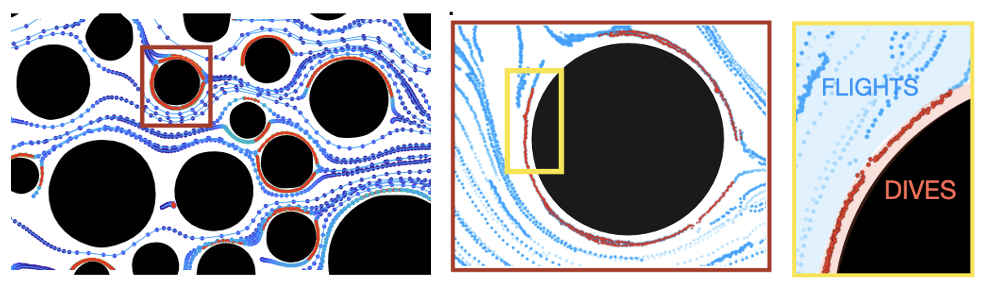}    \caption{Particle tracking images showing particle transport (blue) through porous media. Wall induced hydrodynamic effect enables some particles (in red) to dive towards obstacle that can lead to arrest the particles at obstacle restraining their transport through the media \cite{miele2025flow}.}\label{fig pmf}
\end{figure}

Thus far, we have demonstrated a systematically tunable particle transport manipulation by hydrodynamic interaction in Stokes flow breaking the fore-aft symmetry. Our goal was to unveil the fundamentals of particle-obstacle interaction in an inertialess flow system in maneuvering particles, particularly whether the particles can be guided to a specific direction in transport around an obstacle.  In porous media transport and filtration microfluidics, this type of guided motion towards obstacle leads a particle in a `dive" trajectory (in red as in figure~\ref{fig pmf}) and can restrain their transport through the media by arresting it at the obstacle (see figure~\ref{fig DLDPMF}(c) in chapter~\ref{chap intro}). 
In this chapter\footnote[1]{This chapter is mostly adapted from Das et al. \cite{das2025controlled} and Miele et  al. \cite{miele2025flow}}, we aim to extend our framework to investigate where hydrodynamic transport brings particles sufficiently close to obstacle resulting in localized capture, sticking, and attachment at specific locations on the obstacle surface. This is a qualitatively new type of particle manipulation which is the central of particle filtration applications in porous media where particle mobility is restraining by the effect of short-range inter-molecular attraction. 

Addressing the importance of hydrodynamic force in pushing particles through the electrostatic repulsion barrier between particle and obstacle wall, we quantify the individual hydrodynamic force components acting on a force-free particle that is being transported over an elliptic obstacle, as obtained in the preceding chapters. We show that, such obstacle curvature induced flow geometry substantially amplify these force components compared to a symmetric circular obstacle that can effectively overcome the electrostatic barrier. From this finding, we conceptualize the topology in flow structure, arising naturally from the pore irregularity and the medium heterogeneity in a practical porous structure, can promote a favorable interplay between hydrodynamics and DLVO forces, thereby facilitating particle capture and deposition.

\section{Introduction}

A passive microparticle transport system needs to determine how far particles can travel around obstacles/pores without being captured by the obstacle. Attachment and deposition of transported microparticles to obstacles/collectors is a fundamental microfluidic phenomena in a wide range of separation and filtration processes including colloid transport and aggregation through porous media like soils, sediments, and aquifers, carrying pollutants migrate through fractured rocks and soils \cite{grolimund2005colloid,bradford2013transport,pradel2024role}. Particle transport and deposition is also central to industrial filtration processes, including automotive and chemical applications, where aerosol particles and pollutant-laden solvents are captured and retained within or at the entrance of porous media \cite{tcharkhtchi2021overview}. The key phenomenon governing these system performance is the transition from particle transport to particle attachment and deposition on solid surfaces, which ultimately controls filtration efficiency, clogging dynamics, and long-term permeability. Deposition dynamics plays crucial role in designing microfluidic strategies that either avoid or favor blockage and clogging an obstacle/pore by a single passive object (particle, cell, microbes, bubbles and droplets) \cite{dersoir2015clogging}.

Porous media microfluidics enables direct visualization of particle trajectories, deposition events, and clogging at the pore scales. Bacchin et  al. \cite{bacchin2011colloidal} emphasize that particle deposition is governed by a balance between hydrodynamic forces and potential barrier between the particle-wall surfaces, and that this balance must be examined at the scale of individual pore to achieve predictive understanding. They demonstrated that deposition often initiates at specific locations—typically pore entrances and regions of strong confinement. Dersoir et al. \cite{dersoir2015clogging} investigated colloidal attachment of a single pore to determine the primary feature of clog formation within porous constrictions under inertialess flow conditions. By varying the hydrodynamic conditions at a pore scale, their experiments establish clogging as a hydrodynamically mediated process where flow structure determines where particles are guided to adhere to the wall.

Pradel et  al. \cite{pradel2024role} studied with microplastic attachment to a collector surface and eventual clogging dynamics that showed that particle capture to obstacle is highly localized, often occurring in converging flow regions, pore entrances, and zones of strong flow velocity gradients. Their experiments demonstrated that even force-free particles in inertialess flows can be driven toward solid boundaries through purely hydrodynamic interactions, particularly in geometries that generate wall-normal velocity components or strong shear gradients. Particle deposition is frequently initiated with the capture of a micrometric particle by physical interception between particle-obstacle wall, a mechanism that pushes particle to penetrate the repulsion barrier. While this first microparticle capture on a specific location of the obstacle surface subsequently act as secondary collectors and leads to accelerated aggregation of smaller particles and the growth of porous deposits (ripening), the underlying hydrodynamics of particle-wall interaction responsible for the initial localized attachment remains unresolved.

Motivated by these observations, this chapter focuses on the role of hydrodynamic transport in enabling localized particle sticking and attachment to an obstacle. Building on the hydrodynamic framework developed in the preceding chapters, we examine how flow structure induced by an asymmetric obstacle maneuver particles into close proximity to the obstacle surfaces, where short-range attractive forces can dominate. To understand the physics of macroscopic filtration by a heterogeneous porous system (producing a flow asymmetry though a different kind than that of a single inclined elliptic obstacle) we focus on individual ``dive" trajectories that bring particles into very close to solid pore/obstacle surface $\Delta\ll1$. We explicitly quantify how flow asymmetry amplifies wall-directed hydrodynamic forces, allowing particles to overcome electrostatic repulsion barriers.

\section{Closest approach and sticking}

Our formalism, as developed in the preceding chapters, finds the strongest net deflections of particles when particles follow the obstacle surface very closely, i.e., $\Delta\ll 1$ along a significant part of the trajectory (``dives", \cite{miele2025flow}). Necessarily, the minimum gap $\Delta_{min}$ on such trajectories becomes very small, and with typical microfluidic scales of obstacle and particle sizes, $\Delta_{min}$ can easily translate to sub-micrometer distances. In the presence of short-range attractive interactions between the surfaces (London forces, van der Waals attraction, etc.) this can lead to sticking (capture) of the particles, an effect important in fouling, clogging, and cleaning \citep{gul2021fouling,bacchin2011colloidal,kumar2015fouling,dersoir2015clogging}, porous-media filtration \citep{miele2025flow,mays2005hydrodynamic,pradel2024role}, or elimination of pathogens \citep{nuritdinov2025experimental,uttam2023hypothetical,sande2020new}.   
Particles used for size-based sorting, selecting, and trapping applications are often a few $\mu$m in size, so that a typical $\Delta_{min}=O(10^{-3})$ corresponds to a few nanometer gap, easily close enough for short-range attractions to be important. 



\begin{figure}[t]
    \centering
\includegraphics[width=\textwidth]{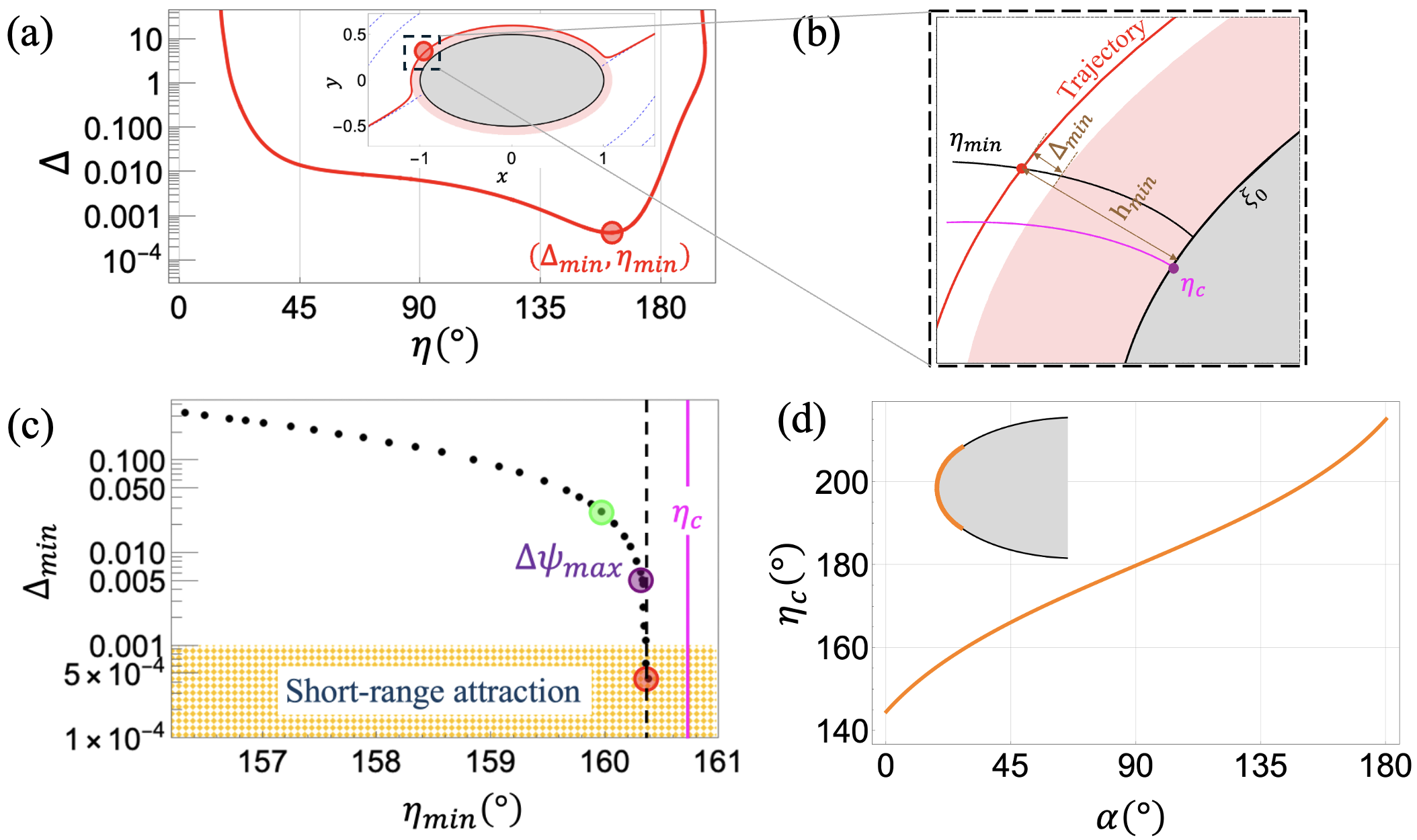}    \caption{(a) Gap size $\Delta$ as a function of $\eta$ along a ``dive" trajectory creeping around the obstacle (inset). 
(b) Close-up sketch near $\eta_c$ demonstrating that the particle center angular coordinate at closest approach $\eta_{min}$ cannot coincide with $\eta_c$ if $a_p>0$.  
(c) Closest approach coordinates $(\Delta_{min}, \eta_{min})$ for different trajectories in the $\psi_i\rightarrow0$ limit. Dashed line represents $\eta_{min}(\Delta_{min}=0)$.  While $\Delta_{min}$ varies by orders of magnitude, $\eta_{min}$ stays close to the value $\eta_c$ where flow curvature $\kappa$ vanishes.
$\Delta_{min}=O(10^{-3})$ represents a few nanometer of particle to surface wall gap for a typical microparticle. Short-range inter-molecular forces activate in this range and restrict particle movement. 
(d) Variation of $\eta_c$ with $\alpha$ for $\beta=1/2$, showing that particle sticking is always most likely near the major-axis tip of the ellipse.}\label{fig min approach}
\end{figure}

Figure~\ref{fig min approach}(a) plots $\Delta(\eta)$ along a dive trajectory, showing a well-defined closest approach point $(\Delta_{min},\eta_{min})$. Changing $\psi_i$, figure~\ref{fig min approach}(b) shows that $\Delta_{min}$ is very sensitive to initial conditions, but $\eta_{min}$ is not. In fact, it follows from the wall-expansion approximation \eqref{eq wall xpnsn} that the closest approach  is always determined by the point on the wall where $\kappa(\eta_c)=0$, a function of background flow only. However, for finite particle size the angular position $\eta_{min}$ of the particle where the dynamics is governed by $\eta_c$ is slightly different, as illustrated in figure~\ref{fig min approach}(c). It is easy to show that
\begin{equation}\label{eq etamin etac connection}
    \eta_c-\eta_{min}\sim a_p^2    
\end{equation}
to leading order, so that even as $\psi_i\to 0$, a finite difference between $\eta_{min}$ and $\eta_c$ remains. For $a_p\ll 1$, practically relevant trajectories nevertheless have a very well-defined point of closest approach, to within a few degrees of $\eta_c$ (figure~\ref{fig min approach}(b)), and thus a well-defined location where sticking is most likely follows directly from the background flow wall curvature $\kappa$. The angular position $\eta_c$ changes with the flow angle of attack $\alpha$ but not very widely. Figure~\ref{fig min approach}(d) demonstrates that this point is located near the major-axis pole of the elliptic cylinder for all $\alpha$. This can be obtained by from solving $\kappa(\eta=\eta_c)=0$, $\kappa(\eta)$ being the background flow curvature at the wall point as defined in chapter~\ref{chap 2} and takes the following analytic form as a function of the flow parameters $\alpha\,,$and $\beta$ based on the analytical formulation as presented in chapter~\ref{chap 4}
\begin{multline}\label{eq kappa}
     \kappa(\eta)=-\frac{\left(\beta ^2-1\right) \cos 2 \eta+\beta ^2+1}{2 \left(\beta ^2-\left(\beta ^2-1\right) \sin ^2\eta\right)^{7/2}}\\ \left(\cos \alpha \cos \eta \left(\beta ^2 (\cos
   2 \eta-3)+2 \sin ^2\eta\right)+\beta  \sin \alpha \sin \eta \left(\left(\beta ^2-1\right) \cos 2
   \eta+\beta ^2-3\right)\right)\,.
\end{multline}

In the case of internal Stokes flow confined by flat walls \citep{liu2025principles}, the approach to minimum gap distance is well described by a single exponential, as predicted by theory \citep{brady1988stokesian,claeys1989lubrication,claeys1993suspensions}. In the present case, the strong  variation in $\kappa(\eta)$ with obstacle topography precludes such a simple behavior. The direct comparison between our full (variable expansion) formalism and a pure wall expansion calculation in the inset of figure~\ref{fig trajectory}(c) confirms, however, that the decrease of $\Delta$ over orders of magnitude in proximity to $\eta_c$ is governed by
$\kappa$. The rapid approach in the presence of a variety of short-range forces will reliably lead to sticking at this predefined location. 

Pradel et  al. \cite{pradel2024role} in their latest research on particle accumulation in transport flow through porous media (similar to the asymmetric set-up of obstacles in a DLD device), emphasize the role of hydrodynamics to arrest a microparticle on the surface of a pore but did not establish any concrete mechanism on defining specific location of single particle capture. In this context, our mechanism offers valuable strategies for microfluidic filtration and particle capture in transport flow through porous media \cite{spielman1977particle,miele2025flow,pradel2024role,bacchin2011colloidal}.

\section{Hydrodynamic forces on inertialess particles transport in symmetric and asymmetric flows}

\begin{figure}[t]
    \centering
\includegraphics[width=0.65\textwidth]{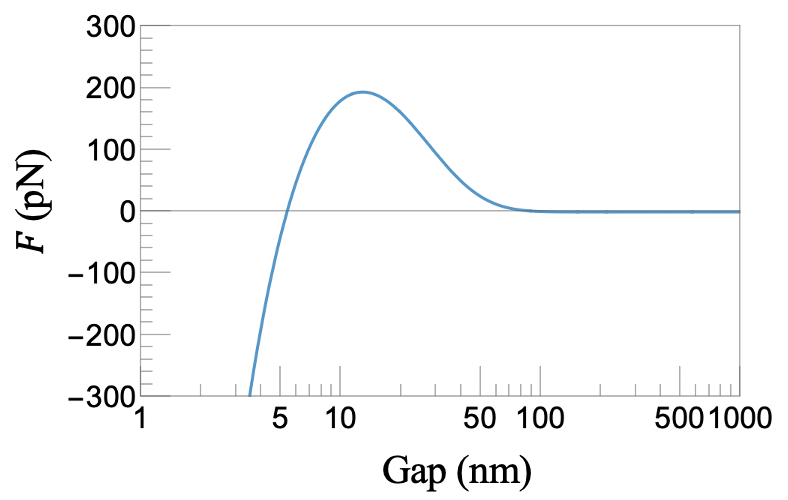}    \caption{DLVO force distribution as a function of gap size between a PS particle (radius $0.5 \mu m$) and plane PDMS substrate based on equations \eqref{eq PhiDLVO} to \eqref{eq PhiVDW}. All the parameters are evaluated for PS particle and PDMS substrate as taken from \cite{kirby2004zeta,yu_evaporative_2017,ruckenstein_adsorption_1976}}\label{fig DLVO}
\end{figure}

When a particle falls into a dive-trajectory (cf section~\ref{subsec trajectory} of chapter~\ref{chap 3}), it experiences a strong wall from the solid surface of the obstacle as well as the presence of Van der Waals and electrostatic interactions. When particles are suspended in pure water (or with low ionic strength I.S. $\approx0.5 \,mM$) a strong repulsive electrostatic force emerges around each obstacle, raising the energy needed for the particle to approach the grain (obstacle) in the region of attractive Van der Waals force. Under these  so-called ``unfavorable conditions", a spontaneous attachment by pure diffusion only is not possible since the energy associated to Brownian motion is of the order of $k_BT$, while the repulsive energy barrier computed from the measured zeta potentials is of the order of several hundreds of $k_BT$.  Miele et  al. \cite{miele2025flow} measured the suspension zeta potential to be about 66 mV, corresponding to an energy barrier of about $300 k_BT$. The associated interaction energy profile for the system PDMS-water-colloids is computed following the DLVO theory \cite{ruckenstein_adsorption_1976, Hogg1966,Elimbook}, with $H_p$ ,$H_w$ and $H_{PDMS}$ as Hamaker constant for latex particles, water and PDMS, respectively (\cite{ruckenstein_adsorption_1976, Hogg1966,yu_evaporative_2017}, that correspond to a repulsive electrostatic force peaked at $\simeq 250$~pN on the order of sub-nanometer gaps . The variation of DLVO force with the gap, $h$ between a polystyrene particle of $a_p=0.5\mu m$ radius and a plane PDMS substrate is drawn in figure~\ref{fig DLVO} that has been derived from the following expressions of double layer potential $\Phi_{DL}$ and van der Waals potential $\Phi_{VDW}$ energies \cite{chrysikopoulos2017cotransport,hogg1994inertial},
\begin{equation} \label{eq PhiDLVO}
    \Phi_{DLVO}=\Phi_{DL}+\Phi_{VDW}
\end{equation}
\begin{equation}\label{eq PhiDL}
    \Phi_{DL}=\pi  \epsilon_0 \epsilon_r a_p \left(\left(\Psi_p^2+\Psi_s^2\right) \log \left(1-e^{-2
   h k}\right)+2 \Psi_p \Psi_s \log \left(\frac{e^{-h k}+1}{1-e^{-h k}}\right)\right)
\end{equation}
\begin{equation}\label{eq PhiVDW}
    \Phi_{VDW}=-\frac{A_{123} \, a_p}{6 h \left(\frac{14 h}{\lambda }+1\right)}
\end{equation}
where the constants are taken from \cite{kirby2004zeta,yu_evaporative_2017,ruckenstein_adsorption_1976} corresponding to PS particle and PDMS substrate. Since this magnitude of repulsive interaction corresponds to an energy barrier on the order of hundreds of $k_{\!B}T$, a purely thermal diffusion (on the order of $k_{\!B}T$) are insufficient to cause attachment. For particles to overcome these repulsive barriers during a dive trajectory, an additional driving mechanism must be present. In the absence of any other external influences, that mechanism is provided by hydrodynamic forces arising directly from the flow field induced by the obstacle.

To understand the contribution of hydrodynamic force arising from the obstacle induced flow in pushing the particles towards the wall against the repulsive electrostatic force, we examine the wall-normal force components on the force-free particle that we have modeled in section~\ref{subsection: wall normal particle} in chapter~\ref{chap 2}. From equation~\eqref{Flambda} we can decompose the individual wall-normal force components as
\begin{equation}\label{eq FA}
    F_\mathcal{A}=-6\pi\mu a_p\,\mathcal{A}\{(\boldsymbol{v}_p-\boldsymbol{u})\cdot\boldsymbol{e_{\perp}}\}\,
\end{equation}
\begin{equation}\label{eq FB}
    F_\mathcal{B}=-6\pi\mu \frac{a_p^2}{a}\,\mathcal{B}\{(\boldsymbol{e_{\perp}}\cdot \nabla\boldsymbol{u})\cdot\boldsymbol{e_{\perp}}\}\,
\end{equation}
\begin{equation}\label{eq FC}
    F_\mathcal{C}=3\pi\mu \frac{a_p^3}{a^2}\,\mathcal{C}\{(\boldsymbol{e_{\perp}}\boldsymbol{e_{\perp}}:\nabla\nabla\boldsymbol{u})\cdot\boldsymbol{e_{\perp}}\}\,
\end{equation}
\begin{equation}\label{eq FD}
    F_\mathcal{D}=3\pi\mu \frac{a_p^3}{a^2}\,\mathcal{D}\{\nabla^2\boldsymbol{u}\cdot\boldsymbol{e_{\perp}}\}\,
\end{equation}
The analysis of these forces elucidates how the hydrodynamic particle-wall interactions compete with external DLVO forces ($\sim 250$ pN) at close proximity ($\Delta=\mathcal{O}(10^{-3})$, cf figure~\ref{fig min approach}) and ultimately enable particle attachment despite the presence of repulsive barriers.

We determine the forces acting on a particle with radius $a_p=0.55 \mu m$ in a representative ``dive" trajectory that has a $\Delta_{min}\ll10^{-3}$ falling in the region suggested by figure~\ref{fig min approach}(c) for favorable attachment. We first obtain the trajectory by evaluating the particle velocity $\boldsymbol{v_p}$ in a force-free condition as we modeled in the preceding chapters. We compute and compare the distribution and the magnitude of the forces for (i) a symmetric flow situation (particle being transported around a circular obstacle) and (ii) a fore-aft asymmetric flow situation (particle being transported around an inclined elliptic obstacle)— the latter being the geometry used earlier to induce net spatial particle displacement via hydrodynamic wall effects. We here aim to compare the hydrodynamic force derived from our  symmetry broken flow model from chapter~\ref{chap 3} with realistic microfluidic scenario with a finite Reynolds number $Re$ and therefore we quantify our flow field and all derived parameters from it (e.g.\ $\partial_\perp u_\perp$) after multiplying the background streamfunction $\psi_B$ in equation~\eqref{eq psiB} with $(-\ln{Re})$.

\begin{figure}
    \centering
\includegraphics[width=0.75\textwidth]{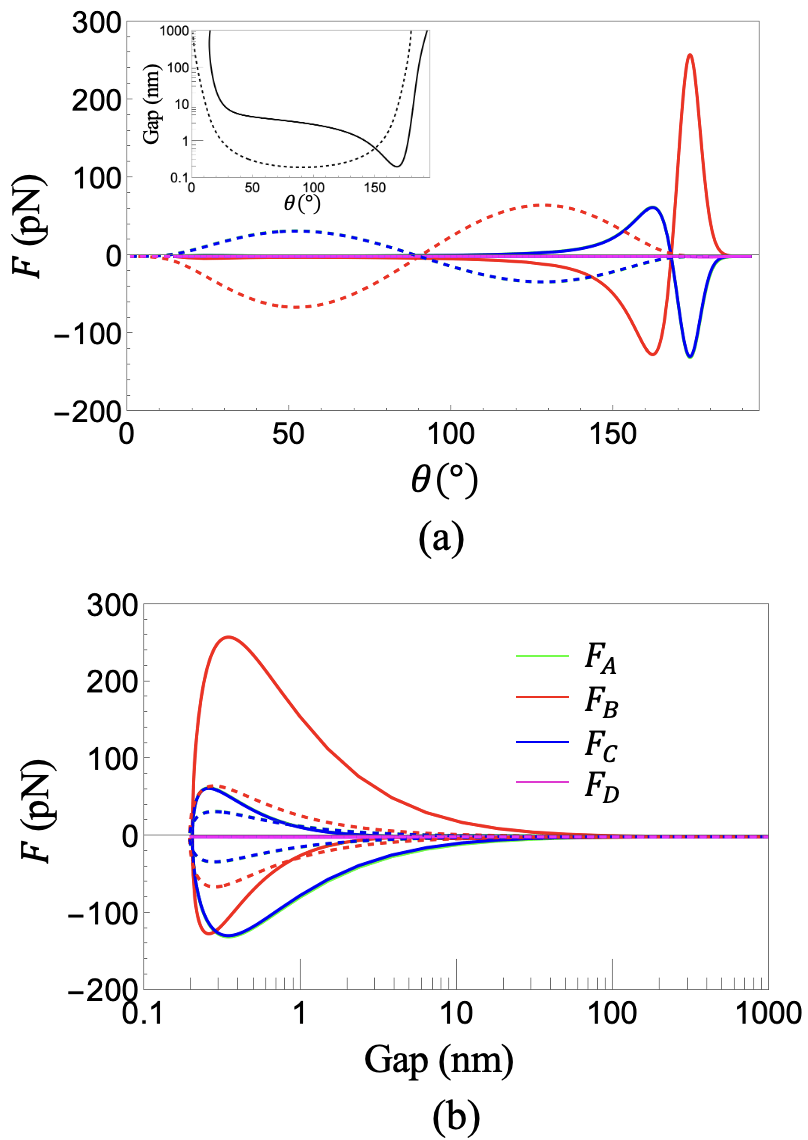}    \caption{(a) Wall-induced wall-normal hydrodynamic forces acting on a ``dive" particle (radius $a_p=0.55 \mu m$) trajectory along its instantaneous angular positions around an obstacle. Inset shows the change in physical gap between particle surface to obstacle surface at different angular positions of the particle. (b) The force variations at instantaneous physical gaps between particle surface to obstacle surface. In both figures dashed line represent a symmetry flow situation (an uniform flow with reference background velocity $U_{ref}$ around a circular obstacle of $5.5 \mu m$ radius) and solid represents an asymmetric flow with the same $U_{ref}$ around a $30^{\circ}$ inclined elliptic obstacle ($a=10a_p=5.5 \mu m$ major radius with aspect ratio $\beta=1/2$). The force component $F_\mathcal{D}$ arising from the Laplacian of the flow field are both negligibly small compared to the other components. The viscosity is taken a for water $\mu=0.001 Pa.s$. We use $U_{ref}=1200 \mu m/s$ corresponding to a Reynolds number $Re=10^{-2}$ with respect to the obstacle size. All the numbers are relevant to our experimental work in \cite{miele2025flow}.}\label{fig hydro force}
\end{figure}

Figure~\ref{fig hydro force} illustrates the distribution of the wall-normal hydrodynamic forces acting on the particle along its instantaneous angular positions around the obstacle and their variation as a function of the instantaneous physical gap between the particle and the obstacle surface. We remarkably find that our asymmetry model amplifies the wall induced hydrodynamic forces particularly those that come from wall induced Stokes drag $(F_{\mathcal{A}})$, the extensional gradients of the background flow $(F_{\mathcal{B}})$, and the curvature of the background flow $(F_{\mathcal{C}})$. For example, in the dive trajectory shown in figure~\ref{fig hydro force}, where the minimum approach is approximately $~0.2\, nm$, the flow around the stretched boundary of the ellipse generates hydrodynamic force magnitudes that are very comparable to the magnitude of the DLVO force barrier depicted in figure \ref{fig DLVO}. These results underscore the critical role of topology Stokes transport flows (porous media) for enabling localized particle attachment to obstacles. The amplification of wall-normal hydrodynamic force components facilitates the motion of particles into the near-surface region, allowing them to overcome repulsive barriers that would otherwise preclude attachment under a single symmetric circular obstacle induced flow conditions.

\begin{figure}
    \centering
\includegraphics[width=\textwidth]{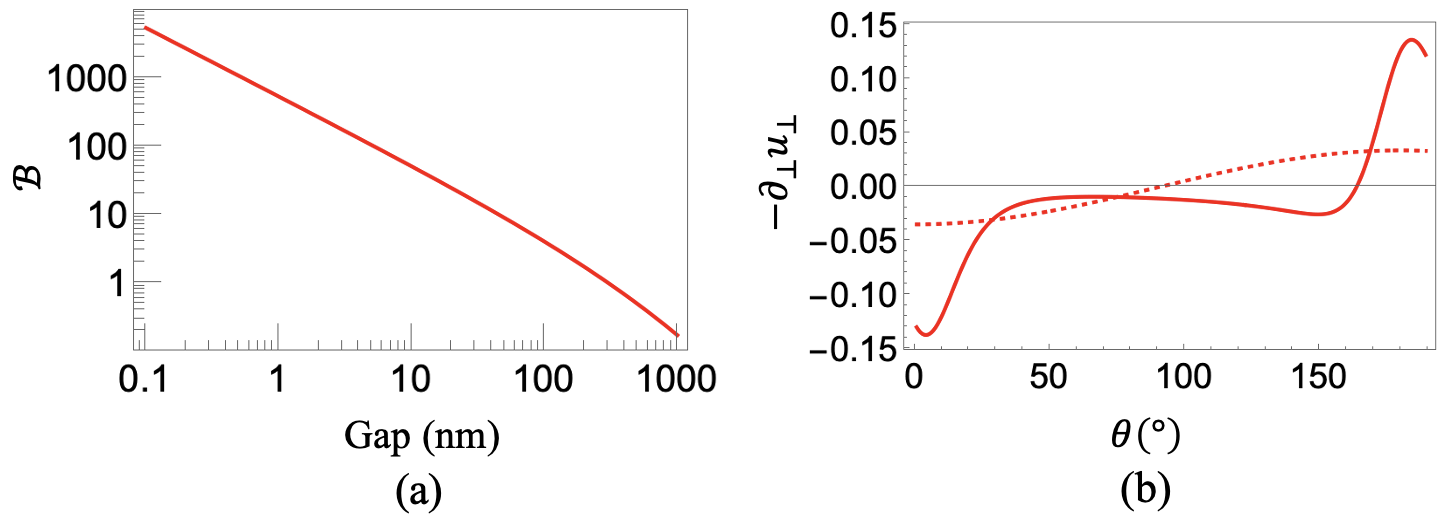}    \caption{(a) Resistance coefficients $\mathcal{B}$ as a function of gap \cite{rallabandi2017analysis}. (b) Normal gradient of the normal component of the background flow along $a_p$ distance from the obstacle $(\Delta=0)$. Here $a_p=0.55\mu m$ and Reynolds number $Re=10^{-2}$. Dashed line represents symmetric circular obstacle of $5.5 \mu m$ radius while solid line is for the flow induced by $30^{\circ}$ inclined elliptic obstacle of $5.5 \mu m$ major radius and $\beta=1/2$ aspect ratio.}\label{fig hydro force gradients}
\end{figure}

Equation \eqref{eq FB} expresses the functional dependence of the dominant hydrodynamic force $F_{\mathcal{B}}$ on the angular position $\theta$ and gap $\Delta$ through $\partial_{\perp}u_{\perp}$ and the wall resistance coefficients $\mathcal{B}=\mathcal{B}(\Delta)$ that changes monotonically only with $\Delta$ as can be inferred from Rallabandi et al. \cite{rallabandi2017hydrodynamic} and as depicted in figure~\ref{fig hydro force gradients}(a). To understand the amplification factor of $F_\mathcal{B}$ because of wall curvature induced flow stretching close to the wall, we thus need to quantify the normal gradient of the normal background flow $(\partial_\perp u_\perp)$ as a function of $\theta$ at $\Delta=0$: a hypothetical path outlining the obstacles curvature from $a_p$ distance away. As can be seen from figure~\ref{fig hydro force gradients}(b), in the fore–aft symmetric configuration around a circular obstacle, the distribution of $\partial_{\!\perp}u_{\perp}$ with respect to $\theta$ is itself symmetric. However, in the fore–aft asymmetric configuration, the profile of $\partial_{\!\perp}u_{\perp}$ becomes skewed, with peaks that are shifted and amplified by nearly the same factor as the corresponding peak in $F_{\mathcal{B}}$ shown in figure~\ref{fig hydro force} relative to the circular case. This confirms that the amplification of the wall-normal hydrodynamic force is predominantly driven by the tangential stretching of the background flow near the obstacle (on particle size distance scale) as a consequence of the geometric eccentricity of the elliptic boundary.

\section{Conclusion}

In this chapter, we extended the hydrodynamic framework developed in the preceding chapters to address particle sticking and attachment in Stokes transport flows. While earlier chapters established that fore–aft asymmetry in an inclined elliptic obstacle geometry produces deterministic lateral displacement of force-free particles, the present analysis demonstrates that the inherent tangential stretching of the flow from the elongated curvature of the elliptic wall can amplify hydrodynamic forces sufficiently to compete with short-range electrostatic repulsion and enable particle attachment. We first identified representative dive trajectories that fall in the relevant physical regime for attachment, characterized by extremely small minimum particle–wall separations in nanometer range. We further analyzed the dominant forces normal to the obstacle in this near-wall region and find that the leading component of hydrodynamic wall-induced force on the particle stems from extensional background flow gradients arising from the tangential stretching of fluid element along obstacle topology. We demonstrated that transport flows over obstacles generate wall-normal hydrodynamic forces that are comparable in magnitude to DLVO repulsive forces for representative dive trajectories and, together with attractive Van der Waals interactions, make attachment possible.

We have demonstrated the effect of a particular tangential stretching in the Stokes flow: by deforming a circular obstacle into an ellipse and found that this type topology can amplify the required hydrodynamic force towards the obstacle for attachment. We derive the amplification effect from the enhanced flow stretching on the scale of small particle size distances from the ellipse. This geometry amplifies the wall-normal component of the hydrodynamic force experienced by a nearby particle, increasing the likelihood that the particle will be driven into close contact with the surface. It is important to emphasize that the type of topological enhancement utilized here does not fully replicate the complexity of flow asymmetry in real porous media, where irregular obstacle orientation, multiple pores, and random confinement produce a more heterogeneous flow asymmetry at larger pore sized length scales. While a systematic quantification of the effect of this flow asymmetry induced by multiple obstacles remains necessary to fully understand porous media filtration applications, the present results confirm the essential role of flow topology in generating a hydrodynamic ``push" towards obstacle for physical interception. Therefore, our work opens new pathway for designing inertialess microfluidic systems in which particle capture, deposition, and accumulation can be tuned through controlled manipulation of flow structure.

Finally we mention here that besides our theoretical work on the hydrodynamics of localized particle sticking to an obstacle boundary in a pure Stokes transport flow which is realized in porous media transport experiments, we experimentally obtained the evidence of particle sticking and deposition to the obstacle wall induced by corner streaming flow which is a shear layer driven Stokes flow. We separately present this finding in appendix~\ref{appen sticking from streaming}. Therefore in Stokes flow, whether it is generated purely as the primary flow (e.g.\ by pressure gradient) or is driven by a boundary layer as an second order effect, our work builds the fundamental understanding of the hydrodynamics that plays the chief role in microfluidic systems for particle capture and deposition.

\chapter{Inertialess hydrodynamics induced by multi-obstacle asymmetry}\label{chap 6}

We have identified geometric heterogeneity and asymmetry as particularly important elements for strong particle-obstacle interaction. Real porous media and microfluidic filters are inherently irregular, with variations in pore size, obstacle curvature, and local flow orientation that break idealized symmetries \cite{miele2025flow,bacchin2011colloidal,mays2005hydrodynamic,pradel2024role}. At the same time, while the present work puts the elementary hydrodynamic effects behind Deterministic Lateral Displacement on a firm footing, the practical DLD set-ups consist of many circular cylinders (pillars) arranged in an array under an inclination angle to the flow direction.

The preceding chapters establish the hydrodynamic role of a particular Stokes flow geometry around a single symmetry-broken obstacle in (1) producing a net deterministic particle displacement of transport particles by breaking the inherent flow symmetry (chapter~\ref{chap 3} and \ref{chap 4}) and (2) generating significantly large hydrodynamic forces to induce localized attachment of particles to the obstacle restraining their transport (chapter~\ref{chap 5}). While this single-obstacle framework isolates the physical mechanism of particle transport and attachment, many practical microfluidic environments (e.g.\ DLD devices, porous media flows) involve multiple nearby obstacles whose collective interaction modifies the background flow and the asymmetry at different scales. 

\begin{figure}  
    \centering
\includegraphics[width=\textwidth]{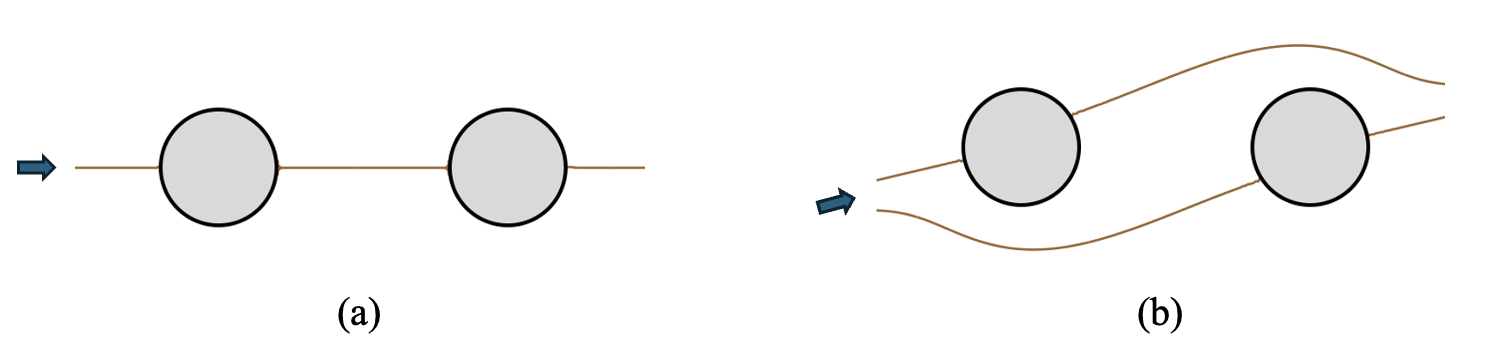}    \caption{Sketches of separating streamlines (in solid brown) originating from the surfaces of two tandem circular cylinders placed in an inclined flow.  Without inclination (a), the separating streamlines merge to a single straight line along the central symmetry axis that separates the flow above and below both cylinders. Crossing the separating streamlines by a passive particle is never possible. With non-trivial inclination with respect to the flow (b), the streamlines separate out and produces a fore-aft asymmetry making particle crossing a possibility because of the simultaneous wall effects from the two obstacles.
}  \label{fig two_cyl}
\end{figure}

A typical DLD mechanism is based on repeated encounters with multiple obstacles. The obstacles are arrayed with an inclination to the flow that effectively breaks  symmetry and particularly shifts the position of separating streamlines between two adjacent obstacles. This is schematically shown in figure~\ref{fig two_cyl}. Breaking this fore-aft symmetry of the streamline structure around two adjacent inclined circular obstacles, transported particles experience a net displacement effect than can even lead to crossing the multiple separating streamlines which is never a possibility for a single obstacle (as described in chapter~\ref{chap 3} and \ref{chap 4}). Therefore, one could approximate the multi-cylinder DLD effects on the flow of groups (e.g. two) of these circular cylinders from the same geometric motivation for symmetry breaking as we have already established for single deformed circular cylinder in the preceding chapters.

In this chapter, we investigate Stokes transport flow around two tandem circular obstacles. The central hypothesis is that symmetry breaking can emerge not only from morphing obstacle shape (i.e.\ elliptic obstacle) but also from obstacle arrangements. Even when each obstacle is individually symmetric, the pair configuration combined with an inclined flow generates asymmetric flow structures that significantly modify particle motion in the flow.

\section{Asymmetry in flow structure around two tandem circular obstacles}

\begin{figure}  
    \centering
\includegraphics[width=\textwidth]{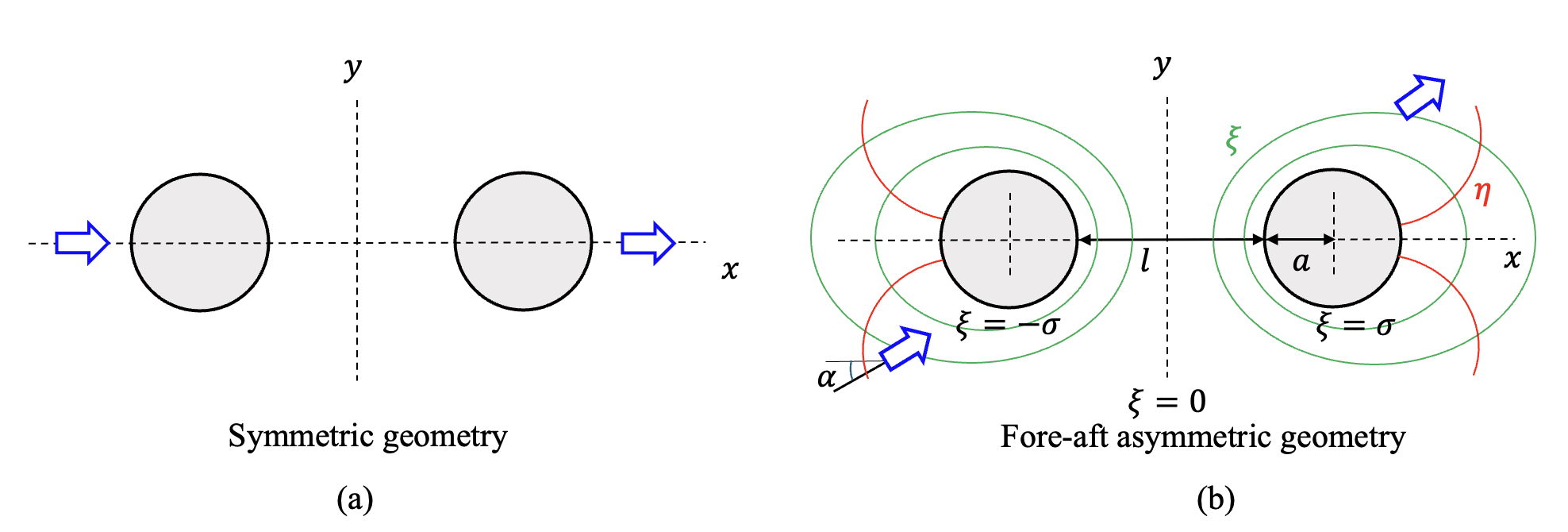}    \caption{Sketches of the principle of fore-aft symmetry breaking in a Stokes flow over two circular obstacles in tandem. (a) shows fore-aft symmetric obstacle-flow geometries by the cylinder pair met head on by the flow. (b) shows the breaking of the fore-aft symmetry flow has non-trivial inclination angle $\alpha$ between the flow direction and the cylinder centerline. Also sketched is the bipolar coordinate system $(\xi,\eta)$ used to determine the flow field as in equations \eqref{eq psiB two cyl} where the obstacle surfaces are represented by $\xi=\pm\sigma=\pm\operatorname{arcosh}\left(\frac{l+2a}{2a}\right)$ and $\xi=0$ represents the $y-$axis. $\eta$ is the angular coordinate in this system.
}  \label{fig setup two cylinders}
\end{figure}

We consider the canonical configuration of two identical circular cylinders in tandem with radius $a$ separated by a finite surface-to-surface gap $l$ and immersed in a background Stokes flow $\psi_{B2}$ with an inclination angle $\alpha$ with respect to the center connector of the cylinders. Similar to the symmetry breaking mechanism  described in section~\ref{sec flow field} for a single obstacle, figure~\ref{fig setup two cylinders} shows how
a non-trivial inclination angle $\alpha$ between the external flow and the axis connecting the cylinder centers breaks the fore–aft symmetry. The corresponding Stokes flow around two cylinders with same size (radius $a$) is known analytically from a derivation in the bipolar coordinate system $(\xi,\eta)$ by Umemura \cite{umemura1982matched} in the form of an infinite series. The stream function is 
\begin{equation}\label{eq psiB two cyl}
    \psi_{B,2}=c \frac{\Phi}{\cosh{\xi}-\cos{\eta}}\,,
\end{equation}
where
\begin{equation}\label{eq Phi}
\begin{aligned}
\Phi
&=
P\Big[
\sinh \xi \ln(\cosh \xi - \cos \eta)
+ \hat{A}\xi(\cosh \xi - \cos \eta)
+ \hat{B}\sinh \xi
+ \sum_{n=1}^{\infty}
\{\hat{b}_n \sinh (n+1)\xi + \hat{d}_n \sinh (n-1)\xi\}\cos n\eta
\Big] \\
&\quad
-
Q\Big[
\sin \eta \ln(\cosh \xi - \cos \eta)
+
\sum_{n=1}^{\infty}
\{\hat{a}'_n \cosh (n+1)\xi + \hat{c}'_n \cosh (n-1)\xi\}\sin n\eta
\Big].
\end{aligned}
\end{equation}
The bipolar coordinates are defined as $x=\frac{c\sinh{\xi}}{\cosh{\xi}-\cos{\eta}}$ and $y=\frac{c\sin{\eta}}{\cosh{\xi}-\cos{\eta}}$ with respect to the cartesian coordinates $(x,y)$. The obstacle surfaces are represented by $\xi=\pm\sigma=\pm\operatorname{arcosh}\left(\frac{l+2a}{2a}\right)$ as sketched in Fig.~\ref{fig setup two cylinders}(b), $\xi=0$ at $x=0$ ($y$-axis), and $c=a\sinh{\sigma}$. The prefactors in equation~\eqref{eq Phi} are given in \cite{umemura1982matched} while the multiplying factors $P=\sum_{m=1}^{\infty}\frac{1}{(\ln{Re})^m}P_m$ and $Q=\sum_{m=1}^{\infty}\frac{1}{(\ln{Re})^m}Q_m$ are functions of $\alpha, a$, and $l$ only and come from the asymptotic matching between inner solution (Stokes solution) and outer solution (Oseen solution) similar to that obtained for an elliptic cylinder (cf. \ref{sec flow field}).  In the limit
$Re\rightarrow0$, the leading order term $m=1$ is dominant, and we scale out the $Re$ dependence by multiplying $\psi_{B2}$ with $(-\ln{Re})$ where $P_1=1/2 \sin{\alpha}$ and $Q_1=1/2 \cos{\alpha}$. We further quantify the reference streamfunction by adding the fax\'en streamfunction according to equation~\eqref{eq faxen flow}. We evaluate the relative qualitative and quantitative error accumulated from taking a finite number $N$ of terms with index $n$ in the infinite spatial sum in equation~\eqref{eq Phi} and found that the solution converges to a stable value at $N\geq10$.  We have taken the obstacle radius $a$ as the reference length scale and the far-field uniform speed $U$ as the reference velocity scale.

\begin{figure}[p]  
    \centering
\includegraphics[width=\textwidth]{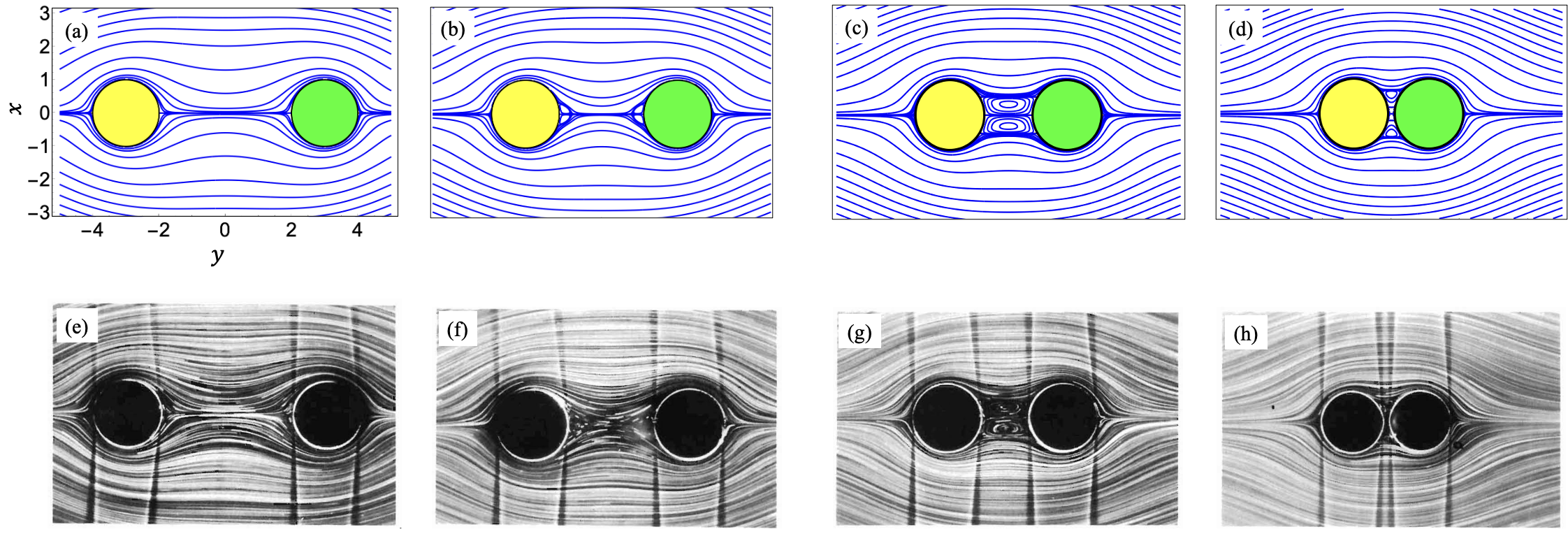}    \caption{Stokes flows met head on by two tandem circular cylinders at different spacing. Analytical streamlines (on top (a-d)) derived from equation~\eqref{eq psiB two cyl} for $n=20$ are in a very good match with the flow field from particle tracking (PT) experiment (on bottom (e-h)) performed by Tatsuno et al. \cite{tatsuno1989steady} for $l=4a$ (a,e), $l=2.5 a$ (b,f), $l=1.4a$ (c,g), and $l=0.2a$ (d,h) where $a$ is the cylinder radius. All the flows are fore-aft symmetric whereas small spacing can induce Stokes recirculation in between region of the cylinders.}  \label{fig l=0 deg}

    \centering
\includegraphics[width=\textwidth]{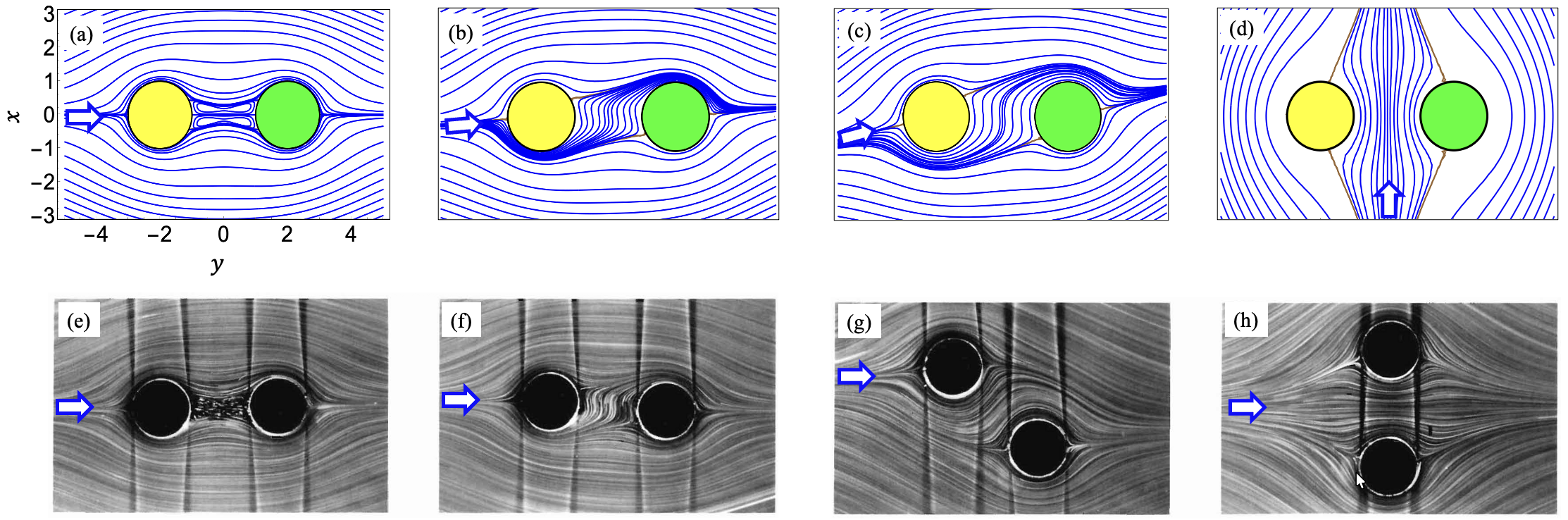}    \caption{Breaking the symmetry of the Stokes flow by inclining the flow with an angle $\alpha=5^\circ$ (b), $\alpha=15^\circ$ (b) whereas $\alpha=0^\circ$ in (a) and $\alpha=90^\circ$ (d) remains symmetric. Cylinder spacing is large $l=2a$. Top figures (a-d) are obtained from equation~\eqref{eq psiB two cyl} for $n=20$ while bottom figures (e-h) are from PT experiment adapted from Tatsuno et al. \cite{tatsuno1989steady}. The region in between the cylinders shows recirculation for $\alpha=0^\circ$ in (a) which evanescence and streamlines get skewed between them.
}  \label{fig l=2a}
 
    \centering
\includegraphics[width=\textwidth]{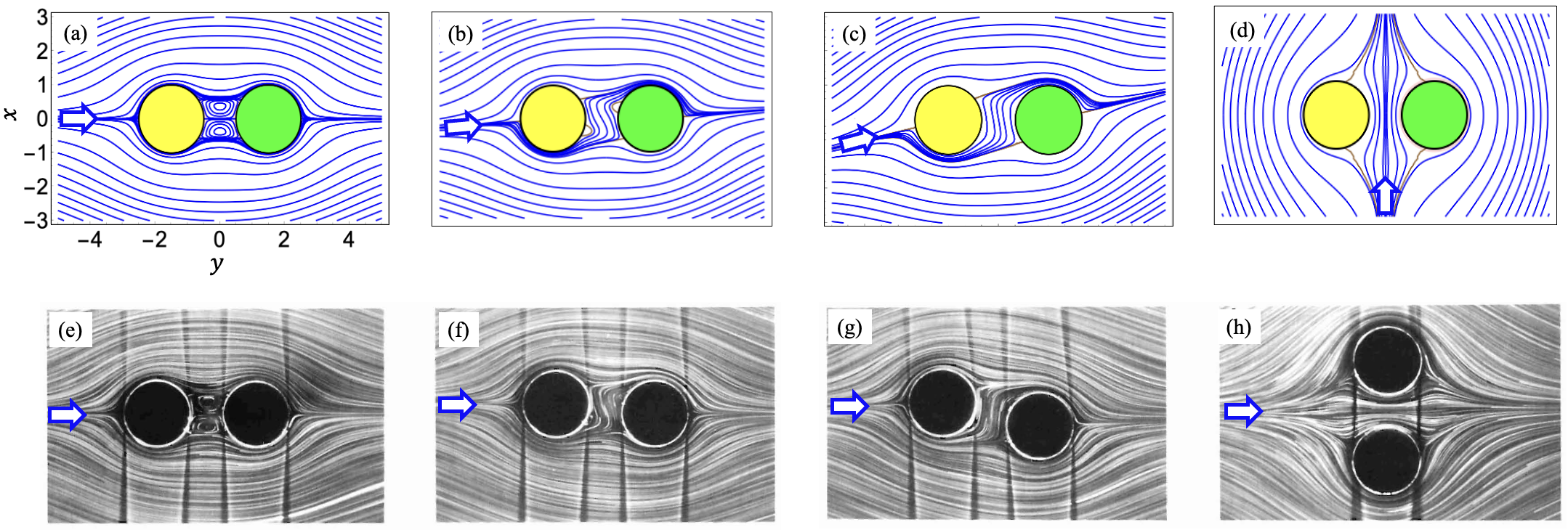}    \caption{Breaking the symmetry of the Stokes flow by inclining the flow with an angle $\alpha=5^\circ$ (b), $\alpha=15^\circ$ (b) whereas $\alpha=0^\circ$ in (a) and $\alpha=90^\circ$ (d) remains symmetric. Cylinder spacing is small $l=a$. Top figures (a-d) are obtained from equation~\eqref{eq psiB two cyl} for $n=20$ while bottom figures (e-h) are from PT experiment by Tatsuno et al. \cite{tatsuno1989steady}.
}  \label{fig l=a}
\end{figure}

The resulting flow structure is governed by the two control parameters: gap size $l$ and inclination angle $\alpha$. The computed flow fields, obtained using the analytical formulation \eqref{eq psiB two cyl} and \eqref{eq Phi} exhibit a rich variety of qualitative regimes particularly between the cylinders including non-intuitive streamline curvature shifting of the stagnation points for non-zero inclination $\alpha$, recirculation zone (Stokes vortices) for small gap size $l$. The flow structure from some combinations of $\alpha$ and $l$ are shown in figures~\ref{fig l=0 deg} to \ref{fig l=2a} where we demonstrate the variation of flow structure for changing inclination $\alpha$ in figures~\ref{fig l=a} and \ref{fig l=2a} and for changing the gap $l$ between cylinders in figure~\ref{fig l=0 deg}. Comparison with experimentally observed flow visualizations from Tatsuno et al. \cite{tatsuno1989steady} shows close agreement in the topology of streamlines, supporting the validity of the modeled flow structure. When the flow meets the cylinder pair head on ($\alpha=0^{\circ}$ and $90^{\circ}$), the overall flow geometry is symmetric following the symmetry of the individual circular cylinders, and for small gap symmetric Stokes vortices can be formed confined by the cylinder walls. This fore–aft symmetry of the flow structure is broken once the flow is inclined. Even though each obstacle remains individually symmetric, the pair configuration with the inclined flow breaks fore–aft symmetry. This asymmetry manifests most strongly in the gap region producing strongly skewed streamlines at small inclination which is accompanied by asymmetric Stokes vortices adjacent to the cylinder walls at small gap. It is interesting to observe that for certain configurations (i.e.\ $l=a$, $\alpha=5^\circ$ in figure~\ref{fig l=a}(b)), streamlines gets skewed in between the cylinders and exhibit non-intuitive turns towards an obstacle leaving a localized recirculation zone adjacent to the obstacle wall.

Under these conditions, the streamline topology breaks mirror symmetry and the spatial distribution of normal velocity gradients and thus the wall normal velocities as developed in chapter~\ref{chap 2} becomes asymmetric between upstream and downstream regions of the pair. Consequently, particles transported through such a symmetric configuration experience a net hydrodynamic effect across streamlines in a similar fashion to the symmetry breaking arguments developed earlier for a single elliptic obstacle.

The key physical observation is that the asymmetry in this configuration is produced by the relative orientation of the symmetric obstacles. The presence of the second cylinder in inclination modifies the spatial recovery of the velocity field generated by the first, thereby creating persistent asymmetry in the flow gradient distribution encountered by a particle. This mechanism provides a configurational route to symmetry breaking that is closely relevant to multi-obstacle microfluidics application (e.g\ DLD, porous media filtration).

\section{Particle motion around two circular obstacles}

We numerically solve the equation of motion of force-free particles as developed in chapter~\ref{chap 2}. Wall effects $\boldsymbol{W}$ originate simultaneously from each of the obstacle boundaries and are added to the Fax\'en flow to obtain the particle velocity according to the formalism described in chapter~\ref{chap 2} This is consistent with the two-wall interaction framework established in by Liu et al. \cite{liu2025principles}. We apply variable expansion model with $\Delta_E=3$ (cf \eqref{eq xE}) to describe particle motion normal and tangential to each obstacle wall. As in the single-obstacle analysis  in chapter~\ref{chap 3}, the obstacle surface is locally approximated as flat, an approximation previously shown to capture both far-field and near-field particle dynamics with good accuracy. Following the same procedure of trajectory calculation in chapter~\ref{chap 3}, we release particles far from obstacle and track their pathline around the two obstacle system to the end their journey at far downstream. The starting point of a trajectory corresponds to an initial streamline. We choose a particle size of $a_p=0.1$ and track its center position to obtain the computational pathline. We present the computation in the subsequent sections for a background flow inclined with $\alpha=5^{\circ}$ perturbing the symmetry of the two circular obstacles placed close to each other (separated by a gap of $l=0.5a$).

\subsection{Particle trajectory}

\begin{figure}  
    \centering
\includegraphics[width=\textwidth]{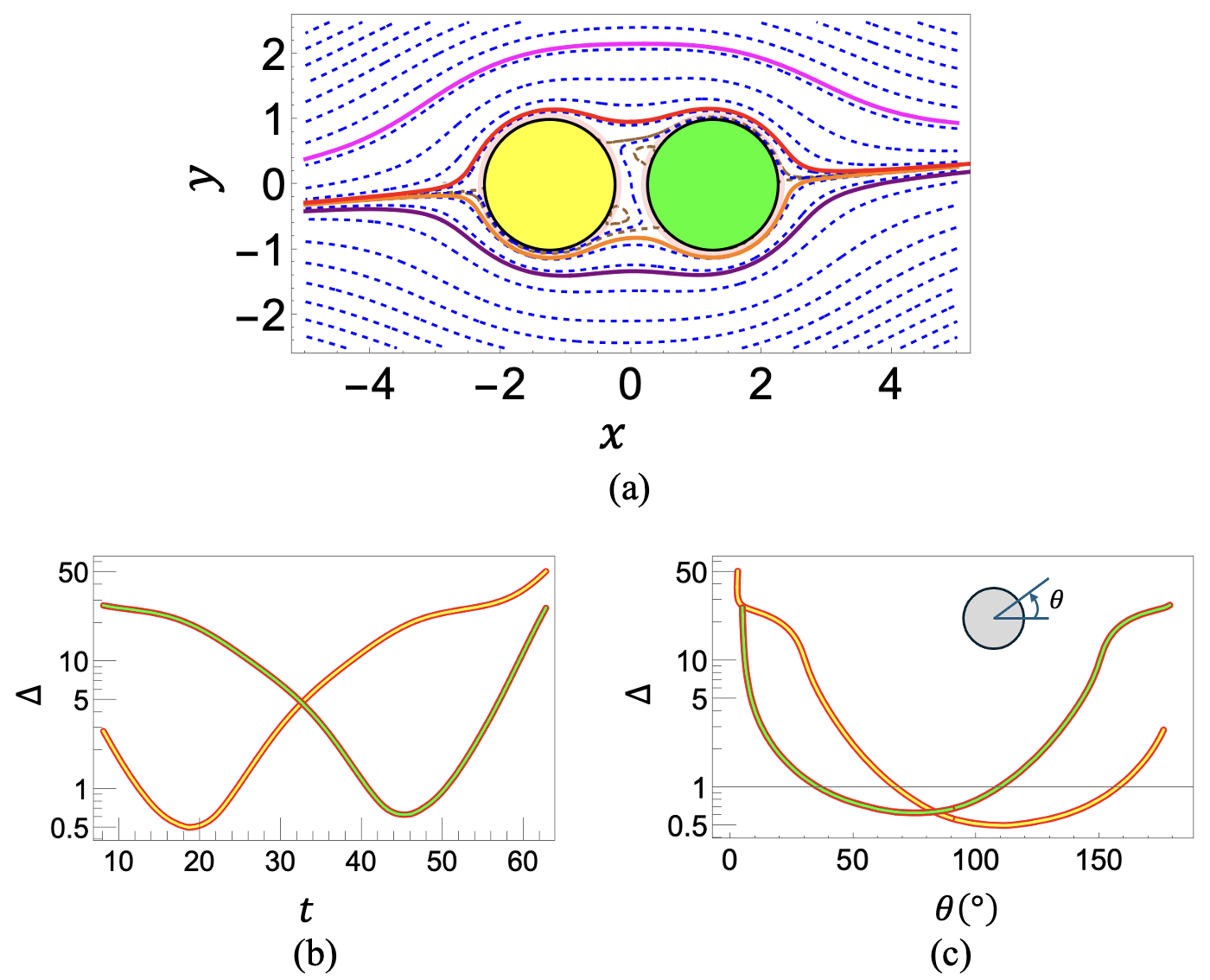}    \caption{(a) Trajectories of particles (in solid lines) being transported by the background Stokes flow (in dashed blue lines) around two $\alpha=5^\circ$ inclined, closely spaced $(l=0.5a)$ circular cylinders. Changes of $\Delta$ along the ``red" type trajectory in (a) over time (b) and over angular position around obstacle (c). In (b) and (c), yellow and green lines are for the approaching and receding obstacles (``yellow" and ``green" obstacle in (a)) respectively. 
}  \label{fig trajec}
\end{figure}

Upon time integration of the particle velocity we can obtain its position as function of time. We show four representative trajectories in figure~\ref{fig trajec}(a) of particles transported over two circular cylinders. Because of the time-reversibility the flow, the trajectories are fully time-reversible and take the same path upon reversing the flow direction. ``Magenta" type trajectory is staying mostly far from the obstacle $(\Delta\gg1)$ and dominated by particle center expansion model (cf chapter~\ref{chap 2} for details) whereas $\Delta_{min}<1$ for the ``red" type trajectory  which can be seen from the $\Delta$ plot over time in figure~\ref{fig trajec}(b) and over angular position around each obstacle in figure~\ref{fig trajec}(c). The two curves in figures~\ref{fig trajec}(b) and (c) corresponding to instantaneous $\Delta$ positions of the particle with respect to each obstacle are not mirror symmetric because of the symmetry breaking by the flow  inclination. 

A key feature emerging from these computations is that the trajectory evolution becomes spatially asymmetric once the flow is inclined. Even though the governing Stokes equation remains time-reversible, the distribution of wall interaction along the trajectory differs between upstream and downstream segments. The particle therefore experiences unmatched near-wall “dive” and “recovery” phases, which generates a net wall affected deviation. This behavior directly resembles the symmetry-breaking displacement mechanism identified earlier for elliptic obstacles but arises here purely from obstacle arrangement. 

Figure~\ref{fig trajec}(c) also shows that the position of minimum approach is shifted towards upstream of the first encountered obstacle (approaching obstacle in ``yellow" in the figure~\ref{fig trajec}(a)) and towards downstream of the second obstacle (receding obstacle in ``green" in the figure~\ref{fig trajec}(a)) which  would require to be exactly at $\theta=90^\circ$ for a single symmetric circular obstacle. This positional shift of $\Delta_{min}$ is also enhanced by the asymmetry from inclination.

\subsection{Hydrodynamic force on a particle}

\begin{figure}  [t]
    \centering
\includegraphics[width=\textwidth]{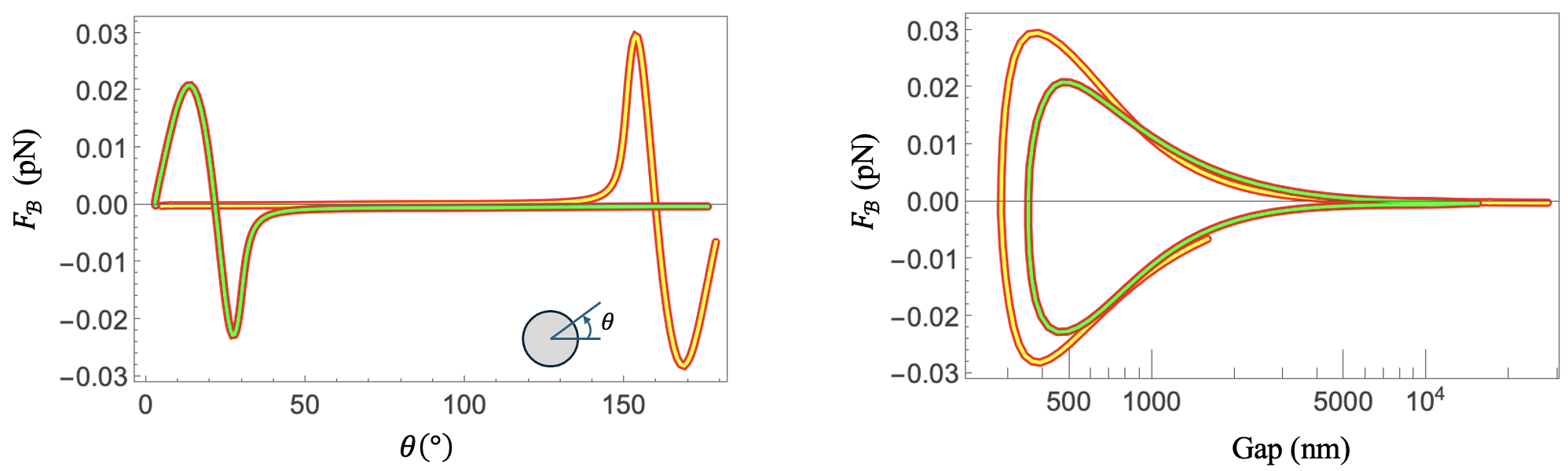}    \caption{Hydrodynamic force $F_{\mathcal{B}}$ originating from the flow extensional gradient according to \eqref{eq FB}. (a) shows its distribution around each obstacle and (b) shows its change at different gaps from each obstacle. The force have calculated from experimentally realized parameters with ($Re=10^{-2}, a_p=0.55 \mu m, a=10a_p=5.5 \mu m$, uniform reference velocity, $U_{ref}=1200 \mu m/s$, and fluid viscosity $\mu=0.01$Pas) \cite{miele2025flow}. Yellow and green lines are for the approaching and receding obstacles (``yellow" and ``green" obstacle in (a)) respectively.}  \label{fig fb}
\end{figure}

In chapter~\ref{chap 5} we have computed the individual hydrodynamic force components acting on the particle normal to elliptic wall that are eventually responsible for diving the particle through the electrostatic barrier and stick eventually. We have found that the flow extensional gradient driven force $F_{\mathcal{B}}$ is amplified and can compete with DLVO forces under experimental condition ($Re=10^{-2}, a_p=0.55 \mu m, a=10a_p=5.5 \mu m$, uniform reference velocity, $U_{ref}=1200 \mu m/s$). We follow the same calculation procedure of quantifying $F_{\mathcal{B}}$ according to equation~\eqref{eq FB} on the trajectory over two circular cylinders, particularly for the trajectory close to the obstacle (``red" type). We present the result of $F_{\mathcal{B}}$ distribution along different angular positions of each obstacle and at different particle-obstacle gaps from each obstacle in figure~\ref{fig fb} for the ``red" type trajectory.

Figure~\ref{fig fb}(a) shows that maximum wall normal force on the particle can be found either in the upstream side of the approaching obstacle (``yellow") or in the downstream side of the receding obstacle (``green") while the receding wall effect is enhanced from symmetry breaking. Figure~\ref{fig fb}(b) shows that $F_{\mathcal{B}}$ varies at different gaps from the walls and maximum effect can be found close to the position of $\Delta_{min}$. Although the computed force magnitude for the present trajectory remains much smaller than typical DLVO forces, it is important to note that this trajectory does not correspond to a dive trajectory (see chapter~\ref{chap 5} for the ``dive" nomenclature), with the minimum separation remaining on the order of hundreds of nanometers. Therefore, the results primarily demonstrate the asymmetry induced modification of hydrodynamic wall normal forces rather than the ultimate force levels achievable during stronger near-wall encounters. Within the wall-interaction framework, these forces arise primarily from gradients of the background flow evaluated at particle-size distances from the obstacle boundaries. 

Because the two-cylinder configuration modifies the spatial distribution of velocity gradients, the resulting hydrodynamic forcing differs qualitatively from the single-obstacle case. Accurate characterization of dive trajectories in this multi-obstacle configuration presents significant numerical challenges, as resolving many number of terms in the background streamfunction (\eqref{eq psiB two cyl} and \eqref{eq Phi}) and requires substantially higher spatial grid resolution in the near-wall region. A systematic investigation of dive trajectories and the associated force amplification is deferred to future work, where improved numerical strategies will be employed to capture near-contact dynamics with sufficient accuracy.

\section{Conclusion}

The symmetry-breaking hydrodynamic force model developed in this dissertation work (chapters~\ref{chap 3}--\ref{chap 5}) implies that particle displacement depends sensitively on particle size, providing the fundamental ingredient required for size-based sorting in DLD devices. In particular, particles of different sizes experience different strength level of wall-interaction asymmetry, leading to distinct trajectory deviations. While we demonstrated this symmetry breaking effect for a single asymmetric obstacle (inclined ellipse), a symmetric obstacle (circular) can not break the fore-aft symmetry of the flow unless there is another obstacle nearby. Two cylinder configuration thus provides the fundamental base of analyzing the asymmetry in the flow structure in practical multi-cylinder deterministic sorting in a DLD device which not only a qualitatively different type of asymmetry but involves multiple particle manipulating flow features which can not be a possibility with a single obstacle.

This chapter presents an elementary result of particle transport in a symmetry breaking Stokes flow around two circular cylinders. The mechanism of symmetry breaking is introduced by inclining the flow with a non-trivial angle. We observe this symmetry breaking effect from simultaneous wall interactions on the particle trajectory while passing over individual obstacle. We locate the maximum hydrodynamic force around each obstacle while particle experience greater wall-induce forcing effect while leaving the obstacles. 

The analytical form of the streamfunction of background Stokes flow over two obstacles is an infinite sum of spatial coordinate system which defines the accuracy of the flow geometry. While we have found that a summation over 10 consecutive terms converge the solution with good accuracy, it demands very high computational requirement to quantify the wall effects with high numerical precision and accuracy especially for a particle passing the obstacles at very small $\Delta$ is strongly dependent on grid refinement near the obstacles. This potential numerical limitation restricts us from tracking particles much closer to the separating streamline which would travel through the complex flow structures in between the cylinders. 

The gap region between the cylinders generates non-intuitive flow structures where streamlines get skewed, stretched, and closed within localized vortices adjacent to the walls. This modifies the local streamline topology affecting particle motion significantly in the region.  Particles from starting very close to separating streamlines and transporting through this region can therefore exhibit trajectory classes that do not occur in the single circular obstacle configuration. Whereas numerical refinement for these particles in such complex background flow region in between the obstacles is time-consuming to resolve with good numerical accuracy, preliminary computations indicate the possibility of streamline crossing within the gap region under symmetry-breaking conditions. This result of streamline crossing in such two-cylinder geometry serves as a motivation to perform a thorough analysis with acceptable numerical accuracy.

A systematic extension of the present two-obstacle configuration is presented in appendix~\ref{appen 3cyl} with three cylinders, where we present preliminary particle tracking experiments  demonstrating that the presence of additional obstacles at different orientations to the others can generate configurational asymmetry and complex streamline topology qualitatively similar to two-cylinder asymmetry. These observations support the interpretation of the two-cylinder framework as the fundamental building block of multi-obstacle transport relevant to DLD arrays and porous-media filtration.

\chapter{Conclusions}\label{chap: conclusions}

In this dissertation, we have presented a systematic quantification of obstacle wall induced hydrodynamics in manipulating density-matched, force-free, spherical particle transport over the obstacle. The background flow is non-inertial (Stokes flow) and particles are passive in nature. The goal of our study has been to isolate the purely hydrodynamic displacement effect on a particle encountering a cylindrical obstacle interface. This work establishes a new fundamental framework for passive particle manipulation in many microfluidic applications by leveraging particle–obstacle interactions independent of external forces, short-range contact effects, or fluid inertia. This is the first time this pure hydrodynamic effect has been rigorously described in external flows, providing a firm physical basis for previously neglected effects important for the operational mechanisms of DLD devices and porous media filtration.

\section{Summary of key results}

The first part of this dissertation work focused on developing a rigorous methodology to model force-free, neutrally buoyant, spherical particle trajectory in the presence of an obstacle in an arbitrary Stokes flow, with both particle and fluid inertia absent. Utilizing the framework of our previous work in modeling particle's velocity in cartesian coordinate system, we employ the modeling approach in extending the framework for general curvilinear coordinate system in order to derive wall affected hydrodynamic velocity corrections in both normal and parallel to an elliptic obstacle wall. These wall induced corrections alter the particle movement beyond just passively following the background flow. By performing a proper matching between already existing descriptions-- particle center expansion model that is appropriate at large distances from the wall and lubrication model which accurately captures the near-wall regime, we developed an improved technique through a variable point expansion model that describes particle motion at any distances from the obstacle. Our work in obtaining these uniformly valid wall-interaction expressions is the first to formulate a closed hydrodynamic based equation of motion for particles suspended in a wall induced Stokes flow.

Having developed a proper description of particle-wall interaction, we then focused on whether a Stokes flow by itself can cause significant particle displacement while transporting particle over an obstacle. While literature exists with descriptions of particle's net displacement over encountering an obstacle in Stokes flow, practically relevant in DLD microfluidics, obstacle interactions are either heuristically developed and/or fully non-hydrodynamic in nature. We aim to displace particles by pure hydrodynamic interaction with obstacle wall in Stokes transport flow. We employ the analytically known Stokes flow expression over an inclined elliptic obstacle where the geometry induced fore-aft asymmetric distribution of velocity corrections result in a net particle displacement. We obtain numerical trajectories of particles being transported around the obstacle (1) mostly keeping large distances from it and therefore experiencing trivial wall effect, (2) mostly being very close to the obstacle and therefore exhibiting significant wall interaction. We quantify the net streamline crossing as the consequence of the wall affect which increases with particle's initial streamline getting closer to the separating streamline as it encounters the obstacle from closer proximity. Obeying the physical constraint, our numerical results show a prominent maxima of the displacement effect signifying an obstacle's capacity in displacing particles hydrodynamically. A comparison with the effect arising from the existing model of roughness based contact highlights the importance of incorporating our symmetry breaking hydrodynamics in modeling particle displacement in many microfluidic applications.

To better understand the displacement capacity of the asymmetric obstacle, we then investigate the wall interaction analytically. We focus on the analysis of wall interaction in near-wall regime where particle motion is more appropriately described by simplified lubrication model facilitating an analytically integrable expression. 
Rather than focusing on trajectory-specific details, the analytical treatment provides direct insight into the scaling structure of the displacement mechanism upon close interaction with wall. A key outcome of this approach is that it captures the existence and scaling of the maximum displacement observed in trajectory computations. The resulting scaling relations express this maximum displacement in terms of particle size and geometric control parameters, including obstacle inclination angle and elliptic aspect ratio. The scaling law demonstrates our symmetry breaking hydrodynamic wall effect as a tunable system in controlled particle manipulation microfluidics.

Building upon our symmetry breaking mechanism developed to induce net spatial displacement on particle, we focus on the resulted hydrodynamic forces acting on the particle and normal to the obstacle wall in the next part of the thesis. This wall-normal forces play the crucial role in porous media filtration microfluidics by forcing the non-Brownian particles towards the obstacle through the repulsive electrostatic barrier that would otherwise prevents close approach. And the eventual fate of the particle once they transported to sufficiently small gap, is to stick to the obstacle by the activation of short-range attractive force. We quantify the minimum approaches by our computed particle trajectories in a force-free system and remarkably obtain that our proposed symmetry breaking method can effectively amplify the hydrodynamic force, as particles dive into their minimum approaches, to be comparable to and overcome the repulsive DLVO forces and lead localized attachment. Our work here establishes the fundamental importance of hydrodynamics in the application of particles' sticking, capture, and accumulation on the obstacle surface while restraining their transportation around it.

Finally we demonstrate the principle of symmetry breaking in Stokes transport arising from the relative arrangement of multiple symmetric obstacles, using the two-cylinder configuration. The interaction between cylinders generates asymmetric streamline topology in the gap region that are not achievable around a single circular obstacle. These results establish multi-obstacle symmetry breaking as a fundamental tool underlying particle manipulation in DLD arrays and porous-media filtration.

\section{Ongoing and future work}

\subsection{Rigorous investigation of particle net displacement in two-obstacle geometry}

Chapter~\ref{chap 6} establishes that symmetry breaking can arise not only from geometric deformation of a single obstacle that we have developed by inclining an eccentric obstacle (ellipse) but also from the arrangement of multiple symmetric obstacles (circle). In particular, the two-cylinder configuration provides the elementary physical model in which fore–aft symmetry is broken through obstacle interaction and flow inclination that fundamentally alter particle trajectories in a multi-cylinder configuration present in a typical DLD device. Therefore, this exemplary approach can be potentially extended to entire DLD arrays where particle transport results from repeated encounters with multiple nearby obstacles. The non-intuitive flow structures formed in the near-gap region of the two cylinders as shown in chapter~\ref{chap 6}— including skewed streamlines, localized recirculation zones, and strongly asymmetric velocity gradients — play a central role in determining whether particles pass between obstacles and/or undergo separating streamline crossing.

While we have presented this fundamental approach of multi-obstacle architecture analogous to a DLD device and potentially to more disprdered porous media, accurately quantifying particle net displacement in the two-obstacle geometry remains computationally challenging. The analytical streamfunction describing Stokes flow around two cylinders is expressed as an infinite summation in bipolar coordinates, and resolving particle motion requires evaluation of many terms with high numerical precision. As highlighted in Chapter~\ref{chap 6}, trajectory segments corresponding to near-wall “dive” motion — particularly for particles traveling through the narrow gap between cylinders — are highly sensitive to spatial resolution and locally demand very fine computational grids. These requirements substantially increase computational cost and limit the systematic exploration of trajectory classes, especially those responsible for meaningful displacement and force amplification.

A rigorous computational technique of net particle displacement in the two-obstacle geometry is a necessary step toward extending the present theory to entire DLD arrays and porous-media transport problems, quantifying the degree of particle displacement as well as that of geometric amplification of hydrodynamic force components necessary for particle attachment and fileterin. Therefore, an important direction of ongoing and future work is the development of more thorough computational strategies for resolving particle trajectories in between two-obstacle environment. Such approaches may include improved numerical evaluation of the streamfunction, adaptive near-wall discretizations capable of capturing near-contact dynamics with controlled accuracy. 

A systematic computation of ``dive" trajectories that involve the gap region between obstacles will enable precise quantification not only of net displacement effects from the configurational asymmetry induced wall effects, but identification of  possible separating streamline crossing.
Note that the ability of DLD to sort particles by size is based on (1) differential displacement effects of  particles encountering a single obstacle, and (2) these differential displacements leading to particles of different sizes following trajectories that traverse obstacles on opposite sides. In order for the latter process to occur, separating streamlines originating from one obstacle have to be crossed through interaction with another obstacle. Thus, this necessary ingredient of DLD sorting can only be studied in a scenario with two or more obstacles.  


\subsection{Transport of non-spherical rigid and flexible particles}

\begin{figure}
\centering
\includegraphics[width=0.7\textwidth]{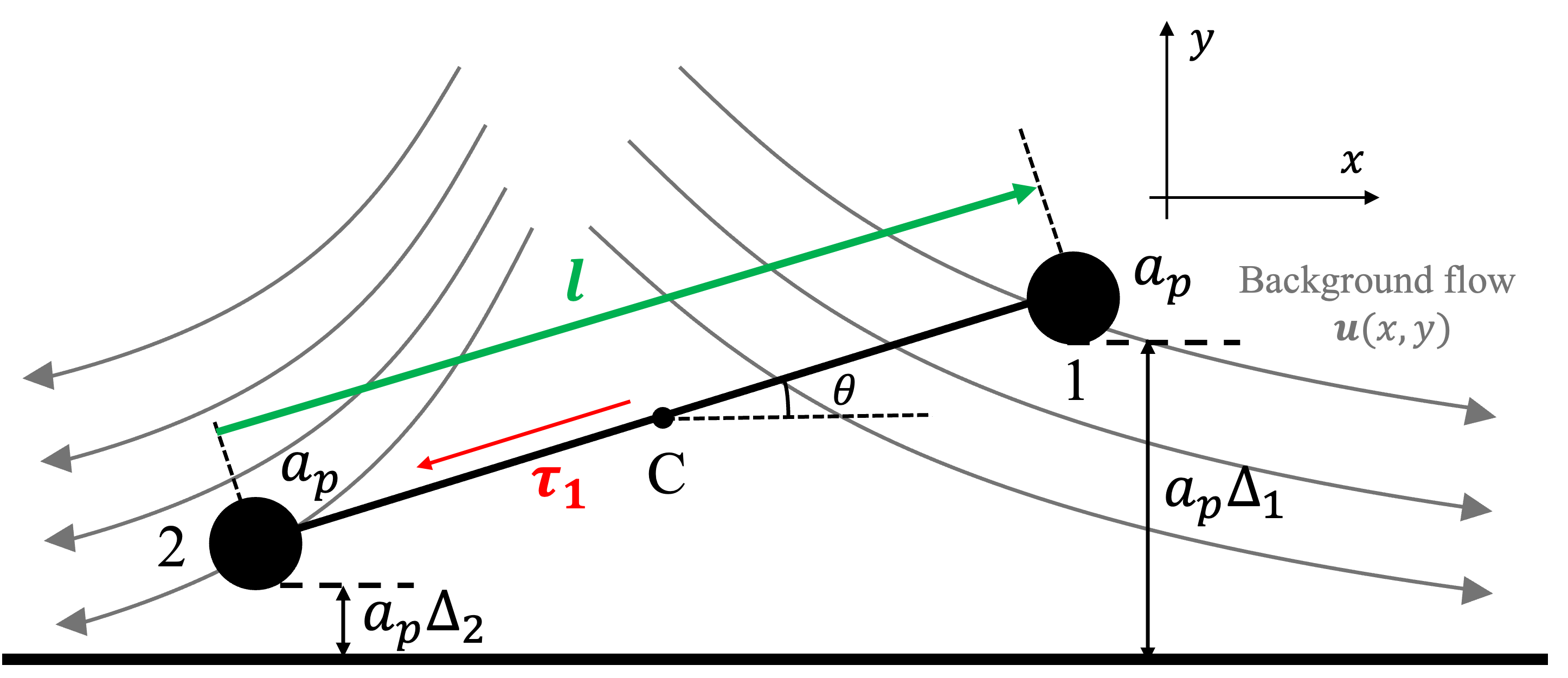}
\caption{Schematic of a rigid dumbbell in an arbitrary flow field near a plane wall adapted from \cite{liu2025particlethesis}.}
\label{fig dumbbell}
\end{figure}

An important direction of the study concerns particle's shape and additional degrees of freedom it introduces. The present dissertation focused on spherical, for which translational motion is the primary degree of freedom modified by the hydrodynamic corrections from wall. Wall correction of rotational velocity of the symmetric spherical particle does not bear importance in this context of the spatial manipulation of non-inertial particle transport. In contrast, non-spherical particles exhibit coupled translational and rotational dynamics and orientation-dependent wall interactions, which can fundamentally create qualitatively new trajectory even without inertia. Experimental and theoretical studies of fiber transport over a triangular obstacle in Stokes flow, as performed by Li et  al. \cite{li2024dynamics} (cf figure~\ref{fig fiber_vortex}) shows fiber orientation playing significant role in the eventual fiber displacement. Very recent work by Liu and Hilgenfeldt \cite{liu2025particlethesis} on a rigid dumbbell shaped non-spherical particle motion in Moffatt eddy flows requires to solve the angular dynamics of the dumbbell orientation along with the translational dynamics of its center. The resulting computation by simultaneously solving these 3 equations of motion demonstrated that dumbbell asymmetry can achieve persistent cross-stream transport transforming from quasi-periodic motion to limit-cycle behavior in internal flow, even in the absence of wall interactions. Motivated by these findings, a natural extension of the present framework is to model the dumbbell dynamics as sketched in figure~\ref{fig dumbbell} in an analogous fashion but in external flow where the dumbbell is being transported over symmetric (circle) or asymmetric (inclined ellipse) obstacle. 

Beyond modeling rigid fibers, Chakrabarti et al. \cite{chakrabarti2020trapping} recently demonstrated transport dynamics of flexible fiber through multi-obstacle porous media system by numerically modeling anisotropic drag on fiber elements that lacks the full description of hydrodynamic interaction between fiber and pillars. As an important extension of their study, Makanga et al. \cite{makanga2023obstacle} modeled the interplay between fiber's elastic force, contact force, and ad-hoc interaction forces with an obstacle. While these works manifest useful qualitative insight of different modes of fiber transport and reassure flexible fiber migration across streamlines in transport, the dynamical behavior demand the complete hydrodynamic formalism of particle-obstacle interaction of the type developed in this dissertation. 

Therefore, the present framework offers a natural foundation for modeling dumbbell, and rigid/flexible fiber dynamics during transport past obstacles. We anticipate that incorporating full hydrodynamic wall interactions into such modeling of non-spherical particles will produce rich dynamical behavior, including orientation-dependent migration modes similar to the pole-vaulting, gliding, and above/below transitions, and trapping reported by \cite{li2024dynamics, makanga2023obstacle,chakrabarti2020trapping} for fibers moving past an obstacle in Stokes flow. The rigorous hydrodynamic wall-interaction formalism developed in this dissertation provides a systematic way to quantify how particle shape and symmetry breaking jointly affect its transport behavior around single and multiple obstacles. This direction offers a pathway toward understanding shape-controlled particle manipulation in inertialess microfluidic transport.

\subsection{Consideration of asperities originating from elasticity and surface roughness}

\begin{figure}
\centering
\includegraphics[width=0.8\textwidth]{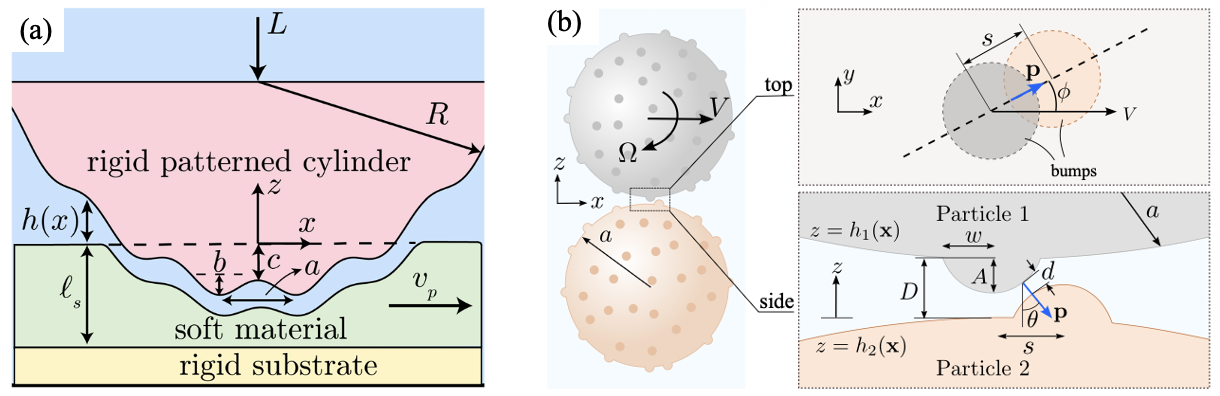}
\caption{(a) Schematic of a rigid particle in near contact gap with a plane substrate coated with soft material and sinusoidally patterned surfaces. (b) Sketches of surface asperities configuration for rough particles in sliding and rotation. (a) and (b) are adapted from \cite{kargar2025non} and \cite{minten2025hydrodynamic}.}
\label{fig soft_rough}
\end{figure}

An alternative way to break the relative fore-aft symmetry is by inducing finite elasticity or introducing surface roughness on the particle and/or obstacle surface. Very recent work by Kargar-Estahbanati and Rallabandi \cite{kargar2025non} has demonstrated that even weak elastic deformation of a boundary can substantially modify near-wall hydrodynamic interactions. By modeling a gently deformable planar surface with small sinusoidal perturbations (amplitude of the deformation is much smaller than particle size and on the scale of gap between the surfaces, see figure~\ref{fig soft_rough}(a)), they obtained the analytical formalism of the effective friction coefficient with its scaling law and showed that softness of the material generates counterintuitively non-monotonic dependence in the fiction coefficient with the relative elastic deformation, revealing that material softness can qualitatively alter the force balance between sliding surfaces. In another most recent work by Minten and Rallabandi \cite{minten2025hydrodynamic} introduced asperities on the scale of small gap $\Delta\ll a_p$ originating from the surface roughness of particles sliding and rotating in near-contact region (see figure~\ref{fig soft_rough}(b)). They showed that the hydrodynamic resistance grows as the the inverse of the gap between the asperities in contrast with the weak logarithm scaling derived from classical smooth surface lubrication theory. 

Introducing these modified hydrodynamic forces into the governing dynamics and recomputing trajectories would allow one to quantify how irreversible boundary effects influence net particle displacement, near-wall migration, and the ultimate fate of trajectories. Such extensions would bridge the gap between idealized symmetry-breaking mechanisms and realistic microfluidic environments where material compliance and surface structure are unavoidable Motivated by these results of modifying hydrodynamic forces by incorporating roughness bumps and elasticity on the particle and obstacle would significantly alter the motion of particle, we suggest to introduce these forces into the equation of motion of the force free particle as in equation~\eqref{Flambda} and recompute the trajectory to explore the effect of these irreversible forces on the net fate of the trajectories. Incorporating elasticity and surface roughness into the present wall-interaction framework thus offers a systematic route to extend the proposed model beyond the smooth, rigid surface condition. Such extensions would enable the investigation of near-wall hydrodynamic forces in more realistic particle–boundary interactions, integrating the hydrodynamic interactions detailed in the present work with the concepts of surface roughness induced breaking of time symmetry without ad hoc assumptions of surface-to-surface contact (see section~\ref{sec roughness} of chapter~\ref{chap 3} for detailed discussion).

\subsection{Incorporating flow inertia}

In practical microfluidic devices Reynolds numbers may be small, but cannot be zero. If $Re$ is small enough, the Reynolds number independent wall effects described in this thesis, while having received comparatively limited prior attention, can be dominant. Finite Reynolds number effect can arise either in steady transport flows confined by channel walls or from oscillatory flows near interfaces. Steady inertial effects of the former type have been well studied whose effect can become comparable with the non-inertial wall effect if the channel wall symmetry can broken effectively (i.e. channel expansion-constriction, flow induced deformation \cite{karan2020flow}) as mentioned in the conclusion section~\ref{sec concl chap 4} of chapter~\ref{chap 4}. A promising direction of incorporating our work is thus to model such channel wall asymmetry making it necessary to compare and combine the inertial and non-inertial effect when predicting particle transport near walls. 

In microfluidics, another type of inertial effect induced by oscillatory flow has attracted considerable attention. Second order rectification effects that persist after time-averaging over the oscillation are inertial in origin. This involves both rectification of fluid element motion (steady streaming flows) and rectification of particle trajectories. These effects are involved in many particle manipulation processes such as size-based sorting \cite{wang2011size}, trapping \cite{lutz2006hydrodynamic}, and particle sticking. Related to these phenomena, we have experimentally demonstrated streaming induced particle sticking and cluster formation around a triangular solid obstacle in appendix~\ref{appen sticking from streaming} where streaming vortices transport particles into the vicinity of the obstacle wall. The wall dependent hydrodynamic forcing that leads to sticking in this situation could involve two simultaneous wall effects: (1) the second order (of relative oscillation amplitude)  inertial rectification effect from the primary oscillatory flow that has been rigorously modeled in certain situations by \cite{agarwal2018inertial,agarwal2021unrecognized,agarwal2023density}, and (2) the steady non-inertial contribution from the particle-wall interaction in the steady rectified streaming flow, amenable to the analysis presented in this thesis. The relative role of these two processes that are simultaneously present in many practically relevant situations has not been investigated.

While the oscillation-induced streaming flow is fundamentally Stokes flow satisfying the biharmonic equation for the streaming streamfunction (see appendix~\ref{appen Rich Variety of Streaming Flow}), it emerges as a second order effect driven by the first order oscillatory shear layer which imposes a slip velocity boundary condition at the edge of a viscous boundary layer from the asymptotic matching between the inner solution and outer solution \cite{longuet1998viscous,davidson1971cavitation}. Therefore, proper modeling of the non-inertial content of hydrodynamic forcing in streaming demands revisiting the resistance coefficients formalism for finite slip velocity boundary condition in \cite{rallabandi2017hydrodynamic} to be used in obtaining the wall influenced equation of motion of particle according to equation~\eqref{Flambda} for a finite slip velocity boundary. 

\begin{figure}
\centering
\includegraphics[width=0.45\textwidth]{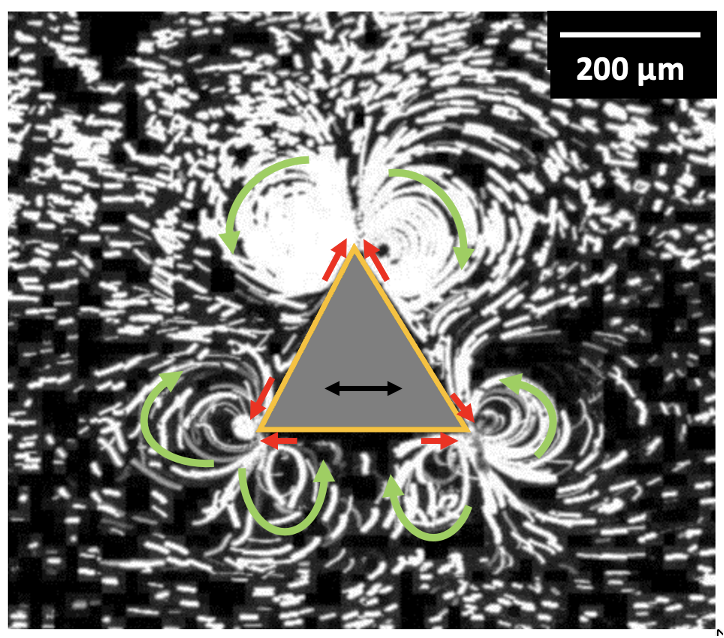}
\caption{Streaming vortices around corners of a no-slip obstacle wall which is fundamentally a Stokes flow driven from the shear boundary (slip velocity indicated by red arrows). Oscillation direction is denoted by black arrow.}
\label{fig slip vel}
\end{figure}

We propose to begin by following the fundamental methodology described in \cite{adamczyk1983resistance, maude1961end, brenner1961slow} where the coefficients are derived for no-slip and no shear boundaries. By rigorously solving the governing equations for the friction force between a sphere and a solid plane wall with finite tangential slip velocity (shown in figure~\ref{fig slip vel}) and vanishing normal velocity (no penetration) at the plane wall, we propose to remodel the hydrodynamic resistance that will describe particle motion from the particle-wall interaction in streaming flow. This remodeling will extend our work in combining and comparing oscillatory inertial and steady non-inertial particle-wall interactions in maneuvering particles across streamlines of streaming vortices and provide a deeper understanding of the hydrodynamic forces leading to  particle sticking in streaming flow scenarios.

\section{Closing remarks}


A central conceptual outcome of this dissertation is that hydrodynamic wall interactions in Stokes flow can generate spatially irreversible particle manipulation even though the governing Stokes equation is time-reversible. In this sense, the results are counterintuitive to the common expectation that “nothing irreversible can happen hydrodynamically” in a purely inertialess flow: the mechanism does not rely on inertia, does not violate time-reversal symmetry, and yet produces a non-zero net spatial irreversible displacement after a single obstacle encounter because the spatial distribution of wall-induced effects lacks fore–aft cancellation once the flow–obstacle configuration is asymmetric. 

This thesis demonstrates that a single particle-wall encounter can indeed have a net displacement effect on particle transport when the geometry of the flow and obstacle breaks fore-aft symmetry. An exemplary case is an elliptic obstacle placed in a uniform flow with non-trivial angle of attack. Our trajectory computation predicts that there is a particular initial condition for given parameters for which the net displacement is maximal. Overall, this study establishes hydrodynamic wall interactions in symmetry-broken Stokes flows as a fundamental and quantifiable mechanism for particle manipulation at low Reynolds numbers. By combining analytical theory with computational simulations, we have shown that controlled symmetry breaking in flow geometry leads to predictable particle displacement, capture, and attachment without relying on inertia or external actuation. The analytical scaling laws, uniformly valid wall-interaction formulations, and trajectory predictions presented here provide new design guidelines for elementary microfluidic systems — including size-based sorting in deterministic lateral displacement arrays and particle capture in porous filtration architectures — and open a pathway toward inertia-less, geometry-driven strategies for particle deflection, sorting, capture, deposition, and controlled transport in next-generation microfluidic devices.

\appendix

\chapter{Streaming induced particle sticking and cluster formation}\label{appen sticking from streaming}

In chapter~\ref{chap 5}, we described the inertialess hydrodynamic mechanism in which a transported particle’s dive toward an obstacle leads to its capture at the obstacle boundary. The framework is analogous to the situation in porous media microfluidics where a Stokes flow through many obstacles is generated purely by pressure gradient. In such transport flow, individual particle trajectory is open, interacts with multiple obstacles where each interaction determines particle's fate (either transport or arrest). In our recent work \cite{liu2025principles}, a Stokes vortex is utilized where a closed particle trajectory repeatedly interacts a boundary accommodating its exponential approach towards obstacle. In either situation the background Stokes flow is purely generated by a driving force (e.g/ pressure gradient) far from the wall.

\begin{figure}
    \centering
\includegraphics[width=\textwidth]{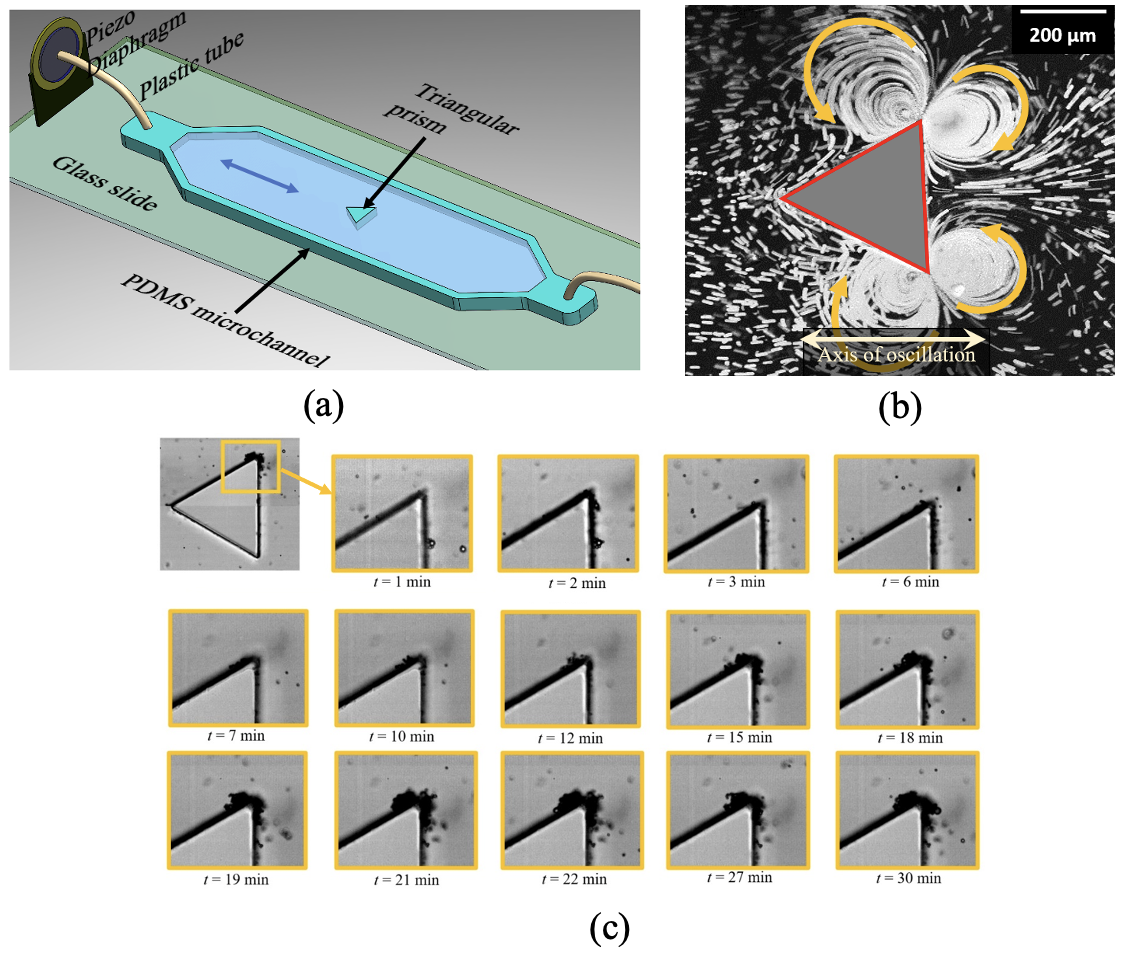}
    \caption{(a) Experimental PDMS microfluidic set-up for translationally oscillating flow around an triangular obstacle to induce streaming, (b) Corner streaming around an equilateral triangle facing in the direction of oscillation, obstacle characteristic length $125 \mu m$ is half of the triangle edge length, oscillation amplitude $10 \mu m$, oscillation frequency $2.7 k Hz$. (c) Aggregated particles at a corner orthogonal to the oscillation axis. Particles swirling in the streaming vortices deposit on both sides of the corner which evolves in time and forms particle cluster.}
    \label{fig sticking}
\end{figure}

A Stokes flow can emerge as the second order time rectification effect of an oscillatory flow around an obstacle. This is an unique steady vortical flow system driven by the oscillatory shear boundary layer which we describe in more details in appendices~\ref{appen robustness and sensitivity of microstreaming} and \ref{appen Rich Variety of Streaming Flow}. Here, we utilize this boundary layer driven Stokes vortical flow close to an obstacle wall for enabling repeated particle-wall interaction that eventually lead to stick. We experimentally produce a streaming flow from translational oscillation of fluid around a triangular obstacle. The experimental set up is depicted in figure~\ref{fig sticking}(a) where a translational oscillation around an equilateral triangular cylinder in a PDMS microchannel is provided by exciting a piezo diaphragm transmitting the back and forth oscillation to the channel through a plastic tube. A steady vortical structure arises from the second order time rectification of the fluid's translational oscillation (see appendix~\ref{appen Rich Variety of Streaming Flow}) and can be visualized at high frame rate imaging as shown in figure~\ref{fig sticking}(b). Frequency and the oscillation amplitude of the piezo diaphragm and thus the flow were controlled by a wave generator through a signal amplifier. Suspension of density-matched Polystyrene particles of varying diameter from $1 \mu m-10 \mu m$ were used after the effect of surfactant was neutralized by adding NaCl to facilitate surface to surface sticking.

We investigated the sticking in such streaming flow around a shape with abruptly varying surface curvature (near the corners of the triangular obstacle). The qualitative corner streaming pattern in such a set-up as shown in figure ~\ref{fig sticking}(b) is characterized by two fountain vortices adjacent to the corners orthogonal to the axis of oscillation. Figure~\ref{fig sticking}(c) shows the time evolution of the particle cluster formation around one of the streaming corner. Particles that are trapped in the vortex recirculate around the side wall of the corner. In each recirculation particles interact with wall and cross streamlines. As we described the wall effect in Stokes flow in this thesis (chapter~\ref{chap 2}, \ref{chap 3}, \ref{chap 4}), we anticipate that this wall effect is attractive in the part of the recirculating trajectory where particle is leaving the wall. After a large number of repeated interactions, we observed particles getting stuck to the obstacle, indicating that an inherent asymmetry was present in the vortical structure that enabled net streamline crossing towards the wall. The first occurrence of particle sticking to the wall act as a secondary source of asymmetry that enhance further particle sticking around it promoting particle accumulation around the corner. The deposition structure of the particle is primarily governed by the geometry of the streaming vortices. 

Our experimental demonstration of particle sticking and deposition induced by corner streaming offers the concept of cluster formation with specific configuration leveraging the streaming pattern induced by corner streaming. While the underlying mechanism that drive the particle towards corner wall has two simultaneous hydrodynamic contributions from (1) inertial force component originating from the primary oscillatory flow and (2) non-inertial component arising from the steady Stokes flow driven by oscillatory shear layer. While the former contribution has been rigorously modeled in recent years \cite{agarwal2018inertial,agarwal2021unrecognized,agarwal2024density}, the fundamental for the later modeling lies in the non-inertial wall-effect in pure Stokes flow which we aim to describe in this thesis.

\chapter{Robustness and sensitivity of streaming flow patterns}\label{appen robustness and sensitivity of microstreaming}

Steady streaming flows result from rectification of periodic flow induced by an oscillating interface, and have been used extensively in microfluidic device design. Even simple objects executing simple motion can give rise to complex streaming patterns that sensitively depend on parameters, such as the prototypical case of a cylinder oscillating translationally. In this chapter \footnote[1]{This chapter is adapted from Das and Hilgenfeldt \cite{das2020robustness}}, we argue that this complexity and sensitivity is not typical for streaming flows encountered in microfluidic applications, chiefly relying on mixed-mode oscillations of deformable objects with pinned contact lines, such as bubbles or droplets. Experiments varying the modality as well as the dynamic boundary condition at the interface (no-stress, tangential stress continuous, no-slip) find an extremely robust vortex-pair streaming pattern independent of frequency or viscosity contrast. Comparing and contrasting the theoretical modeling of these flows with that of classical single-mode streaming patterns, we identify the conditions under which robust or sensitive streaming is expected. These results allow for the design of microfluidic devices guided by physical principles and tailored to applications that either require unvarying, robust flows or easily tunable changes in the streaming. 

\section{Introduction}\label{sec streaming theory}

\begin{figure}
    \centering
\includegraphics[width=\textwidth]{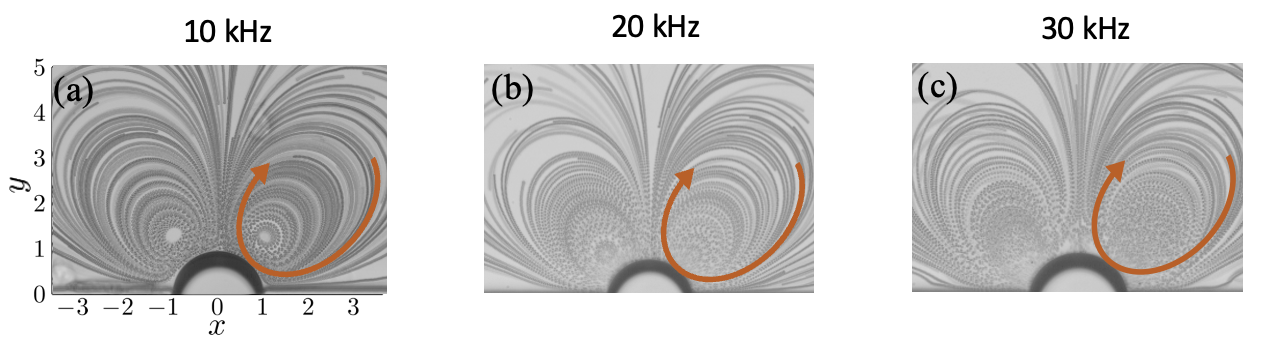}
    \caption{(a)\text{-}(c) Streak images of fountain bubble microstreaming pattern is robust against frequency. \cite{wang2013frequency}}
    \label{fig freq stream rallabandi}
\end{figure}

Classical studies of streaming around rigid boundaries — in particular, translationally oscillating cylinders \cite{riley2001steady,riley1998acoustic,elder1959cavitation,wang1968high} — reveal high sensitivity to frequency and boundary condition; the structure, magnitude may change dramatically with modest parameter variation. In contrast, microbubble or droplet-based streaming, which involves more complex interfacial motions, often displays remarkably stable, repeatable vortex patterns even as frequency vary \cite{wang2013frequency}. Oscillating microbubbles have long been recognized as effective streaming generators. When a gas bubble pinned to a microchannel wall is acoustically driven with a very low amplitude related with the radius, it undergoes mixed mode oscillation-- coupling of volume mode and shape mode oscillations and can be experimentally visualized at a very high sampling rate. The rectification of this leading order oscillatory flow will give us a second order steady flow which can be experimentally visualized at low sampling rate. The streak image of this steady flow exhibits fountain vortex pair and the pattern as can be shown in figure~\ref{fig freq stream rallabandi}, commonly known as microstreaming. 

Figure~\ref{fig freq stream rallabandi} also shows that this microstreaming structures are qualitatively identical and quantitatively similar against frequency changes which can be commonly expressed in terms of AC boundary layer thickness $\delta_{AC}/a=\sqrt{\nu/\omega a^2}$ and generally kept much less than one. Here $a$ is the bubble radius, $\nu$ is the surrounding fluid viscosity, and $\omega$ is the angular frequency of the bubble interface oscillation. While this robust nature of micro-bubble streaming is very good for applications of size based particle sorting \cite{thameem2017fast}, microfluidic mixing \cite{rallabandi2017analysis}, but there is an essential issue there, working with bubbles-- bubble interface is diffusively unstable whereas a solid or droplet interface can be found to be more stable while inducing different boundary conditions at the interface. This experimental observation motivates the central question of this chapter: what physical mechanisms determine the robustness or sensitivity of streaming flow patterns in microfluidic systems, and how are these mechanisms influenced by frequency, viscosity, and kinetic boundary conditions? 

Answering this requires a combination of experimental observation and analytical insight into the underlying flow structure. We compare and contrast streaming generated by translational oscillation of rigid bodies with streaming generated by oscillating fluid interfaces such as droplets. We focus particularly on the role of the DC boundary layer, a steady component near the oscillating surface that can mediate sensitivity, and we show how its presence or absence governs the qualitative stability of streaming patterns.

\section{Droplet streaming}\label{sec droplet streaming}

\subsection{Classical translational streaming patterns by varying control parameters} \label{subsec classical single mode streaming}

\begin{figure}[t]
    \centering
\includegraphics[width=0.75\textwidth]{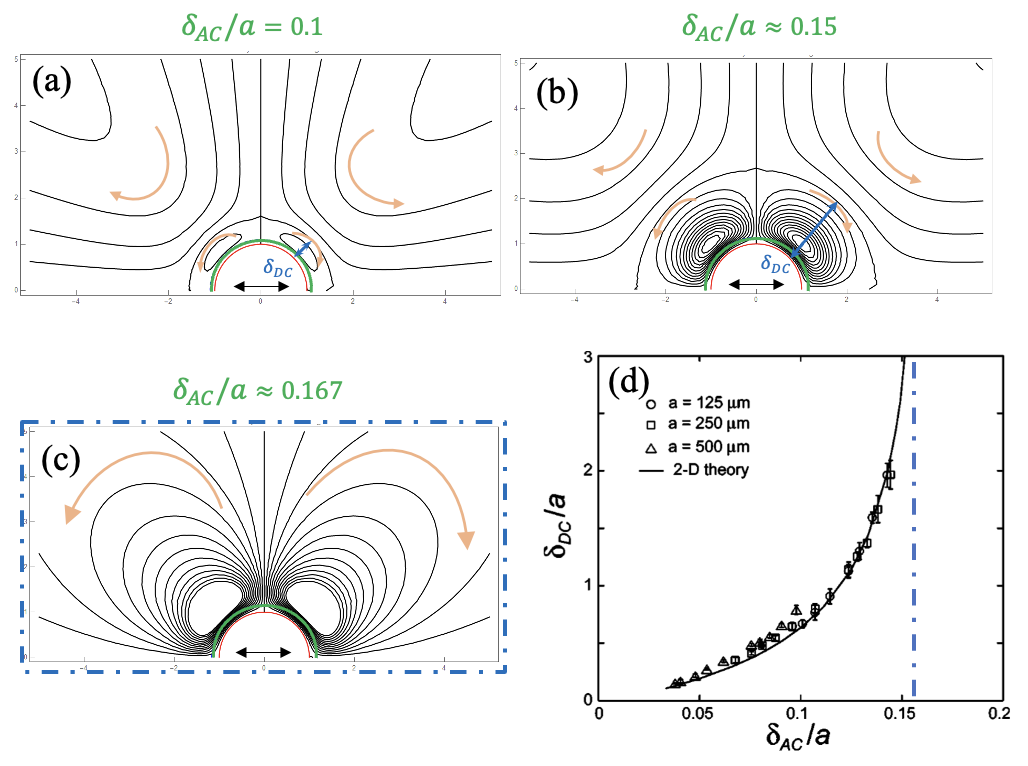}
    \caption{Single mode streaming from translational oscillation is sensitive to frequency. (a) $\delta_{AC}/a\approx 0.1$
in (a), $\approx0.15$ in (b), and $\approx0.167$ in (c) constructed from Bertelsen et al. \cite{bertelsen1973nonlinear}. (d) shows the diverging
nature of DC boundary layer with AC boundary layer from experiments with solid body streaming by Lutz et al. \cite{lutz2005microscopic}.}
    \label{fig transltioanl classical}
\end{figure}

The most classical examples of single-mode streaming, like in a transitional mode of oscillation in a solid body which is far more complicated than macro-bubble streaming. It is characterized by the existence of two different pair of vortices, one of which is a fountain pair of vortices at the neighborhood of the solid surface and another pair of anti-fountain vortices surrounding it. And the thickness of these fountain vortices is known as DC boundary layer thickness $\delta_{DC}$. This DC boundary layer changes with the change in AC boundary layer, $\delta_{AC}$ and it can be also experimentally found (figure~\ref{fig transltioanl classical}) that this DC boundary layer is a function of AC boundary layer. At some particular value of $\delta_{AC}$, this DC boundary layer actually diverges. This dependency highlights a fundamental characteristic of single translational mode rigid-body streaming: it is inherently sensitive to system parameters such as oscillation frequency and viscosity.

\begin{figure}[t]
    \centering
\includegraphics[width=\textwidth]{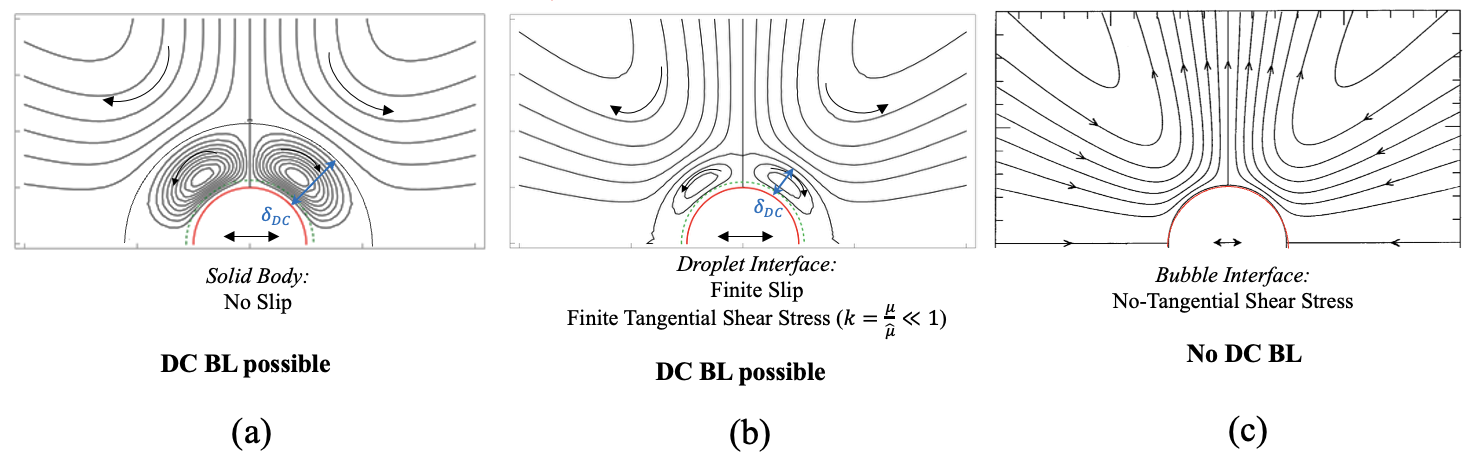}
    \caption{Single mode streaming from translational oscillation is sensitive to boundary conditions where (a) and (b) DC
boundary layer possible for no slip (a) \cite{elder1959cavitation} and finite shear plus slip boundary (b) constructed from Zhao et  al. \cite{zhao1999internal}, and as in (c) no DC boundary layer for bubble interface \cite{longuet1998viscous}}
    \label{fig streaming boundary}
\end{figure}

To further understand the role of boundary conditions in streaming, we replace the solid boundary (no slip) with a droplet interface which provides a finite slip and finite tensile shear stress at the interface. Under these conditions, droplet translational streaming can still exhibit a DC boundary layer similar to that seen in rigid-body streaming. However, as the boundary condition is relaxed further toward a no-shear stress condition, representative of a clean gas–liquid interface such as a microbubble, the DC boundary layer vanishes (see figure~\ref{fig streaming boundary}). This observation demonstrates that the sensitivity of streaming flows—specifically the existence and structure of the DC boundary layer—depends fundamentally on the nature of the boundary condition. These findings highlight the need to identify the physical mechanisms responsible for the formation of DC boundary layers as well as to observe whether they will occur in microbubble streaming while changing the boundary conditions.

\subsection{Experimental set-up for droplet microstreaming}

The experimental set-up used in this study shares key features with previously reported designs\cite{wang2013frequency,wang2012efficient} for bubble microstreaming. The microfluidic channel was fabricated from polydimethylsiloxane (PDMS) using standard photolithography techniques. A $100 \mu m$ thick layer of SU-8 photoresist was spin-coated onto a silicon wafer, and a chrome photomask was used to define the channel geometry through ultraviolet exposure and development. The resulting SU-8 master served as a mold for replica casting. PDMS prepolymer (Sylgard 184, Dow Corning) and curing agent were mixed at a $10:1$ mass ratio, thoroughly degassed to remove entrapped air, and poured over the SU-8 mold. After cross-linking at room temperature for $\sim 24$ hours, the fully cured PDMS was peeled from the mold and bonded to a flat PDMS layer using oxygen plasma treatment. The assembled microfluidic device was then irreversibly bonded to a glass slide following plasma activation. Ultrasound excitation of the bubble was provided by a piezoelectric transducer (Physik Instrumente, Germany; 1 mm thickness, 10 mm diameter) bonded to the glass substrate. Sinusoidal drive signals with frequencies $f=1-100$ kHz were generated by a function generator (7075, Hioki, Japan) and amplified (7500, Krohn-Hite, USA) before application to the transducer. Illumination for bright-field microscopy was achieved using a halogen light source (TH4-100, Olympus, USA). Top-view imaging was performed on an inverted microscope (Olympus IX71) equipped with 20x or 40x objectives and a high-speed camera (Phantom v310, Vision Research, USA). 

\begin{figure}[t]
    \centering
\includegraphics[width=\textwidth]{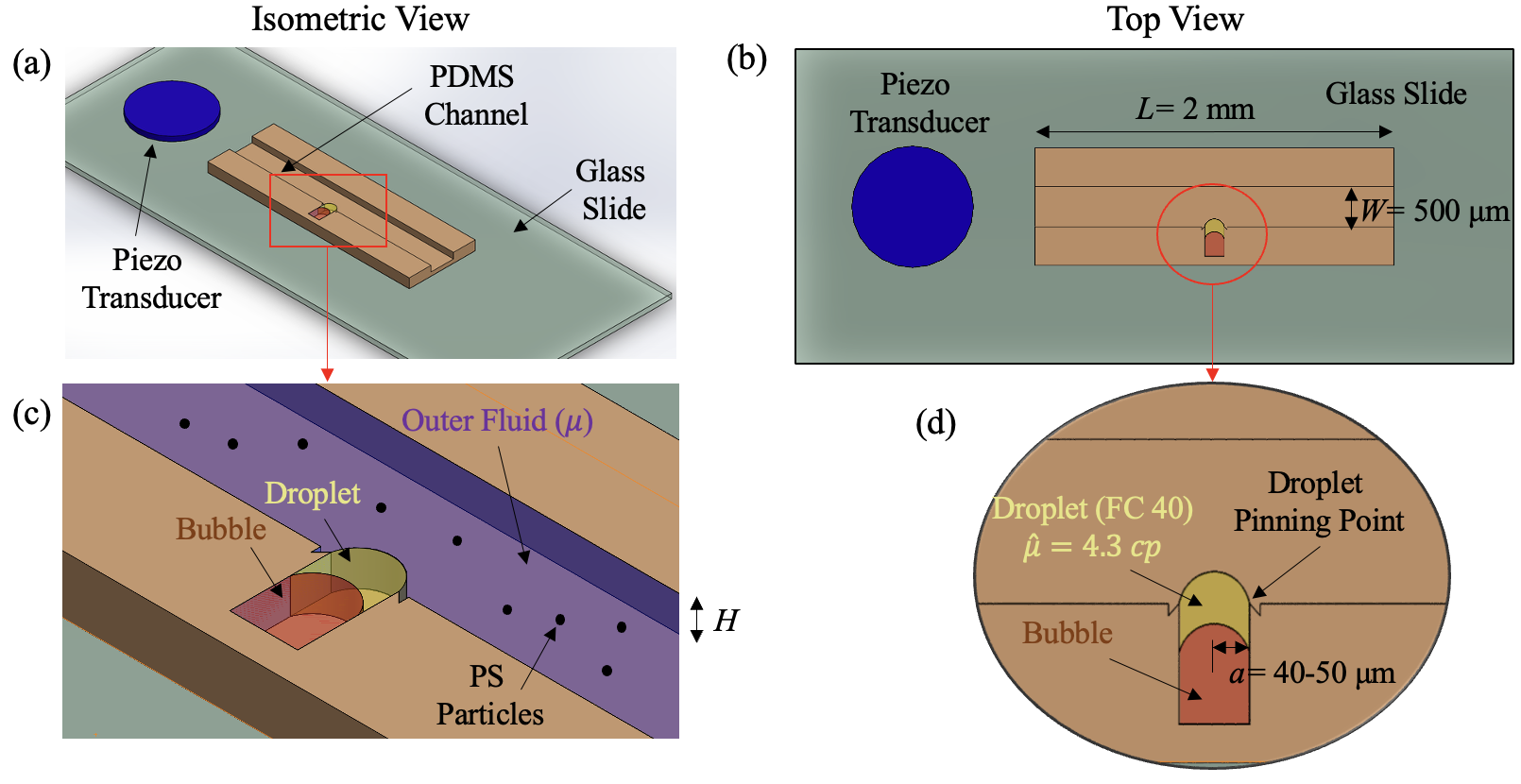}
    \caption{Experimental set-up of microstreaming induced by droplet oscillation driven by a bubble underneath.}
    \label{fig droplet streaming setup}
\end{figure}

In such a bubble microstreaming device, a trapped hemispherical bubble is oscillated in an ultrasound field and undergoes shape mode oscillation \cite{wang2012efficient}. Because bubbles are diffusively unstable — they tend to dissolve or migrate over time whereas it with a droplet of a similar scale can sustain oscillations with more stability. However since PDMS is oleophilic in nature, pinning a droplet in a PDMS channel and inducing shape mode oscillation needs some extra care. We work out with a simple but effective solution of pinning a hemispherical droplet in a microchannel. We prepare an unique experimental setup using PDMS microchannel, which has a side channel to trap microbubble. In order to modify the interfacial condition of this bubble and make the bubble stable as well, we layer it up with a FC-40 droplet (viscosity $\hat{\mu}=4.3\,$cp). The droplet thus provides a flexible tangential shear stress, which depends on the viscosity contrast $k$, which is the ratio of the viscosity of the outer fluid to the droplet $(k=\mu/\hat{\mu})$. Fabricating a careful discontinuity in the plane wall at the droplet contact points with notch-shaped protrusions, we make the droplet pinned on the side channel whereas the unpinned bubble underneath undergoes volume oscillation upon activating the ultrasound. Since the pinned droplet actually follows the bubble volume oscillation the resulting oscillation of the droplet interface is actually a mixed mode oscillation. We have used piezo transducer to oscillate the bubble, and also we have used neutrally buoyant polystyrene particles (by adding $23\% $\,glycerol by weight into distilled water) with radii $a_p=0.5\text{–}10\,\mu\text{m}$ to act as passive tracer particles. A very small amount of Tween 20 surfactant ($1\%$ w/w)  is added in order to prevent particle agglomeration. Figure~\ref{fig droplet streaming setup} illustrates the details of the experimental set-up. The experiments presented in this work extend microbubble streaming to droplet streaming resulting from the mixed-mode oscillations of the pinned droplet driven by the bubble breathing underneath.

\subsection{Result and discussion}

\begin{figure}
    \centering
\includegraphics[width=0.8\textwidth]{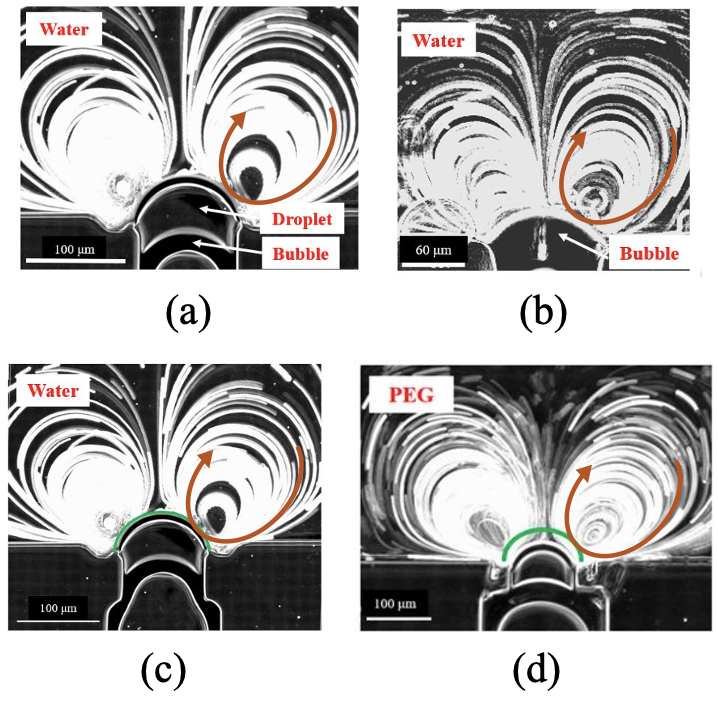}
    \caption{Robustness and qualitative similarity of fountain streaming patterns against tangential shear stress boundary condition as in (a) finite tangential stress on oscillating interface (droplet streaming) and (b) zero tangential shear stress on oscillating interface (bubble streaming). (c)--(d) Droplet streaming with (c) viscosity contrast $k\approx0.25$ for water outer fluid and (d) $k\approx7.32$ for PEG outer fluid. Green envelops in (c) and (d) depict the AC boundary thickness $\delta_{AC}\approx4\mu m$ and $\approx20\mu m$, respectively. Bubble was oscillated with an amplitude $\approx1.5 \mu m$ and droplet was oscillated with an amplitude $\approx2.5 \mu m$ by an ultrasound field with oscillation frequency, $f=11 kHz$ for (a)--(c) and, $=19 kHz$ for (d). }
    \label{fig droplet streaming}
\end{figure}

\begin{figure}
    \centering
\includegraphics[width=0.85\textwidth]{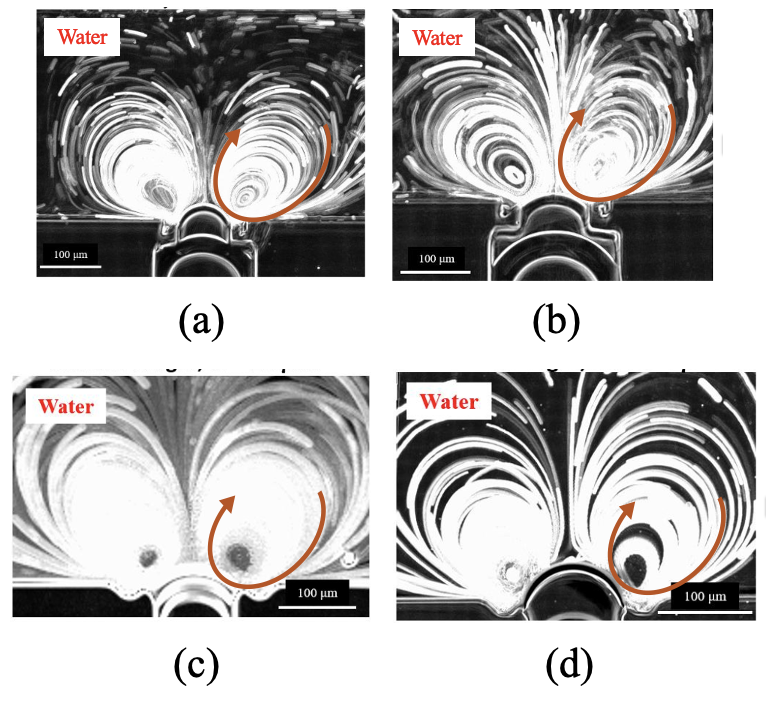}
    \caption{Qualitative similarity of fountain streaming patterns in droplet streaming driven by an ultrasound field with oscillation frequency (a) $f=19kHz$ and (b) $f=24 kHz$, and (c) channel height $\approx25\mu m$ and (d) channel height $\approx100\mu m$ keeping the frequency same at $f=11kHz$. }
    \label{fig droplet streaming 2}
\end{figure}

We investigated the qualitative sensitivity and robustness of bubble microstreaming presented by designing bubble driven droplet induced streaming. We here present the streak images of fountain streaming patterns in figures~\ref{fig droplet streaming} and \ref{fig droplet streaming 2}. In figure~\ref{fig droplet streaming} (a) and (b), we showed the qualitative comparison of droplet streaming with conventional bubble streaming. We found that the streaming patterns are qualitatively identical showing a pair of nice fountain vortex pattern. To quantify the strength of the streaming, we measured the steady tangential velocity field induced by both droplet and bubble interfaces. We confirmed that the maximum tangential streaming velocity, $u_{\theta,\max}$, follows the theoretical scaling prediction, $u_{\theta,\max}=\epsilon^2\omega a$, where $\epsilon=A/a\ll1$ is the dimensionless oscillation amplitude, $a$ is the droplet or bubble radius, $A$ is the amplitude of bubble or droplet interface oscillation, and $\omega$ is the angular frequency of oscillation.

We further investigated the role of finite tangential shear stress at the droplet interface by varying the viscosity of the outer fluid. We prepared a solution of Polyethylene Glycol (PEG) and distilled water to vary the viscosity of the suspension while keeping the density the same of distilled water. We show the resulted streaming in figure~\ref{fig droplet streaming}(c) and (d). By controlling the outer fluid viscosity, we effectively tune the interfacial shear condition as well as the thickness of the AC boundary layer — a parameter known to influence the DC boundary layer in classical single-mode solid body streaming (see Section~\ref{subsec classical single mode streaming}). We also vary the oscillation frequency to see any effect on the streaming pattern as we have presented in figures~\ref{fig droplet streaming 2}(a) and (b). We further our investigation to examine if there is any effect of channel height perpendicular to the plane, result of which is presented in figures~\ref{fig droplet streaming 2}(c) and (d). Across all parameter variations — interfacial shear condition, driving frequency, and geometric confinement — the qualitative characteristics of the microstreaming flow remained consistent. Specifically, no evidence of complex streaming patterns associated with a DC boundary layer was observed on the length scale of AC boundary layer.

\begin{figure}[t]
    \centering
\includegraphics[width=0.95\textwidth]{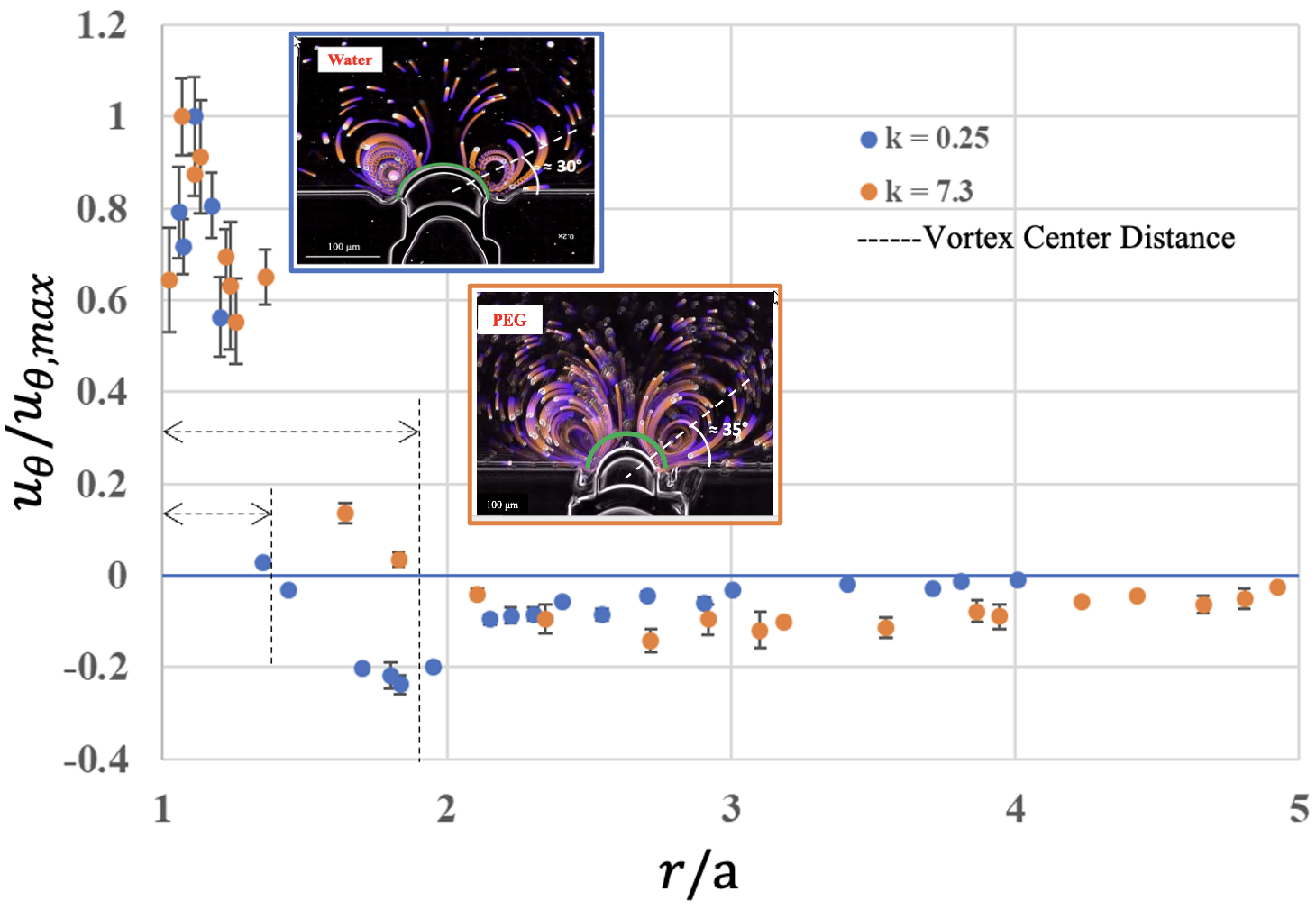}
    \caption{Tangential velocity profile for droplet streaming along different vortex center for viscosity contrasts $k=0.25$ and $k=7.3$ corresponding to $\delta=\delta_{AC}/a=0.08$ and $0.42$ respectively. }
    \label{fig streaming velocity}
\end{figure}

We quantify the tangential velocity along the vortex center as shown in figure~\ref{fig streaming velocity} for two different viscosity contrasts $k$. In both cases, the profile exhibits the expected characteristic structure: velocity rises from a finite slip velocity to a maximum close to the interface, drops to zero at the vortex center which can be pushed little far by an enhanced AC boundary layer, and later decays to zero far from the interface. Together, these results confirm that the qualitative structure of microstreaming flows is robust and insensitive to the range of control parameters, and that patterns involving DC boundary layers do not arise in this parameter regime. 

\section{Conclusion}

We examined the robustness and sensitivity of steady streaming patterns generated by oscillating interfaces, comparing classical single-mode rigid-body streaming with mixed-mode bubble microstreaming. While translational streaming around rigid bodies is highly sensitive to frequency, and boundary conditions due to the presence of a DC boundary layer, our results show that mixed-mode interfacial streaming produces a remarkably robust fountain-type vortex pair against the change in frequency or dynamic boundary conditions. The robust nature of macro stream brings up the curiosity to investigate the reason behind the complex streaming pattern-- a presence of DC boundary layer in single mode oscillation which leads us to investigate the streaming through the lens of theoretical analysis.

The analysis of robustness and sensitivity of streaming flow to frequency, viscosity, or boundary condition is critical for practical device operation as it determines how the flow responds to changes in system parameters and directly impacts the tunability and reliability of microfluidic performance. Our experimental work provides important design principles for bubble microfluidics. In particular, the bubble driven droplet microstreaming configuration, as we have demonstrated, exhibits high tolerance in microfluidic device design where bubbles can be shielded from diffusion.

\chapter{Characterizing the rich variety of streaming}\label{appen Rich Variety of Streaming Flow}

Recent studies find that inertial forces can play a key role in the positional deviation of suspended particles even on the small scale. Particularly in flows over an obstacle, the necessary inertial forces are almost always obtained from flow oscillation and, in parallel to causing the force on the particle, cause the 2nd order rectification of the Navier-Stokes equation which is commonly known as steady streaming. Morphological modifications and different boundary conditions of the obstacle have direct control on streaming quality and therefore on the inertial force induced from it. In this chapter \footnote[1]{This chapter is adapted from Das and Hilgenfeldt \cite{das2021characterizing}} we performed a thorough analytical investigation on the rich variety of steady streaming patterns originating from different interfacial conditions and its multiple mode of oscillation. We carefully examine the signature in the analytical formulation of streaming that arises DC boundary layer in single translational mode streaming (as defined in appendix~\ref{appen robustness and sensitivity of microstreaming}). We further derive the streaming patterns characterizing localized vortices or transport by coupling volume mode with translational mode of oscillation. 

\section{Introduction}

Periodic forcing of a viscous fluid from a relative oscillation of an interface, at an amplitude $A$ much smaller than its size $a$ ($\epsilon=A/a\ll1$), generates an oscillatory flow with a small Reynolds number $Re_{osc}=O(1)$ around it. In the absence of inertia this leading order oscillatory solution consists of irrotational free stream and doublet terms and an exponential decaying rotational term where all the vorticity is confined in the length scale of Stokes boundary layer (AC boundary layer, $\delta_{AC}$) near the interface \cite{davidson1971cavitation}. Perturbation of this leading order unsteady solution through the nonlinear inertial term in the Na-St equation gives the solution of to the second order. Time rectification of the second order flow, arising from the small but finite inertial contribution of the leading order oscillatory flow, is commonly know as the steady streaming whose flow velocity is $\epsilon$ times smaller than that of the oscillatory flow component $(Re_{s}=\epsilon Re_{osc})$.

The most classical mode of streaming is solid body streaming as shown in appendix~\ref{appen robustness and sensitivity of microstreaming} (see figure~\ref{fig transltioanl classical} and references \cite{lutz2006hydrodynamic,riley1966sphere,davidson1971cavitation}) where single mode translational oscillation of fluid around a no slip boundary generates a fountain vortex pair close to the boundary (as known as DC boundary layer) which is surrounded by a anti-fountain vortex pair. While the thickness of this localized vortices, $d_v$ depends on the AC boundary layer thickness $\delta_{AC}$ and diverges at some particular value of $\delta_{AC}$, it does not appear for translational oscillation around a no-shear interface (bubble interface, see figure~\ref{fig streaming boundary}). This brings the curiosity in understand the fundamental of the existence of  DC boundary layer in single mode translational streaming and its dependence on the AC boundary layer and boundary conditions of the interface.

Beyond boundary conditions, the modal composition of the oscillation plays a central role. Deformable interfaces support a spectrum of oscillation modes. A single mode of oscillation (e.g.\ 1-1 interaction in pure translational motion) produces one class of streaming response, while mixed-mode oscillations — involving, for example, coupling between volume and translational modes (0-1 interaction) — produce rich and more complex streaming patterns that combine localized vortices with directional flow. These mixed modes introduce additional control parameters, such as the amplitude ratio of the interacting modes and their relative phase, which influence the strength, orientation, and structure of the resulting streaming. 

\section{Analytical formulation of streaming in single mode translational oscillation}\label{sec single mode}

We here carefully construct the analytical formalism of steady streaming by mostly following the asymptotic matching procedure described in Longuet-Higgins \cite{longuet1998viscous}, Davidson and Riley \cite{davidson1971cavitation}, and Spelman and Lauga \cite{spelman2017arbitrary}. We start from the non-dimensional Na-St equation defined in axisymmetric spherical coordinate $(r,\theta)$

\begin{equation}\label{eq NaSt full}
    \frac{\partial (D^2 \psi)}{\partial t}
+ \frac{1}{r^2}
\left[
\frac{\partial (\psi, D^2 \psi)}{\partial (r,\mu)}
+ 2 \mathcal{L}\psi \, D^2 \psi
\right]
=
\frac{\delta^2}{2} D^4 \psi
\end{equation}
where $D^2 \equiv \frac{\partial^2}{\partial r^2}+ \frac{1 - \mu^2}{r^2} \frac{\partial^2}{\partial \mu^2}$, $\mathcal{L}\equiv\frac{\mu}{1 - \mu^2}\frac{\partial}{\partial r}
+ \frac{1}{r}\frac{\partial}{\partial \mu}$, $\mu=\cos \theta$, $\delta$ is the non-dimensionalized AC boundary layer thickness, and $\psi$ is the Eulerian streamfunction which takes the following asymptotic expression in the power of $\epsilon$ as 
\begin{equation}\label{eq psi asymp}
\psi=\epsilon\psi_1+\epsilon^2\psi_2+O(\epsilon^3)\,.
\end{equation}
All the parameters are non-dimensionalized by the reference lengthscale of object radius $a$ and the reference velocity scale $U_\infty$ which is the magnitude of far-field flow oscillation. In \eqref{eq psi asymp}, $\psi_1$ is the leading order solution of the oscillatory streamfunction that can be obtained by solving \eqref{eq NaSt full} to the leading $\epsilon$ order:
\begin{equation}\label{eq leading order NaSt}
    \frac{2}{\delta^2}
\frac{\partial (D^2 \psi_1)}{\partial t}
=
D^4 \psi_1\,.
\end{equation}
Solving this with appropriate boundary conditions for no-slip and no-penetration boundary condition at $r=1$ and uniform oscillatory stream at $r\rightarrow\infty$ (see \cite{longuet1998viscous}), we obtain the leading order oscillatory solution of $\psi_1$ for solid and bubble interface
\begin{equation}\label{eq psi1 solid}
    \psi_{1,solid}=\left[
\left( \frac{1}{2}r^2 - \frac{1}{2r} \right)
- \frac{3\delta}{2\sqrt{2}}
\left(1+\frac{\delta}{\sqrt{2}}\right)\frac{1}{r}
+ \frac{3\delta}{2\sqrt{2}}
\left(1+\frac{\delta}{\sqrt{2}r}\right)
e^{\frac{\sqrt{2}}{\delta}(1-r)}
\right]
(1-\mu^2)e^{it}
\end{equation}
\begin{equation}\label{eq psi1 bubble}
    \psi_{1,bubble}=\left[
\left(\frac{1}{2}r^{2}-\frac{1}{2r}\right)
-3\left(\frac{\delta^{2}}{2+3\sqrt{2}\,\delta}\right)
\left(1+\frac{\delta}{\sqrt{2}}\right)\frac{1}{r}
+3\left(\frac{\delta^{2}}{2+3\sqrt{2}\,\delta}\right)
\left(1+\frac{\delta}{\sqrt{2}\,r}\right)
e^{\frac{\sqrt{2}}{\delta}(1-r)}
\right](1-\mu^{2})e^{it}
\end{equation}
Perturbing this leading order solution into the non-linear term, \eqref{eq NaSt full} and time rectification of the expression becomes the following governing equation of the steady part of the solution of $\psi_2$.
\begin{equation}\label{eq Nast 2nd order}
    D^{4}\,\overline{\psi}_2=\frac{2}{\delta^{2} r^{2}}\left[\overline{\frac{\partial\!\left(\psi_{1},\,D^{2}\psi_{1}\right)}{\partial(r,\mu)}}\;+\;2\,\overline{\mathcal{L}\,\psi_{1}\,D^{2}\psi_{1}}\right]
\end{equation}
Solution of \eqref{eq Nast 2nd order} gives the inner solution in the region of $\delta$ where the all the vorticity $D^2\psi_1$ confines and decays to zero exponentially. Outside this boundary layer, the flow becomes irrotational and incompressible where \eqref{eq Nast 2nd order} essentially becomes Stokes equation:
\begin{equation}\label{eq Stokes}
    D^{4}\,\overline{\psi}_2=0
\end{equation}
By applying a rigorous asymptotic matching technique through the definition of inner boundary coordinate (see \cite{longuet1998viscous,davidson1971cavitation}) and applying proper dynamic boundary conditions at the interface, we obtain the universally valid solution of steady streaming driven by single mode translational oscillation around a solid boundary and bubble interface
\begin{equation}\label{eq solid streaming}
    \overline{\psi}_{2,solid}=\left[
\frac{45}{32}\left(1-\frac{1}{r^{2}}\right)
+ 9\delta
\left(
\frac{5}{8} e^{-\eta}\cos\eta
+ \frac{3}{8} e^{-\eta}\sin\eta
+ \frac{1}{4}\eta e^{-\eta}\sin\eta
+ \frac{1}{32} e^{-2\eta}
- \frac{21}{32}
\right)
\right]
\mu(1-\mu^{2})
+ \mathcal{O}(\delta^{2})
\end{equation}
\begin{equation}\label{eq bubble streaming}
    \overline{\psi}_{2,bubble}=\left[
-\frac{27}{40}\,\delta\left(1-\frac{1}{r^{2}}\right)
+ 9\delta^{2}
\left(
e^{-\eta}\cos\eta
+ \frac{1}{4}\eta e^{-\eta}\cos\eta
+ \frac{1}{4}\eta e^{-\eta}\sin\eta
- 1
\right)
\right]
\mu(1-\mu^{2})
+ \mathcal{O}(\delta^{3})
\end{equation}
to the leading orders of $\delta$ and thus only applicable for thin AC boundary layer $(\delta\ll1)$. It is to be noted that, a no-slip boundary condition produces a steady streaming flow that is $O(\delta)$ stronger than that generated by a no-shear interface. For a no-shear boundary, the leading-order time-dependent vorticity transport arises at $O(\delta^2)$ as can be seen in equation \eqref{eq psi1 bubble}, and this higher-order dependence on the shear layer thickness propagates directly into the steady streaming solution in \eqref{eq bubble streaming}. In contrast, for a no-slip boundary, vorticity is generated at lower order within the Stokes boundary layer (see equation \eqref{eq psi1 solid}), resulting in a stronger $O(\delta)$ contribution to the streaming field $\overline{\psi}_2$ as in \eqref{eq solid streaming}.

To meaningfully compare our theoretical predictions with experimentally observed particle drift, it is necessary to compute not only the Eulerian mean streamfunction, $\psi$, but also the Lagrangian streamfunction, $\Psi$. The Eulerian mean describes the time-averaged velocity field at fixed spatial locations, whereas the Lagrangian mean accounts for the actual trajectories of fluid particles. The difference between these two descriptions gives rise to the Stokes drift velocity represented by the Stokes drift streamfunction 
\begin{equation}\label{eq stokes drift}
    \psi_S=\frac{1}{r^{2}}
\int
\overline{
\frac{\partial \psi_{1}}{\partial r}
\frac{\partial \psi_{1}}{\partial \mu}
}\, dt.
\end{equation}
Altogether from equations \eqref{eq solid streaming}, \eqref{eq bubble streaming}, and \eqref{eq stokes drift}, the Lagrangian definition of the steady streaming $\overline{\Psi}_1+\psi_S$ for solid and bubble interface become
\begin{equation}\label{eq solid streaming lagrange}
    \overline{\Psi}_{2,solid}=\left[
\frac{45}{32}\left(1-\frac{1}{r^{2}}\right)+\frac{9}{8}\delta\frac{1}{r^2}
+ 9\delta
\left(
\frac{3}{8} e^{-\eta}(\cos\eta+\sin\eta)
+ \frac{5}{32} e^{-2\eta}
+\left(- \frac{21}{32}\right)
\right)
\right]
\mu(1-\mu^{2})
+ \mathcal{O}(\delta^{2})
\end{equation}
\begin{equation}\label{eq bubble streaming lagrange}
    \overline{\Psi}_{2,bubble}=\left[
-\frac{27}{40}\,\delta\left(1-\frac{1}{r^{2}}\right)+\frac{9}{4}\delta^2\frac{1}{r^2}
+ 9\delta^{2}
\left(\frac{3}{4}
e^{-\eta}\cos\eta +(-1)\right)\right]\mu(1-\mu^{2})
+ \mathcal{O}(\delta^{3})
\end{equation}

\subsection{Signature of DC boundary layer}

Equations \eqref{eq solid streaming lagrange} and \eqref{eq bubble streaming lagrange} consists of exponentially decaying terms coming from the inner solution of \eqref{eq Nast 2nd order} and confined to the shear boundary layer $\delta$. In the outer layer region the non-exponential terms dominate and determines the geometric patterns of the streaming flow. A closer look to the equations \eqref{eq solid streaming lagrange} and \eqref{eq bubble streaming lagrange} shows that the relative sign of the non-exponential terms controls the existence of DC boundary layer which is a zero radial velocity at a finite $r$. Therefore, from the asymptotic expressions of equations \eqref{eq solid streaming lagrange} and \eqref{eq bubble streaming lagrange} and replacing $r=1+d_v$ where $d_v$ is the thickness of DC boundary layer, we can write

\begin{equation}\label{eq DCBL solid}
\frac{45}{32}\left(1-\frac{1}{(1+d_{v,solid})^{2}}\right)+\frac{9}{8}\frac{\delta}{(1+d_{v,solid})^2}
- \frac{188}{32}\delta=0
\end{equation}
\begin{equation}\label{eq DCBL bubble}
-\frac{27}{40}\,\delta\left(1-\frac{1}{(1+d_{v,bubble})^{2}}\right)+\frac{9}{4}\frac{\delta^2}{(1+d_{v,bubble})^2}
- 9\delta^{2}=0
\end{equation}
One can easily find that equation \eqref{eq DCBL bubble} does not provide a real positive solution of $d_{v,bubble}$ wheres equation \eqref{eq DCBL solid} can be solved for $d_{v,solid}$ which is
\begin{equation}\label{eq DCBL}
    d_{{v,solid}}=\frac{-\sqrt{84 \delta^2-125 \delta+25}-21 \delta+5}{21 \delta-5}
\end{equation}

\begin{figure}
    \centering
\includegraphics[width=\textwidth]{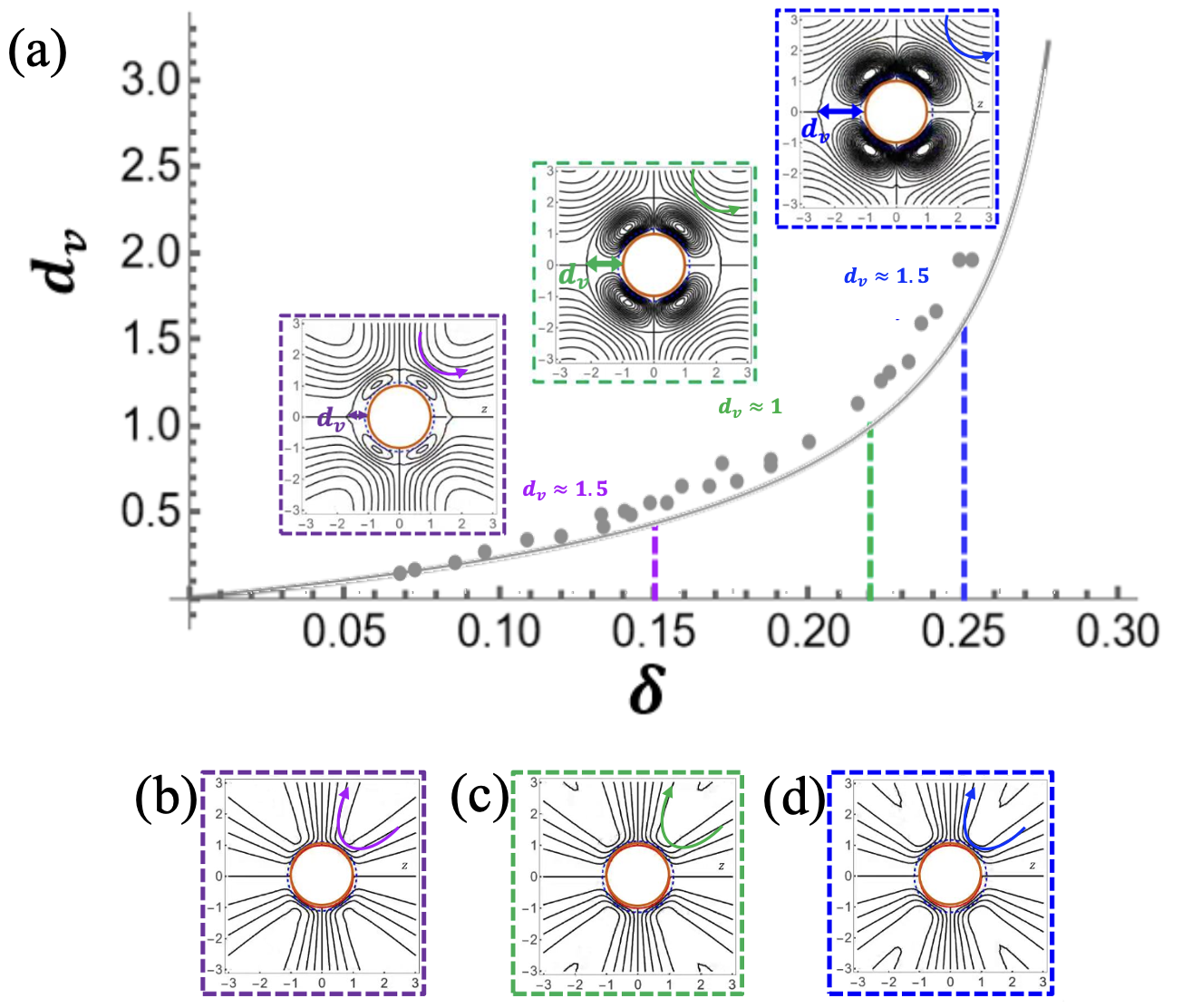}
    \caption{Qualitative variety of streaming from single mode (translational) oscillation around (a) no-slip (solid, equation~\eqref{eq solid streaming lagrange}) and (b)--(d) no-shear (bubble, equation~\eqref{eq bubble streaming lagrange}) interface. Solid body streaming exhibits localized recirculation (DC boundary layer) whose spatial extend $d_v$ depends on the shear layer (AC boundary layer, $\delta$) and diverges for $\delta\approx0.3$. Equation~\eqref{eq DCBL} is shown in solid line which is in good agreement with the experimental data (dots in (a)) from oscillating flow around a cylinder performed by Lutz et al. \cite{lutz2005microscopic}. In contrast, a no-shear boundary produces robust anti-fountain streaming patterns ((b)--(d)) and does not exhibit any evidence of flow recirculation on the scale of $\delta$. The dashed around the interface (in red) represents the AC layer $\delta$.}
    \label{fig DCBL}
\end{figure}

Along with the qualitative structure of solid body streaming patterns from equations~\eqref{eq solid streaming lagrange} which exhibit localized flow recirculation associated with a DC boundary layer, we quantify its variation by plotting $d_v$ as a function of the AC boundary layer thickness $\delta$ using the analytical expression from \eqref{eq DCBL}. The resulting curve is shown in figure~\ref{fig DCBL}(a) where it demonstrates  a good agreement with the experimental measurement reported by Lutz et  al. \cite{lutz2005microscopic} whereas figure~\ref{fig DCBL}(b)--(d) shows no evidence of DC boundary layer in the streaming structure around the no-shear interface across the entire range of $\delta$ examined.

\section{Mixed mode streaming in volume plus translational mode of oscillation}\label{sec Mixed mode}

\subsection{Analytical formulation}

We impose volume mode oscillation (``breathing") on the interface on top of translational mode to obtain a mixed mode streaming. This analogous the scenario when an object (bubble, droplet) undergoes small radial pulsations in addition to its lateral movement at the same frequency. Thus the surface of the object is defined by
\begin{equation}\label{eq radial profile}
r = R(t) = 1 - i \epsilon_0 e^{i(\omega t + \phi)}\,,
\end{equation}
where $\epsilon_0$ is the non-dimensional amplitude of volume mode which is kept very small $(\epsilon_0\ll1)$ and $\phi$ is the phase of the pulsation. The appropriate asymptotic series expansion of $\psi$:
\begin{equation}\label{eq psi asymptotic}
\psi =
\left( \epsilon_1 \psi_{10} + \epsilon_0 \psi_{01} \right)
+
\left( \epsilon_1^{2} \psi_{20}
+ \epsilon_1 \epsilon_0 \psi_{11}
+ \epsilon_0^{2} \psi_{02} \right)+HOT.
\end{equation}
$\epsilon_1$ is the non-dimensional amplitude of translational mode which we also assume much less than 1 (as in previous section~\ref{sec single mode}). Here,
$\psi_{10}$ is the leading-order solution for the translational mode only, $\psi_{01}$ is the leading-order solution for the volume (breathing) mode only, $\psi_{20}$ is the second-order solution for the translational mode only, $\psi_{02}$ is the second-order solution for the volume (breathing) mode only, $\psi_{11}$ is the second-order mixed solution corresponding to the interaction between the volume and translational modes. An interface in volume mode oscillation can not generate any second order effect due to the cancellation of Reynolds stress in that order \cite{longuet1998viscous}, $\psi_{02}$ vanishes. We already derived $\psi_{10}$ and $\psi_{20}$ in the previous section \ref{sec single mode} and $\psi_{20}$ takes the identical form of equation~\eqref{eq psi1 solid} or \eqref{eq psi1 bubble} depending on the boundary condition. $\psi_{01}$ is the streamfunction of the flow generated by the pure radial oscillation of the interface. Now from the equation of continuity,
\begin{equation}
\nabla \cdot \mathbf{u}
= \frac{1}{r^2} \frac{\partial}{\partial r} \left( r^2 u_r \right)
= 0.
\end{equation}
This gives the expression of the velocity in radial pure radial oscillation as
\begin{equation}\label{eq ur}
u_r = \frac{1}{r^2} f(t).
\end{equation}
At $r = R(t)$, the radial velocity must match the interface velocity because of no penetration boundary condition.
\begin{equation}\label{eq uR}
u_R = \dot{R}.
\end{equation}
Thus from \eqref{eq ur} and \eqref{eq uR},
\begin{equation}
f(t) = \dot{R} R^2.
\end{equation}
Altogether \eqref{eq ur} takes the following form,
\begin{equation}
u_r = -\epsilon' \frac{R^2}{r^2} e^{i(\omega t + \phi)}.
\end{equation}
Using the streamfunction relation $u_r = -\left( \frac{1}{r^2} \right) \frac{\partial \psi_v}{\partial \mu}$, 
\begin{equation}\label{eq psivvvv}
\psi_v = \epsilon' \mu R^2 e^{i(\omega t + \phi)}.
\end{equation}
Expanding $\psi_v$ after replacing $R = 1 - i\epsilon' e^{i(\omega t + \phi)}$ from \eqref{eq radial profile},
\begin{equation}
    \psi_v=\epsilon_0 \mu
\left(
1 + 2i\epsilon_0 e^{i(\omega t + \phi)}
- \epsilon_0^2 e^{2i(\omega t + \phi)}
\right)
e^{i(\omega t + \phi)}\,.
\end{equation}
Therefore, we obtain the expression of the leading order oscillatory flow for volume mode to lowest order,
\begin{equation}
\psi_{01}
= \mu e^{i(\omega t + \phi)}.
\end{equation}
Upon establishing this, the only term remains to evaluate in the asymptotic expression of the streamfunction equation~\eqref{eq psi asymptotic} is $\psi_{11}$ which is the second order streamfunction in volume+translation. Following the asymptotic matching framework of Longuet-Higgins~\cite{longuet1998viscous}, and using the same uniformly valid construction employed in Section~\ref{sec single mode} for single-mode streaming, we derive the mixed-mode steady streaming solution $\overline{\psi}=\epsilon_1^2\overline{\psi}_{20}+\epsilon_0\epsilon_1\overline{\psi}_{11}$ by enforcing appropriate boundary conditions systematically order by order. This procedure yields a total expression for the Lagrangian mixed-mode streaming streamfunction $\overline{\Psi}=\epsilon_1^2\overline{\Psi}_{20}+\epsilon_0\epsilon_1\overline{\Psi}_{11}$ by explicitly accounting for the Stokes drift contribution associated with volume–translation coupling, for both no-slip and no-shear boundary conditions.

\begin{equation}\label{eq Psi mixed solid}
\begin{aligned}
\overline{\Psi}_{solid}
&=
\epsilon_1^{2}
\Bigg[
\Bigg\{
\frac{45}{32}\left(1-\frac{1}{r^{2}}\right)
+ \frac{9}{4}\delta\frac{1}{2r^{2}} 
+ 9\delta
\left(
\frac{3}{8}e^{-\eta}(\cos\eta+\sin\eta)
+ \frac{5}{32}e^{-2\eta}
- \frac{21}{32}
\right)
\Bigg\}
\mu(1-\mu^{2})
\\
&\quad
+ \frac{\epsilon_0}{\epsilon_1}
\Bigg\{
-\frac{1}{2}
\left(
\frac{15}{4}r-\frac{13}{4r}-\frac{1}{2r^{4}}
\right)\sin\phi
+ \frac{3\delta}{8}
\left(
-17r+\frac{7}{r}-\frac{2}{r^{4}}
\right)(\cos\phi+\sin\phi) \\
&\qquad
+ \frac{3}{2}
\Big\{
3\delta e^{-\eta}
\big(\sin(\eta+\phi)+\cos(\eta+\phi)\big)
\Big\}
\Bigg\}
(1-\mu^{2})\Bigg]
+ \mathcal{O}(\delta^{2})
\end{aligned}
\end{equation}

\begin{equation}\label{eq Psi mixed bubble}
\begin{aligned}
\overline{\Psi}_{bubble}
&=
\epsilon_1^{2}
\Bigg[
\Bigg\{
-\frac{27}{40}\delta
\left(1-\frac{1}{r^{2}}\right)
+ \frac{9}{4}\delta^{2}\frac{1}{r^{2}} 
+ 9\delta^{2}
\left(
\frac{3}{4}e^{-\eta}\cos\eta - 1
\right)
\Bigg\}
\mu(1-\mu^{2})
\\
&\quad
+ \frac{\epsilon_0}{\epsilon_1}
\Bigg\{
-\frac{1}{2}
\left(
r-\frac{1}{2r}-\frac{1}{2r^{4}}
\right)\sin\phi
+ 3\delta\left(r-\frac{1}{r}\right)(\cos\phi+\sin\phi) \\
&\qquad
+ 9\delta^{2}
\left(
e^{-\eta}\cos(\eta+\phi)-\cos\phi
\right)
\Bigg\}
(1-\mu^{2})\Bigg]
+ \mathcal{O}(\delta^{3})
\end{aligned}
\end{equation}

\subsection{Results and discussion}

\begin{figure}[t]
    \centering
\includegraphics[width=0.9\textwidth]{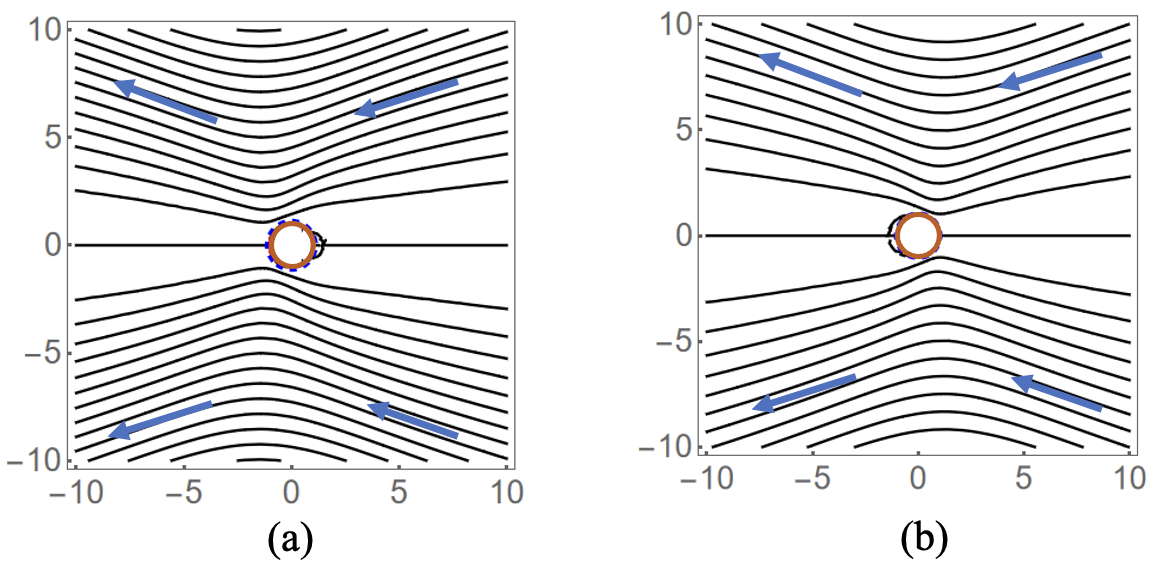}
    \caption{Time rectification of second order streaming solution in mixed mode oscillation (volume mode plus translation) can generate directional transport in the far field resulting from Stokeslet term in \eqref{eq Psi mixed solid} for solid in (a) and in \eqref{eq Psi mixed bubble} for bubble in (b). The interface is shown in red.}
    \label{fig Stokeslet}
\end{figure}

The resulting second-order streaming flow $\overline{\Psi}$ mixed mode of oscillation as in equations~\eqref{eq Psi mixed solid} and \eqref{eq Psi mixed bubble} contains interactions from the pure translational streaming (derived for single-mode streaming in previous section~\ref{sec single mode}) and the volume plus translational streaming . In the far field, the dominant contribution is a Stokeslet, implying the emergence of a net, directional transport flow as shown in figure~\ref{fig Stokeslet}. Mixed-mode oscillation therefore can provide a mechanism for generating throughput without imposing an external pressure gradient.

\begin{figure}[p]
    \centering
\includegraphics[width=0.9\textwidth]{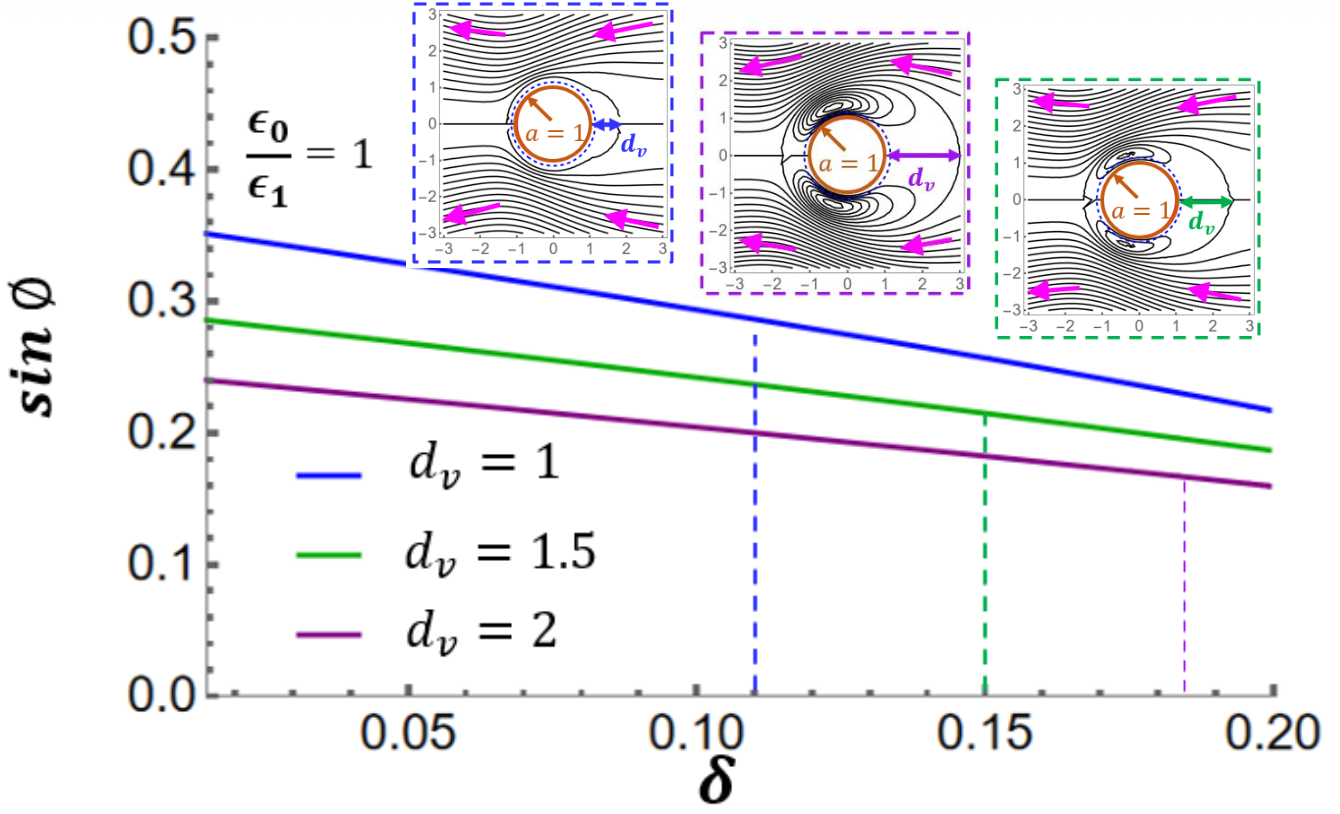}
    \caption{Mixed mode streaming patterns for solid body in volume and translational mode of oscillation where individual modes are of same strength $(\epsilon_0/\epsilon_1=1)$. Localized vortices can be formed in the upstream whose extend depends on the balance between $O(\delta^0)$ terms in \eqref{eq Psi mixed solid} and thus weakly depends on $\delta$. }
    \label{fig No slip 1}

    \centering
\includegraphics[width=0.9\textwidth]{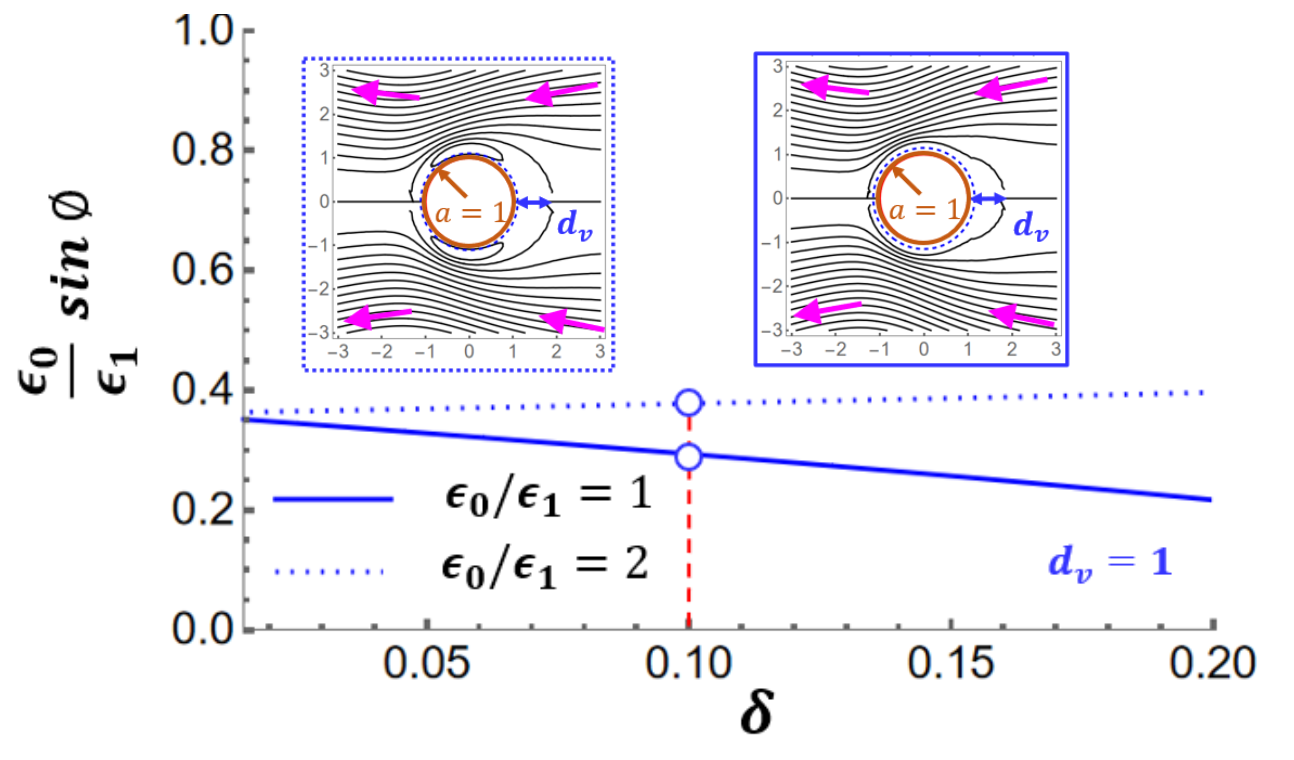}
    \caption{Mixed mode streaming patterns for solid body in volume and translational mode of oscillation creating object size localized vortex $d_v=1$ for $(\epsilon_0/\epsilon_1=1)$ and $(\epsilon_0/\epsilon_1=2)$. At very low $\delta$ (e.g.\ very high frequency), $\epsilon_0/\epsilon_1 \sin\phi$ becomes the only control parameter.}
    \label{fig No slip 2}
\end{figure}

A central question is whether this mixed-mode streaming can simultaneously generate localized recirculating vortices, analogous to those found in single-mode streaming. To address this, we investigate the dominant balance among the non-exponential terms and how this balance scales with the Stokes boundary layer $\delta$, the relative strength of the modes $\frac{\epsilon_0}{\epsilon_1}$, and their phase difference $\phi$.  As can be seen from equation~\eqref{eq Psi mixed solid} for a no-slip boundary, the dominant balance is from the similar leading outer terms which are $\mathcal{O}(1)$ terms arising from the pure translational interaction $(\overline{\Psi}_{20})$ and volume plus translational interaction $(\overline{\Psi}_{11})$. Because the competing terms are independent of $\delta$, localized vortex structures, which has a greater extent to the upstream, can arise without relying on $\delta$, and the phase requirement becomes effectively independent of $\delta$. This is shown in figure~\ref{fig No slip 1}. Moreover as shown in figure~\ref{fig No slip 2}, strong streaming persists even as $\delta \to 0$, so that in this regime the primary control reduces to the mixed-mode parameter $\epsilon_0/\epsilon_1$ together with $\sin\phi$. This identifies mixed-mode oscillation of a no-shear interface as a particularly effective route to achieving coexisting directional transport and localized vortices even under high-frequency (very low $\delta$) conditions.

\begin{figure}
    \centering
\includegraphics[width=0.9\textwidth]{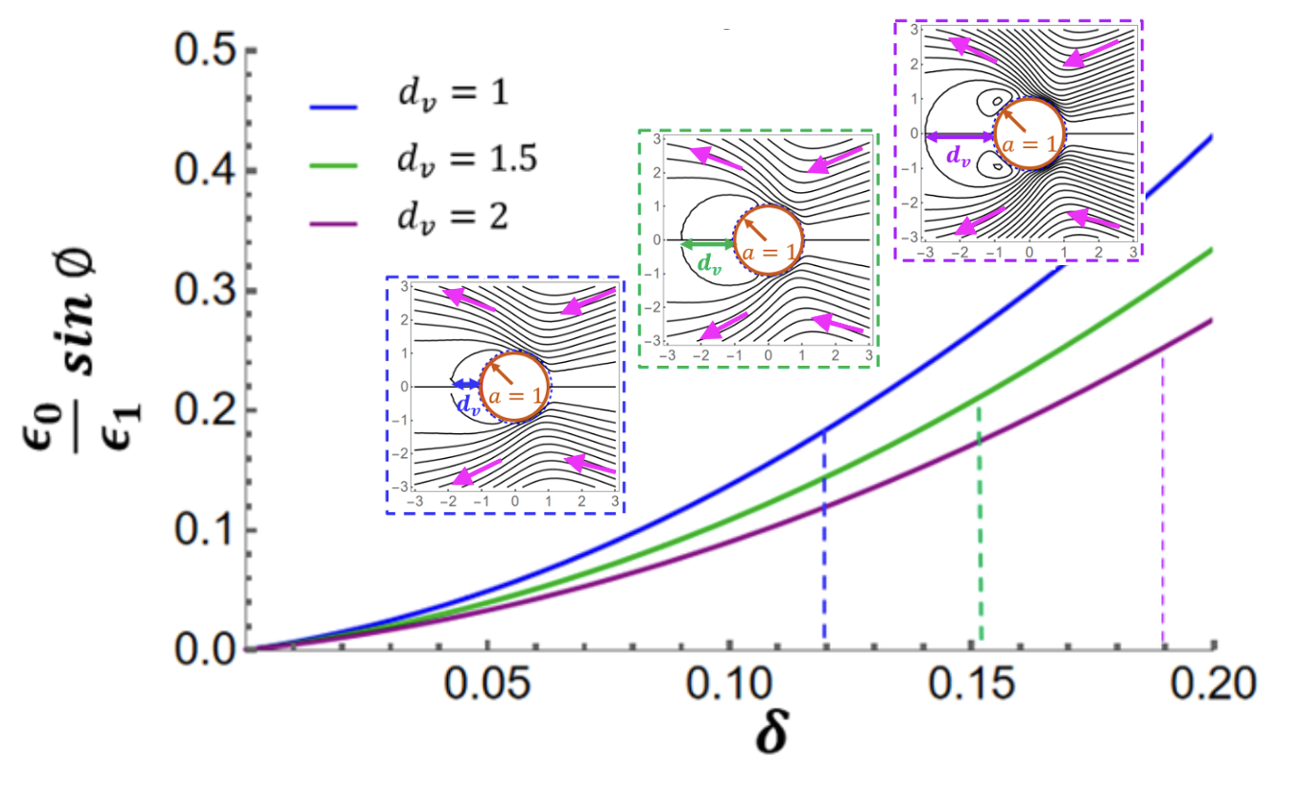}
    \caption{Mixed mode streaming patterns for no shear stress interface (bubble) undergoing volume and translational mode of oscillation where individual modes are of same strength $(\epsilon_0/\epsilon_1=1)$. Localized vortices can be formed in the downstream whose extend can be tuned by many combinations of $\delta$ and $\epsilon_0/\epsilon_1 \sin\phi$.}
    \label{fig No shear}
\end{figure}

In contrast, for a no-shear interface as in equation~\eqref{eq Psi mixed bubble}, the dominant balance is set by $\mathcal{O}(1)$ contributions from $\overline{\Psi}_{11}$, and $\mathcal{O}(\delta)$ contribution arising from pure translational interaction $\overline{\Psi}_{20}$. Because these terms are of different orders of $\delta$, a localized vortex pattern in the downstream of the interface can occur only when parameters are tuned such that the weaker $\mathcal{O}(\delta)$ contribution becomes dynamically relevant and can compete with the $\mathcal{O}(1)$ contribution. This can be achieved over a range of $\delta$ and $\epsilon_0/\epsilon_1\sin\phi$ combinations, including the small $\delta$ limit. However as shown in figure~\ref{fig No shear}, enforcing the necessary balance as $\delta \to 0$ requires $\epsilon_0/\epsilon_1\sin\phi \to 0$, which simultaneously drives the overall streaming amplitude toward zero. Thus, although localized vortices are mathematically permissible in this regime, they are generally accompanied by weak streaming intensity, limiting practical usefulness.

\section{Conclusion}

We have formulated the analytical description of steady streaming flow induced by oscillating an interface in pure translation (single mode) and in translation coupled with radial pulsation (mixed mode). An order by order asymptotic matching technique has been employed to obtain the universally valid expression of the time rectified second order solution of streaming flow. The dominant balance between the relative orders produces rich variations in the characteristic streaming geometry containing closed localized vortices and/or open streamlines generating directional transport. Mixed interaction between radial (volume) and translational oscillation produces these desired features of streaming for applications such as selective particle/cell trapping \cite{lutz2006hydrodynamic}, microorganism feeding \cite{gilpin2017vortex}, and directional transport \cite{bhosale2022multicurvature}. 

We impose radial pulsation (volume mode) on the interface to obtain a mixed mode oscillation on the interface that can be either a no-shear (bubble) or no slip interface. From practicality, pulsating no-slip interface can be on formed either by adding surfactant on a deformable interface (bubble or droplet) that will create an slip free layer on the interface through surfactant adsorption or by tuning the viscosity contrast in droplet streaming (see appendix~\ref{appen robustness and sensitivity of microstreaming}) to effectively increasing shear resistance and inducing large tangential stress on the droplet interface. Additionally, the analysis is performed by oscillating a fluid in translation around a fixed interface. An identical streaming response is obtained if, instead, the interface itself is oscillated with the same amplitude and frequency in a stationary fluid-- resolving the controversy raised by Doinikov et al. \cite{doinikov2019acoustic} confirming that steady streaming depends only on the relative motion between the fluid and the boundary, not on the specific reference frame used to describe the oscillation.

\chapter{Asymptotic behavior of hydrodynamic resistance coefficients for wall induced wall-normal motion}\label{appen res coeff asymp}
The scalar quantities $\mathcal{A}$, $\mathcal{B}$, $\mathcal{C}$ and $\mathcal{D}$ are dimensionless hydrodynamic resistances depending on $\Delta$. The analytical expressions for $\mathcal{A}$, $\mathcal{B}$, $\mathcal{C}$ and $\mathcal{D}$ are given (as infinite sums) in \citep{rallabandi2017hydrodynamic}. For large separations($\Delta \gg 1$), one obtains four hydrodynamic resistances at leading order:
\begin{equation}
    \mathcal{A}_{large} = 1+\frac{9}{8}\Delta^{-1}\,,\quad \mathcal{B}_{large} = \frac{15}{16}\Delta^{-1}\,,\quad \mathcal{C}_{large} = \frac{21}{32}\Delta^{-3} \,, \quad
    \mathcal{D}_{large} = \frac{1}{3}+\frac{3}{8}\Delta^{-1}
    \label{large ABCD}
\end{equation}
The ratios used in the equations for velocity corrections are, to leading order,
\begin{equation}
\frac{\mathcal{B}_{large}}{\mathcal{A}_{large}}\approx \frac{15}{16}\Delta^{-2}\,,\quad 
\frac{\mathcal{C}_{large}}{\mathcal{A}_{large}}\approx \frac{21}{32}\Delta^{-3}\,, \quad
\frac{\mathcal{D}_{large}}{\mathcal{A}_{large}}\approx \frac{1}{3}\,.
    \label{large delta ratios}
\end{equation}
For small separations($\Delta \ll 1$), one obtains four hydrodynamic resistances to leading order:
\begin{equation}
 \begin{aligned}
    \mathcal{A}_{small} &= \Delta^{-1}+\frac{1}{5}\log\Delta^{-1}+0.9713\,,\quad \mathcal{B}_{small} = \Delta^{-1}-\frac{4}{5}\log\Delta^{-1}+0.3070\,,\quad\\
    \mathcal{C}_{small} &= \Delta^{-1}-\frac{14}{5}\log\Delta^{-1}+3.7929 \,, \quad
    \mathcal{D}_{small} = \log\Delta^{-1}-0.9208\,.
    \label{small ABCD}
  \end{aligned}
\end{equation}
These results can be used to obtain the numerical prefactor of the wall-expansion limit equation \eqref{eq wall xpnsn}.

\chapter{Derivation of wall parallel correction factor $f(\Delta)$}\label{appen wall-parallel}
To get the shear coefficient of the particle parallel velocity $f(\Delta)$, we start with the form \cite{pasol2011motion}: 
\begin{equation}\label{pasoldelta}
    f(\Delta)=1-\left[1-a\log(1-\frac{1}{1+\Delta})-b_1\left(\frac{1}{1+\Delta}\right)-b_2\left(\frac{1}{1+\Delta}\right)^2 \right. \left. -b_3\left(\frac{1}{1+\Delta}\right)^3-b_4\left(\frac{1}{1+\Delta}\right)^4\right]^{-1}
\end{equation}
We take a series expansion of $f(\Delta)$ as $\Delta\to\infty$ to the order $\Delta^3$:
\begin{equation}
 \begin{aligned}
&f(\Delta)=-\frac{-a+b_1}{\Delta}-\frac{a^2+a(\frac{1}{2}-2b_1)-b_1+b_1^{2}+b_2}{\Delta^2}\\&-\frac{-a^3+b_1-2b_1^2+b_1^3+a^2(-1+3b_1)-2b_2+2b_1b_2-\frac{1}{3}a(1-9b_1+9b_1^2+6b_2)+b_3}{\Delta^3}+H.O.T.
 \end{aligned}
\end{equation}
According to \cite{goldman1967slow2} as $\Delta\to\infty$, $f(\Delta)\simeq 1-\frac{5}{16}\Delta^{-3}$, which means that the order $\Delta^{-1}$ and $\Delta^{-2}$ must vanish, resulting in 
\begin{equation}
a-b_1=0\, \qquad a^2+a(\frac{1}{2}-2b_1)-b_1+b_1^{2}+b_2=0\,,
\end{equation}
so that $b_1=a$ and $b_2=\frac{a}{2}$.
With $b_1$ and $b_2$ substituted, 
matching with 
Goldman's large $\Delta$ asymptotic expression obtains
\begin{equation}
b_3=\frac{5}{16}+\frac{a}{3}\,.
\label{b3}
\end{equation}
We then take a series expansion of $f(\Delta)$ as $\Delta\to 0$ to leading order:
\begin{equation}
f(\Delta)\approx 1+\frac{2}{3a+2(-1+b_3+b_4)+2a\log(\Delta)}
\end{equation}
According to \cite{williams1994particle} as $\Delta\to0$, $f(\Delta)\simeq 1-\frac{1}{0.66-0.269\log(\frac{\Delta}{1+\Delta})}$. By matching all the parameters, we obtain $a=0.269$ and
\begin{equation}
\frac{1}{2}\left(-3a-2(-1+b_3+b_4 \right)=0.66
\label{b4}
\end{equation}
Combined with equations \ref{b3} and \ref{b4}, this determines
$b_3=-0.223$ and $b_4=0.159$. All parameters of \eqref{pasoldelta} are now specified, and the result is 
Eq.~\eqref{f}.

\chapter{Extension of the symmetry breaking principle to three-cylinder geometry} \label{appen 3cyl}

Size-based particle path bifurcation is not hydrodynamically achievable with a single symmetric obstacle, where trajectory differences remain continuous but do not produce directional separation. In chapter~\ref{chap 6} we introduced modeling two tandem cylinders that effectively breaks the symmetry in in inclination required for net particle displacement. While the modeling approach is the systematic first step to understand the asymmetric mechanism of displacing particles in multi-cylinder situation used in a DLD set-up, We extend this analysis by introducing a third cylinder at different orientation. Introducing an additional obstacle creates the necessary configurational asymmetry even without necessitating an inclination with respect to the flow field that enables particles of different sizes to be steered along different sides of the obstacle sequence. This establishes the elementary framework of multi-obstacle devices in which symmetry-broken hydrodynamic interactions can generate true routing differences, providing a mechanistic basis for deterministic sorting in DLD-type architectures.

\begin{figure}[t]
    \centering
\includegraphics[width=\textwidth]{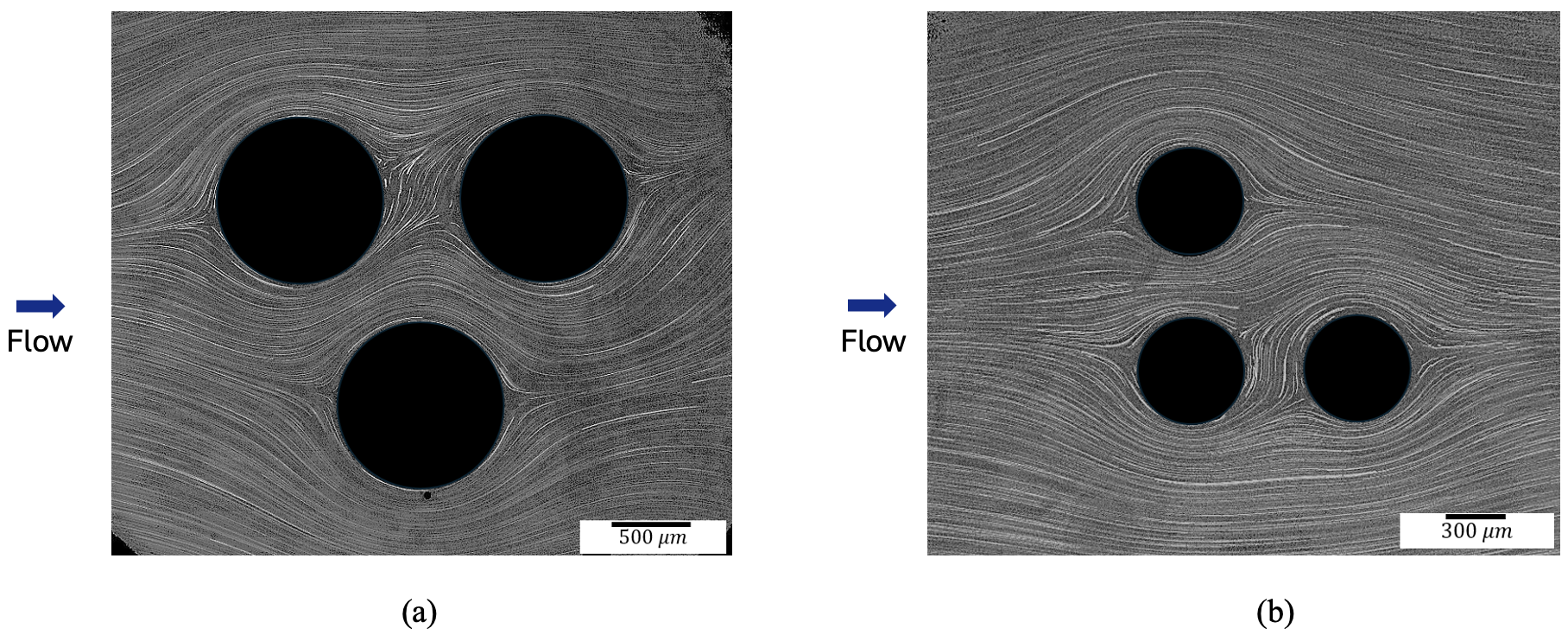}
    \caption{Streak images of flow field around three circular obstacles from particle tracking experiments. Flow rates are tuned to obtain a Reynolds number $Re\approx10^{-3}$ with respect to obstacle size ensuring operation within the Stokes flow regime.}
    \label{fig 3cyl}
\end{figure}

While there is no know analytical modeling technique for the Stokes flow stream function around three cylinder configuration, we experimentally obtained the flow field around three cylinders placed in different configurations close to each other. Figures~\ref{fig 3cyl}(a) and (b) show the streak images of particle tracking experiments with micron-sized particles (polystyrene particles with $1.1 \mu m$ in radius) for two such orientations. We used a 3D printed mold to manufacture PDMS channels with three circular cylinders. PDMS prepolymer (Sylgard 184, Dow Corning) and curing agent were mixed at a 10:1 mass ratio, thoroughly degassed to remove trapped air, and cast over the SU-8 mold. After curing at room temperature for approximately 24 hours, the PDMS layer was peeled from the mold and bonded to a flat PDMS layer using oxygen plasma treatment. The assembled microfluidic device was subsequently irreversibly bonded to a glass slide following plasma activation. Bright-field illumination was provided by a halogen light source (TH4-100, Olympus, USA). The particle suspension was prepared density-matched by adding a controlled amount of glycerol in water as described in the streaming experiments (see the details in \ref{appen robustness and sensitivity of microstreaming}). The flow rate is chosen such that it corresponds to a very small Reynolds number $Re \approx10^{-3}$ to validate the Stokes flow approximation.

The flow visualizations in figures~\ref{fig 3cyl}(a) and (b) show that introducing a third cylinder in the nearby region of two-cylinder pair fundamentally alters the local flow topology. The presence of a third cylinder can induce an asymmetric configuration similar to a inclined two-cylinder configuration. The separating streamlines originating from individual cylinders show inclination and shift along with skewed streamlines in between cylinders which would not be possible for a two-cylinder arrangement without a flow inclination. Such behavior highlights an important physical distinction between geometric symmetry of a single circular obstacle and the configurational asymmetry modifying streamline topology from the collective arrangement of multiple obstacles. This sets mechanism for generating directional displacement effects fundamentally similar to the effect of obstacle shape anisotropy that we have developed in this dissertation work. These observations therefore confirm that multi-cylinder architectures inherently produce asymmetry and non-intuitively complex flow structures important for particle routing across obstacle arrays in DLD devices and transport through heterogeneous structures in porous media.

\backmatter

\bibliographystyle{unsrt}
\bibliography{thesis}

\end{document}